\documentclass{aastex}
\usepackage{emulateapj5}
\pdfoutput=1

\usepackage{natbib}

\shortauthors{Chiboucas,Jacobs,Karachentsev,Tully}
\shorttitle{New Dwarf Galaxies in M81}

\begin{document}

\title{Confirmation of faint dwarf galaxies in the M81 Group}

\author{Kristin Chiboucas\altaffilmark{1}, Bradley A. Jacobs\altaffilmark{2}, 
R. Brent Tully\altaffilmark{2}, and Igor D. Karachentsev\altaffilmark{3}}
\email{kchibouc@gemini.edu, bjacobs@ifa.hawaii.edu, tully@ifa.hawaii.edu, ikar@luna.sao.ru} 

\altaffiltext{1}{Gemini Observatory,  670 N. A'ohoku Pl, Hilo, HI 96720, USA}
\altaffiltext{2}{Institute for Astronomy, University of Hawaii, 2680 Woodlawn Dr., Honolulu, HI 96821, USA}
\altaffiltext{3}{Special Astrophysical Observatory, Russian Academy of Sciences, Nizhnij Arkhyz, Karachai-Cherkessian Republic 369167, Russia}

\begin{abstract}
We have followed up on the results of a 65 square degree CFHT/MegaCam imaging survey of the 
nearby M81 Group searching for faint and ultra-faint dwarf galaxies.  The original survey 
turned up 22 faint candidate dwarf members.  Based on two-color HST ACS/WFC and WFPC2 
photometry, we now confirm 14 of these as dwarf galaxy members of the group. 
Distances and stellar population characteristics are discussed for each.  To a 
completeness limit of  $M_{r^{\prime}} = -10$, we find a galaxy luminosity function slope 
of $-1.27\pm0.04$ for the M81 group.  In this region, there are now 36 M81 group members known, 
including 4 blue compact dwarfs, 8 other late types including the interacting giants M81, 
NGC 3077, and M82, 19 early type dwarfs, and at least 5 potential tidal dwarf galaxies.
We find that the dSph galaxies in M81 appear to lie in a flattened distribution, similar to 
that found for the Milky Way and M31.  One of the newly discovered dSph galaxies has 
properties similar to the ultra-faint dwarfs being found in the Local Group with a size 
R$_e \sim 100$ pc and total magnitude estimates M$_{r^{\prime}} = -6.8$ and M$_I \sim -9.1$.

\end{abstract}


\section{Introduction}\label{intro}

As is often stated near the top of any paper on dwarf galaxies, these small, faint galaxies, following
a continually rising galaxy luminosity function to faint magnitudes, are the most
abundant galaxy type in the universe.  However, at the faintest magnitudes, these objects are
only readily visible in the local universe.  Comprehensive searches for such faint objects
are ongoing in the Local Group \citep[see][]{mcc}, with dwarfs as faint as M$_V \sim -2$ being 
discovered.  Such extremely faint galaxies, with magnitudes fainter than M$_V \sim -8.0$ \citep{brown12}, 
have been termed ultra-faint
dwarf galaxies (UFD) and can have magnitudes and surface brightnesses significantly lower than
the previously established classical dwarfs.  These new Local Group members are currently turning up at such 
high rates with wide field surveys 
such as SDSS that the number of known Local Group dwarfs has more than doubled in the past 15 years.
However, we are far from achieving a complete census of galaxies in any environment, and
even further from obtaining complete samples over a range of environments.  Nearby,
deep and very wide field coverage is required to identify faint Andromeda satellites while full sky coverage,
which includes the zone of obscuration, is required for the Milky Way. More distant dwarf 
galaxy populations quickly suffer from incompleteness at faint magnitudes and low surface brightnesses.

Motivations for studying dwarf populations go well beyond these galaxies simply being the most plentiful
in the universe.  The lowest mass dwarfs may have been the first galaxies to form and 
may constitute relic building blocks within the hierarchical structure formation paradigm \citep{ric09}.
The faintest dwarf spheroidals may contain pure old stellar populations \citep{dolphin05,dJ08,brown12}.  
Measurements of the stellar velocity dispersions suggest these may be the most dark matter 
dominated objects in the universe \citep{simon07,mattw12}. 
The relative abundance of these low mass dwarfs with respect to more massive galaxies directly relates 
to cosmological structure formation models while abundance as a function
of environment probes environment-dependent processes such as stripping, harassment, strangulation,
merging, and pressure containment.  Studying how star formation proceeds in galaxies with such small 
sizes and low gas masses and densities enables an understanding of star formation efficiencies and 
regulation processes.  
Investigating the different dwarf classes, namely dwarf spheroidals (dSph) and dwarf irregulars (dI) 
may provide insight into when, how, and why star formation shuts
off, along with an understanding of the relationship between early and late type galaxies.

In addition, dwarf galaxies impose strong constraints on cosmological models. Although the $\Lambda$CMD
model satisfies most criteria on large scales, it continues to fail at faint galaxy magnitudes,
where predictions of the galaxy abundance as a function of mass and of the internal structure of these
small objects do not agree with observations of the galaxy luminosity function or galaxy physical 
structure \citep{moore99,klypin99}.  In low density environments, this discrepancy is even larger.
Simulations predict relatively
larger fractions of small, faint galaxies, while observationally a flatter faint-end luminosity
function slope is usually found \citep{tsb05, bls05}.  For 
the faintest dwarfs, explanations abound, with various mechanisms such as reionization and feedback
used to explain a reduction of baryonic matter residing in the dark matter halos and limiting star
formation in low mass halos \citep{tw96, larson74, ds86}. At slightly 
brighter magnitudes, it becomes harder to justify differences from theory \citep{bk12}, although
\citet{wang12} find some differences may be accounted for by assuming a slightly lower halo mass for
the host galaxy.  

Thus, dwarf galaxy observations inform cosmological formation and galaxy evolution theories.
Studying dwarf properties - sizes and structures, colors, metallicities, stellar density and
surface brightness, number counts as a function of magnitude, star formation rates and histories,
gas content, stellar and halo masses, and environmental differences - along with the relations between
luminosity, size, surface brightness, metallicity, velocity dispersion, and dynamical mass are
critical for constraining models of galaxy formation and evolution.

In an effort to obtain a sample of dwarf galaxies complete to faint limiting 
magnitudes in an environment outside the Local Group, we undertook a large survey to search for 
dwarf galaxies in the nearby M81 Group.  At a distance of 3.7 Mpc this group is near enough to identify 
faint and small galaxies through just visible resolved stellar populations while at the same
time is distant enough that obtaining complete areal coverage is feasible.  

The M81 Group consists of a core dominated by M81.  It is inferred from the distribution and velocities
of satellites that M81 and its immediate satellites lie in a halo characterized by a current second
turnaround radius (closely related to the virial radius) of $\sim230$ kpc or 3.65 degrees in projection
\citep{tully10}.   A secondary clump of galaxies dominated by NGC 2403 that lies $\sim1$ Mpc to the foreground
of M81 is falling away from us toward the M81 halo.  The smaller galaxies NGC 4236 and VII Zw 403 that lie 
$\sim1$ Mpc behind M81 are also falling toward M81.  
\citet{K2002} find the total mass of the group within the
zero velocity surface to be $1.6 \times 10^{12}$ M$_{\odot}$.  
This mass estimate is uncertain.  A numerical action orbit reconstruction model gives a mass
of $1.0 \times 10^{13}$ M$_{\odot}$ \citep{bj12}.
In the immediate vicinity of M81, the 3 major galaxies M81, M82, and NGC 3077
have undergone significant gravitational encounters within the past 300 Myr producing, through tidal 
disruption, HI bridges connecting all three galaxies, and possibly inducing elevated
star formation rates in these late type galaxies (in particular affecting NGC 3077 and the strongly 
starbursting galaxy M82) \citep{yun94}.  
Within this tidal debris reside a number of suspected tidal dwarfs.  The population around
M81 also contains a large fraction of late galaxy types.  Compared with the Local Group, to M$_V < -10$,
the late type fraction is higher: the Milky Way and Andromeda have $\sim$ 40\% and 30\% late types, respectively, while for
M81 it is 55\%.  All of this suggests that this system is dynamically less evolved than the Local Group.

Using MegaCam on CFHT, a total of 65 square degrees centered around M81 were
imaged, extending beyond the second turnaround radius of the group.  
Within this region are 22 previously known group members. In addition to recovering all the previously 
known systems, our MegaCam imaging survey turned up a total of 21 new candidate dwarf
galaxy group members and one other previously suspected member \citep[][hereafter Paper 1]{Ch1}, 
potentially doubling the number of galaxies in the region around M81.  
We have since followed up with HST WFPC2 and ACS/WFC two-band imaging
to generate color-magnitude diagrams (CMDs) from which to establish membership based on tip of the red giant 
branch (TRGB) distances and investigate the stellar populations of these
faint and small dwarf galaxies.  The results of this follow-up work are presented here.

In Section \ref{obs} we describe our original survey, the new HST observations, and the stellar photometry 
leading to TRGB distance measurements.  Results, including confirmation of new M81 Group dwarf galaxy
members are presented in Section \ref{members}.  Structural and photometric properties for these 
new members can be found in Sections \ref{structpar} - \ref{magpar}. 
The stellar populations of each new member are discussed in Section \ref{stellpop}.  A discussion of these new
detections is found in Section \ref{disc} with summary and conclusions in Section \ref{conc}.

\section{Observations and Data Reduction}\label{obs}

\subsection{Initial Imaging Survey}\label{cfhtsurvey}

The initial CFHT/MegaCam $r^{\prime}-$band imaging survey was carried out over
three semesters 2005B-2006B.  A contiguous 65 square degree region centered on 
M81 was observed in a mosaic pattern which resulted in each region being imaged 
twice with combined total integration times of 1096s.  Full details
of this survey are presented in \citet{Ch1}.   
The M81 Group has a 230 kpc second turnaround radius of spherical collapse \citep{tully10}
which, at the group distance of 3.7 Mpc, corresponds to 3.65 degrees on the sky. 
The MegaCam survey
therefore extends beyond this radius and provides nearly complete areal coverage of
this nearby galaxy group.  The survey limiting magnitude for resolved stars
was $r^{\prime} \sim 25.0-25.5$, dependent on seeing variations throughout the
survey.  Given the high level of extinction in this part of the sky \citep[][see Table \ref{tabX}]{sfd98,schlafly},
we found we could resolve stars down to $M_{r^{\prime}} \sim -2.9\pm 0.5$.
The tip of the red giant branch in r$^{\prime}$ is $-3.1\pm0.1$.  
Stars at the tip of the red giant branch are just at the edge of detection while
young main sequence, red supergiants, and intermediate aged AGB stars are easily resolved
over the low surface brightness diffuse late type main sequence population light.  

Exploiting this, we searched the imaging for previously undetected faint 
dwarf galaxies. Our expectation was that any dwarf galaxy, no matter how 
faint or diffuse, should have been detectable, as long as it had built up
a significant enough red giant branch.  In retrospect from results discussed further along,
this requirement of a well developed red giant branch is satisfied for galaxies brighter
than M$_{r^{\prime}} \sim -9$.  Using two methods (a two-point correlation routine and 
visual inspection by two of the authors) to search 
for concentrations of resolved stars which would indicate potential dwarf galaxies,
we identified a total of 22 candidate dwarf group members. This conceivably
doubles the number of known group members, and includes 3 blue compact 
dwarf (BCD), a tidal dwarf, and a number of dI and dSph candidates.  Three of the
candidates were previously known galaxies. However, d0958+66 (KUG 0945+670) and d1012+64 (UGC 5497) 
had no velocity or distance measures that would associate them 
with the M81 Group.  These two galaxies are compact star forming objects which
were found to resolve into individual stars in our MegaCam imaging and thus considered
strong candidates for group membership.  A third object, d0959+68, was originally noted
by \citet{durrell}.  In a survey for resolved red giant stars near the core of the 
M81 Group, they discovered a concentration of blue stars which they suggested 
was a likely tidal dwarf within the group.  

Artificial galaxies having a wide
range of sizes, total magnitudes, and stellar surface densities were inserted in the 
imaging to determine survey completeness and recovered using the same methods described
above.  Results indicated that limitations did exist for this survey,
due in part to gaps between the CCDs and regions with very high extinction, and 
due to what are effectively total magnitude and surface brightness limits.  
Out to a distance beyond the
2nd turnaround radius (roughly equivalent to the Virial radius), \citet{Ch1} contend 
that the survey is $\sim90$\% complete to $M_{r^{\prime}} = -9.8$.   
 
\subsection{HST Observations}\label{obssuvey}

In order to securely establish group memberships, we need to measure distances to these
candidate galaxies.  Previous work has shown that it is possible 
to measure tip of the red giant branch distances for resolved stellar populations in
dwarf galaxies out to $\sim10$ Mpc \citep[see e.g.][]{mmrt06,luca07,brad09}.
The M81 group is nearby at $\sim 3.7$ Mpc, a distance where the brighter members of galaxy 
stellar populations, including red giant stars, are resolved in single orbit space based-imaging.  
We therefore make use of the high resolution of HST to 
obtain two color photometry of resolved stars in these candidates. From the resulting
CMDs, we can measure the tip of the red giant branch and use this to measure distances and 
confirm group membership.  With the ACS on HST, we can reach $V = 28.0$ and $I = 27.0$
with integrations taken together in a single orbit.  At the distance of the M81 group,
the TRGB is found at $I \sim 23.8$, assuming $M_{I,TRGB} = -4.04\pm0.12$ for
metal poor populations \citep{bell01}.   ACS imaging probes deeper than 3 mag below the
TRGB assuring secure determations of the TRGB and, with uncertainties in the tip
magnitude typically 0.1 mag, distance determinations accurate to 200 kpc. 
In addition, with effective radii $\le 30$ arcsec, the $202\times202$ arcsec field of view of 
ACS provides essentially full coverage for these candidates.
With WFPC2, the data only reach $\sim 1.5$ mag below the TRGB, uncertainties in the distance 
measurement are larger, and bigger galaxies extend beyond the field of view.

\begin{figure}[t]
\begin{centering}
\includegraphics[angle=0,totalheight=5.in]{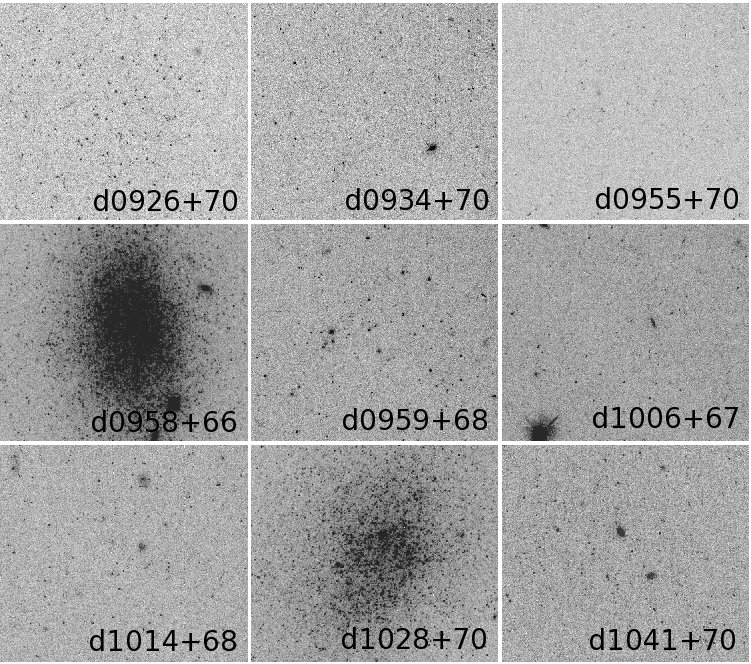}
\caption{$F814W$ thumbnails for 9 candidates observed with WFPC2.  Images
are $34 \times 39$ arcsec on each side.
\label{wfpc2bw1}}
\end{centering}
\end{figure}

\begin{figure}
\begin{centering}
\includegraphics[angle=0,totalheight=3.34in]{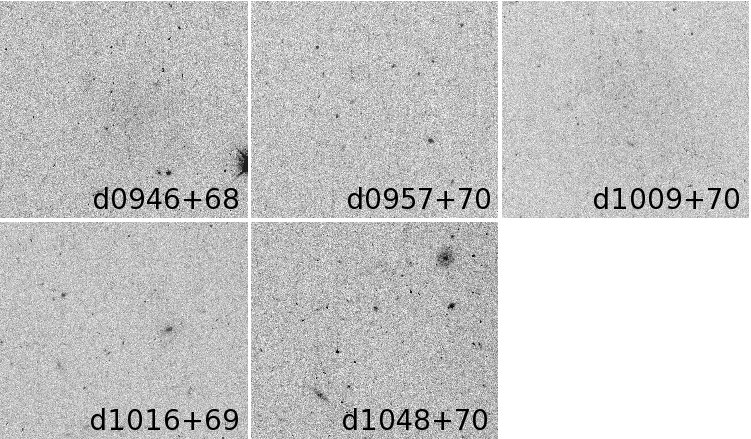}
\caption{$F814W$ thumbnails for another 5 candidates observed with WFPC2.  Images
are $34 \times 39$ arcsec on each side.
\label{wfpc2bw2}}
\end{centering}
\end{figure}

\begin{figure}[t]
\begin{centering}
\includegraphics[angle=0,totalheight=2.in]{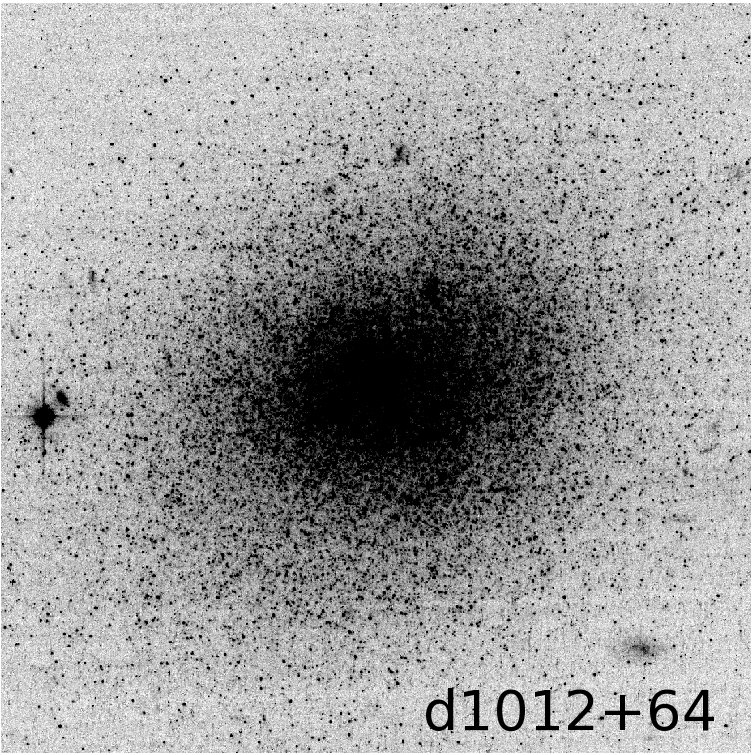}
\includegraphics[angle=0,totalheight=2.in]{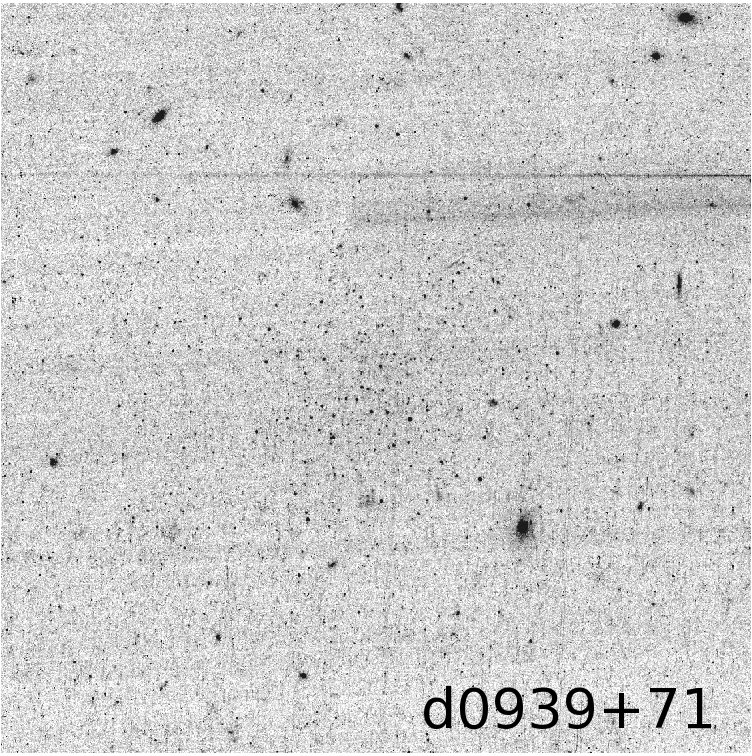}
\includegraphics[angle=0,totalheight=2.in]{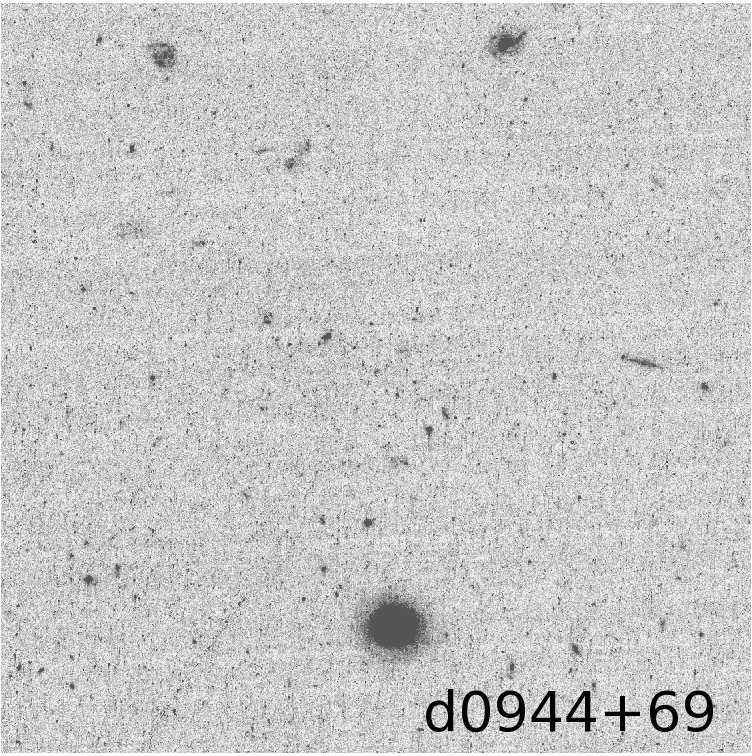}
\includegraphics[angle=0,totalheight=2.in]{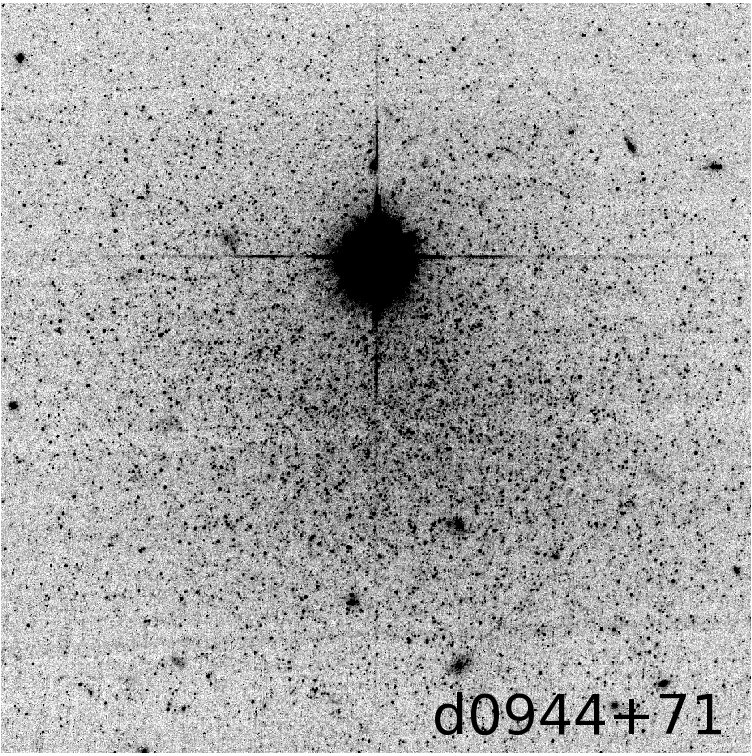}
\includegraphics[angle=0,totalheight=2.in]{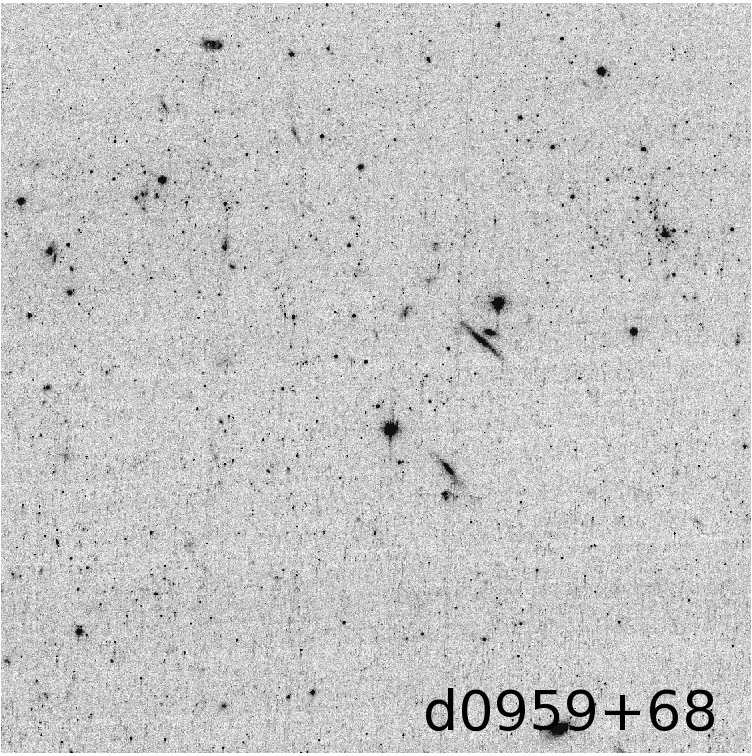}
\includegraphics[angle=0,totalheight=2.in]{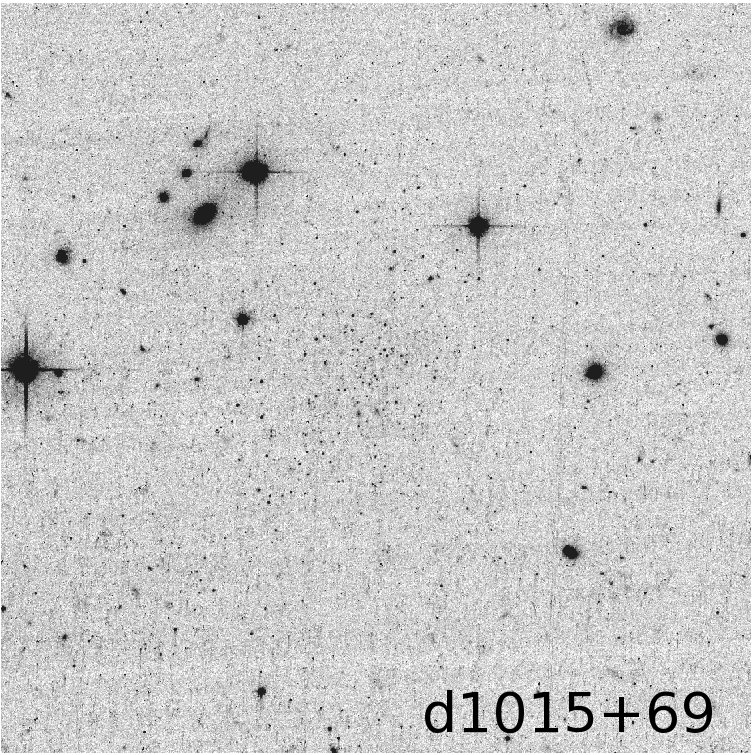}
\includegraphics[angle=0,totalheight=2.in]{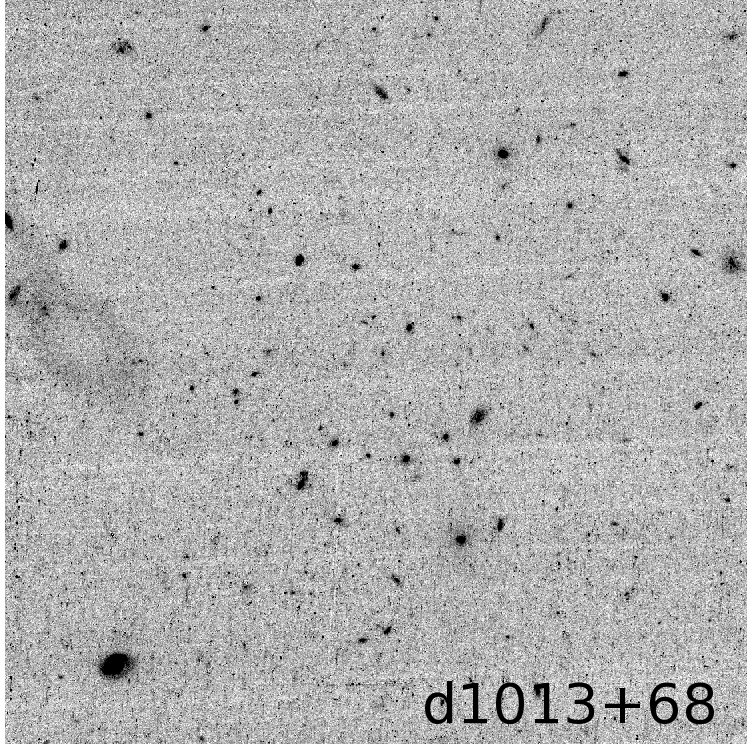}
\includegraphics[angle=0,totalheight=2.in]{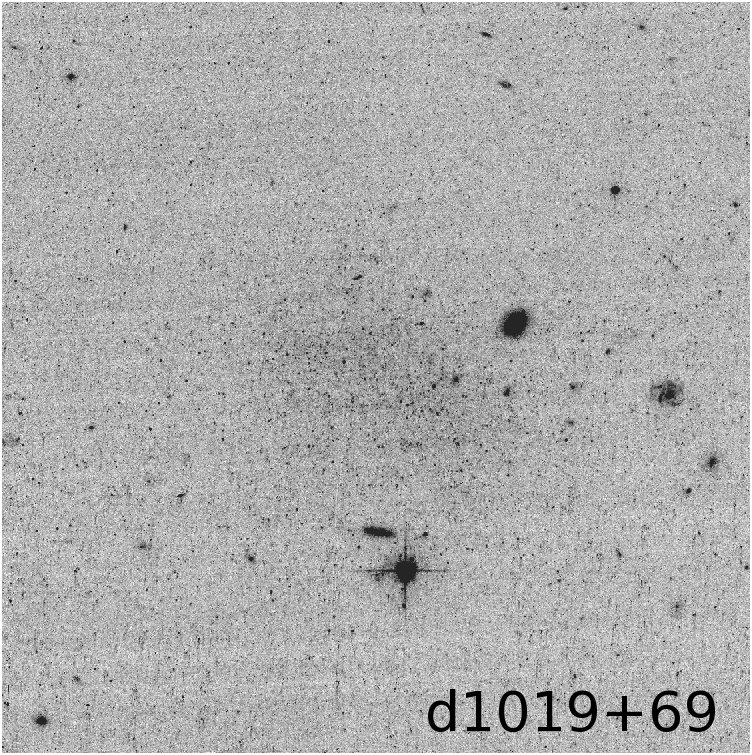}
\includegraphics[angle=0,totalheight=2.in]{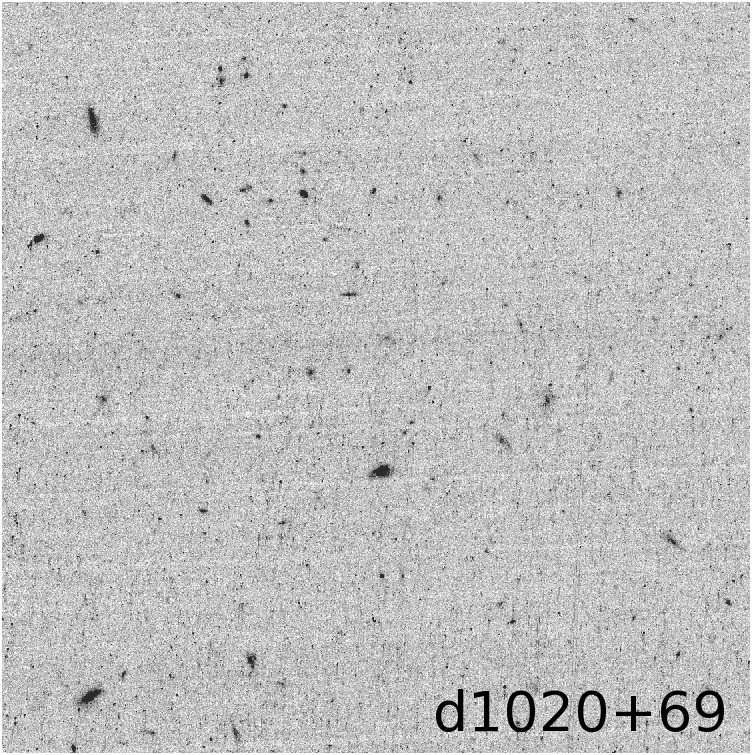}
\caption{$F814W$ thumbnails for 9 candidates observed with ACS.  Images
are 65 arcsec on each side.
\label{acsbw}}
\end{centering}
\end{figure}

\begin{figure}[t]
\begin{centering}
\includegraphics[angle=0,totalheight=2.5in]{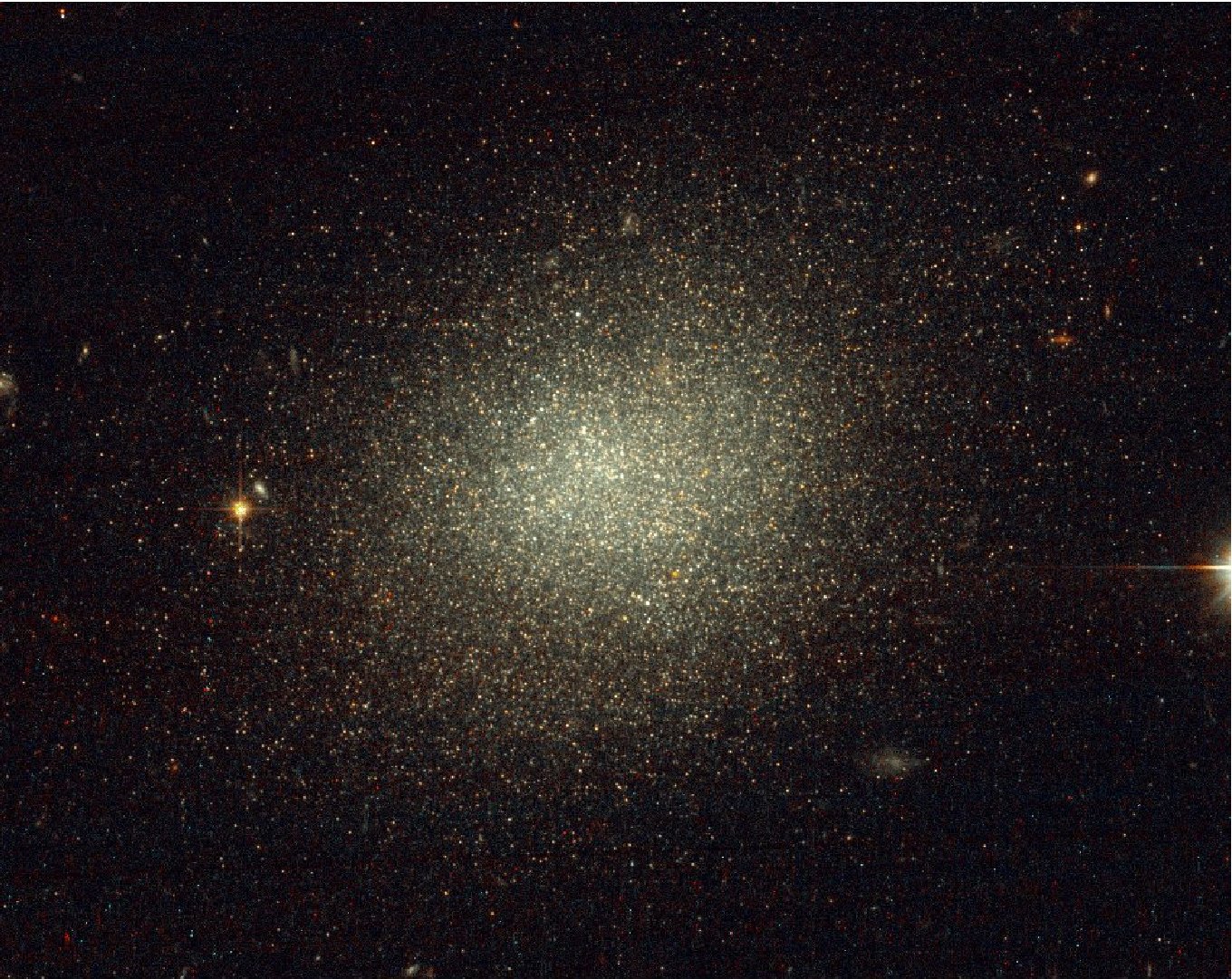}
\includegraphics[angle=0,totalheight=2.5in]{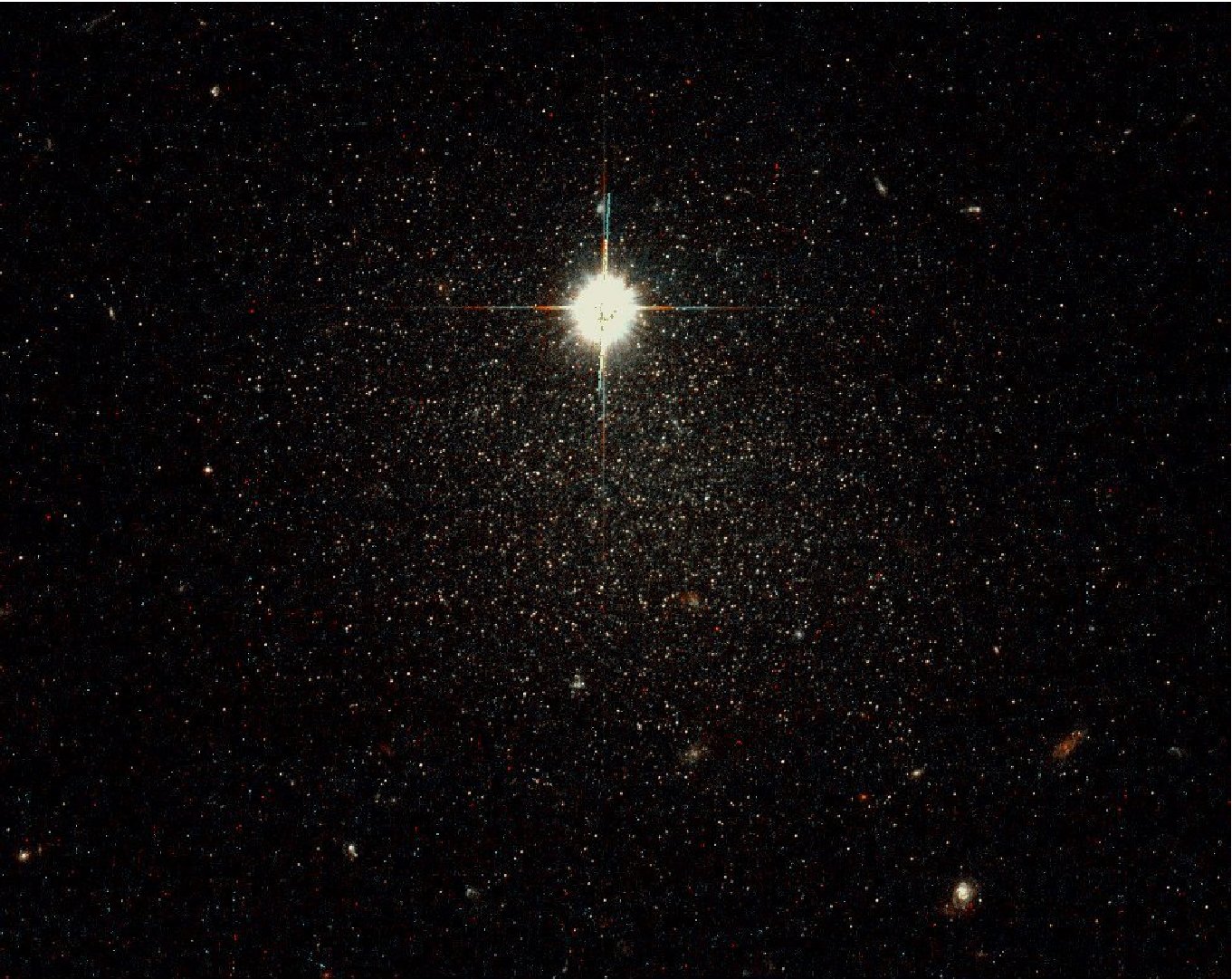}
\includegraphics[angle=0,totalheight=2.5in]{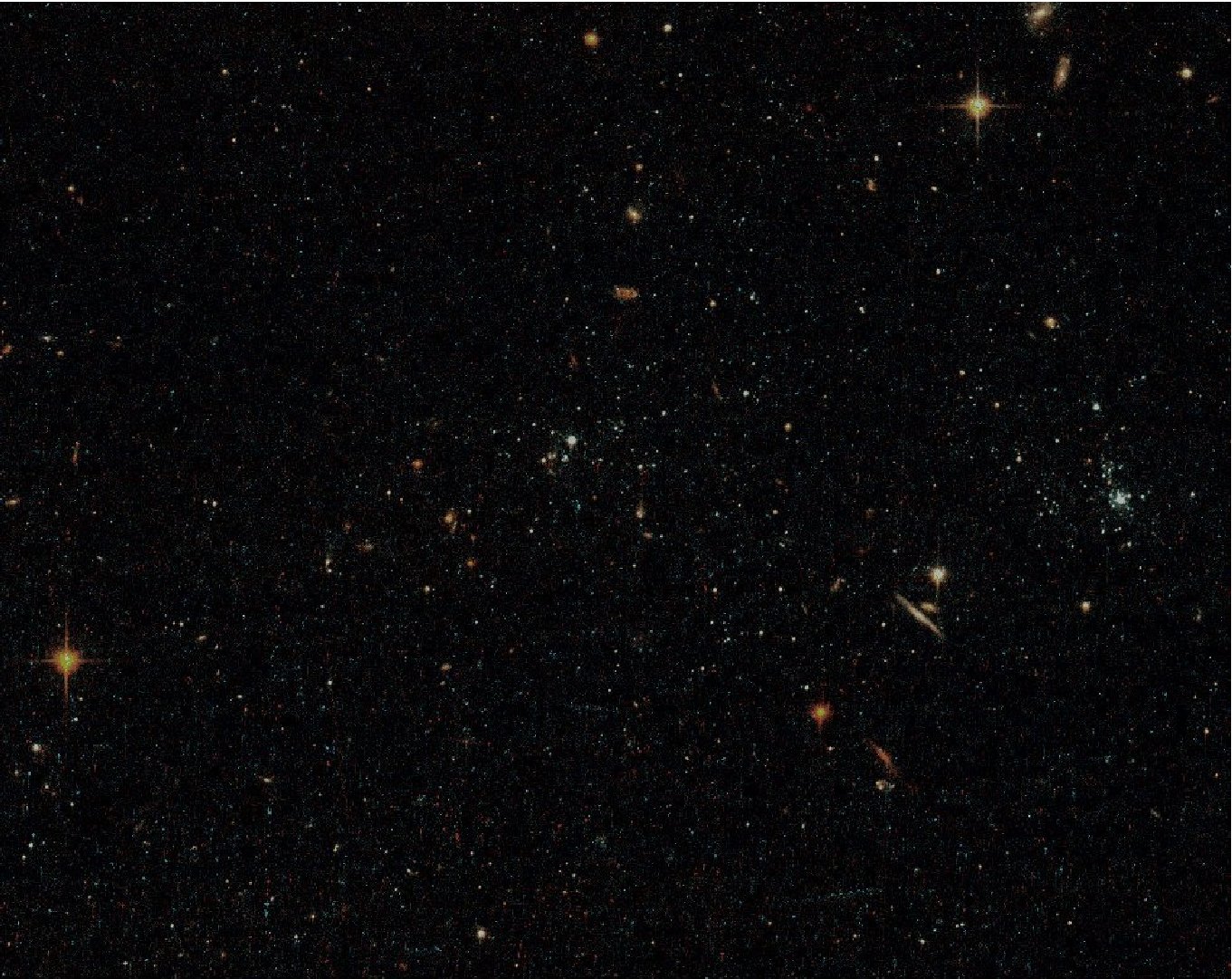}
\includegraphics[angle=0,totalheight=2.5in]{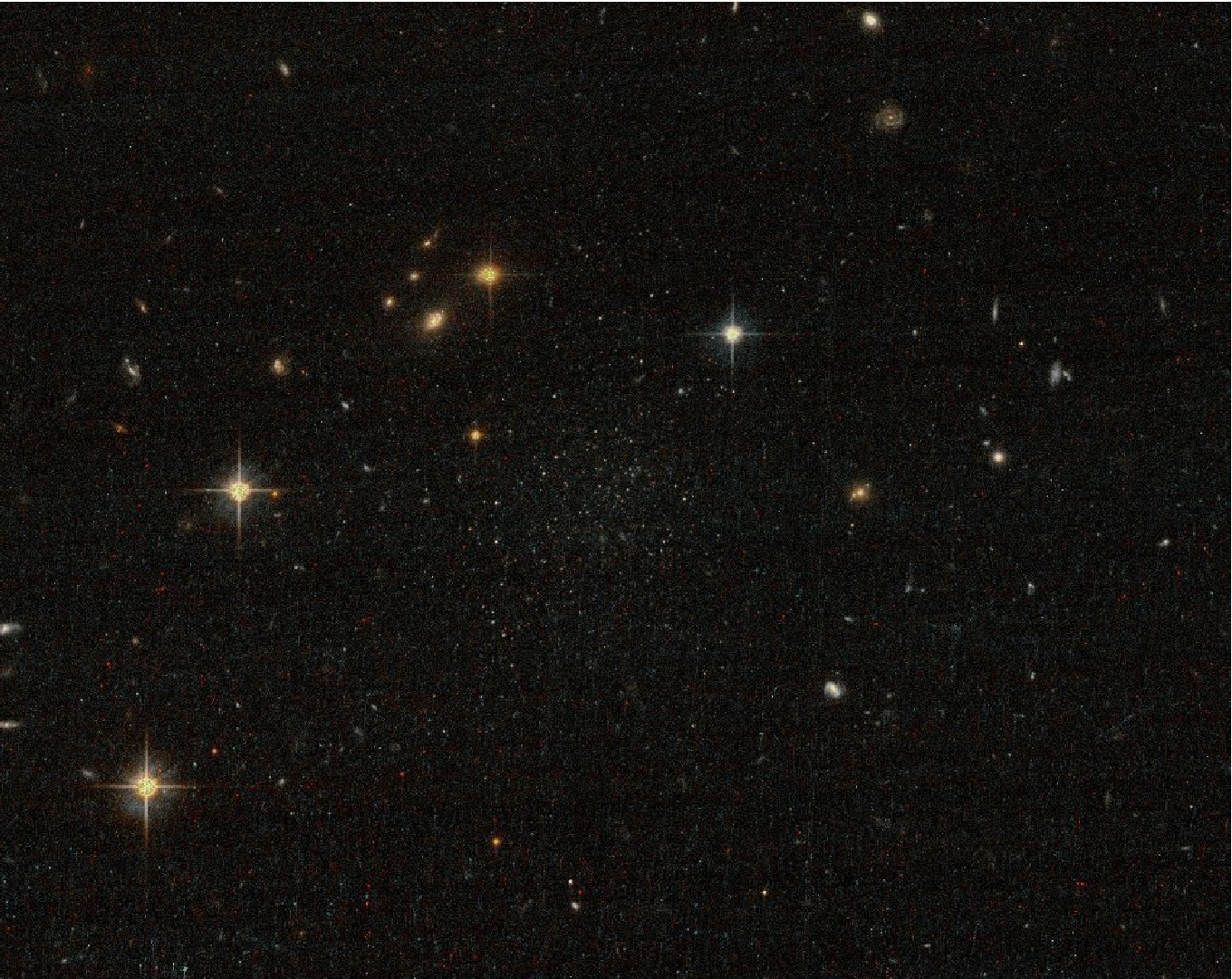}
\caption{Clockwise from upper left, color images of d1012+64, d0944+71, d1015+69, and d0959+68 with log scaling.  Color
images were produced using the ACS $F814W$ and $F606W$ bands as red and green channels, respectively. 
Blue channel images were created by taking $2 \times F606W - F814W$ images.  
Regions shown are 1.6 x 1.3 arcmin.
\label{acscol}}
\end{centering}
\end{figure}

\begin{figure}[t]
\begin{centering}
\includegraphics[angle=0,totalheight=2.5in]{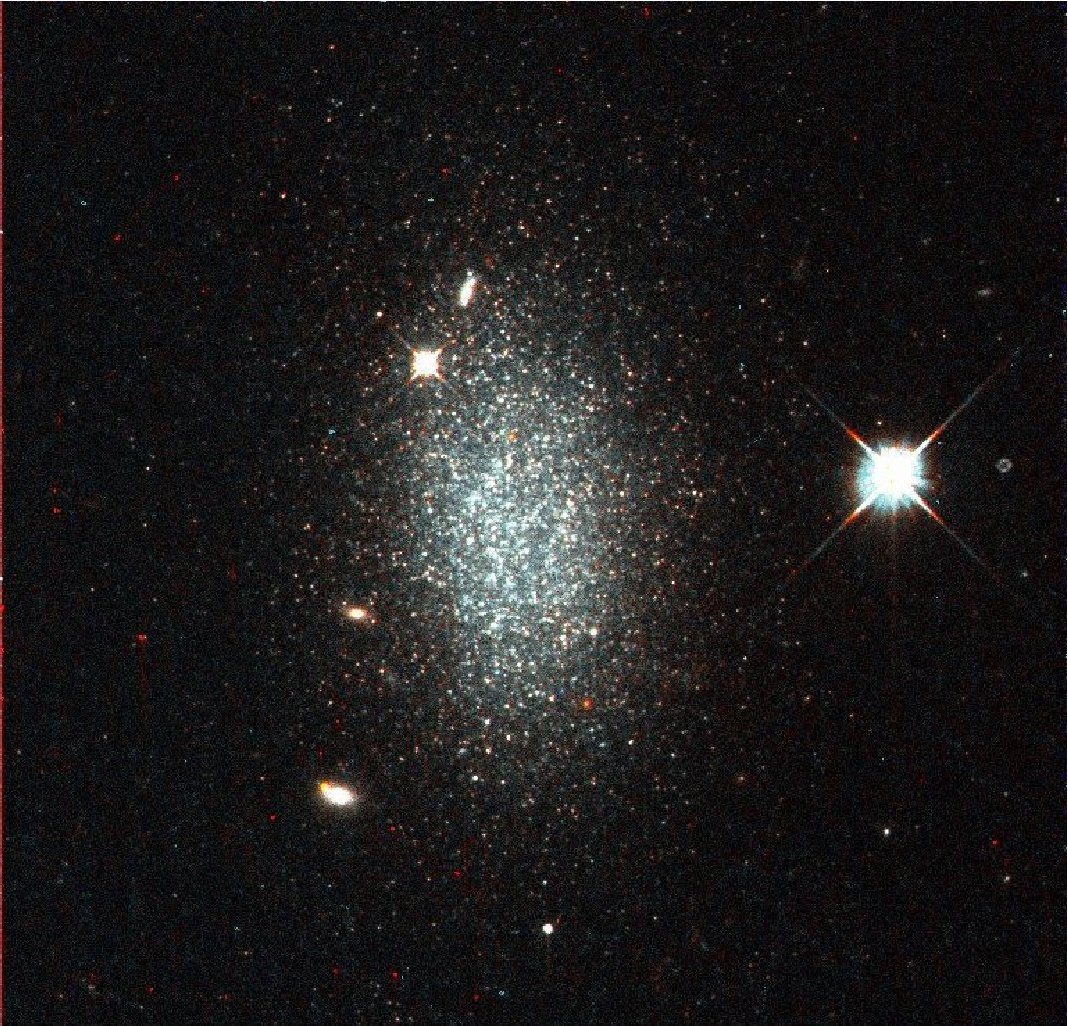}
\includegraphics[angle=0,totalheight=2.5in]{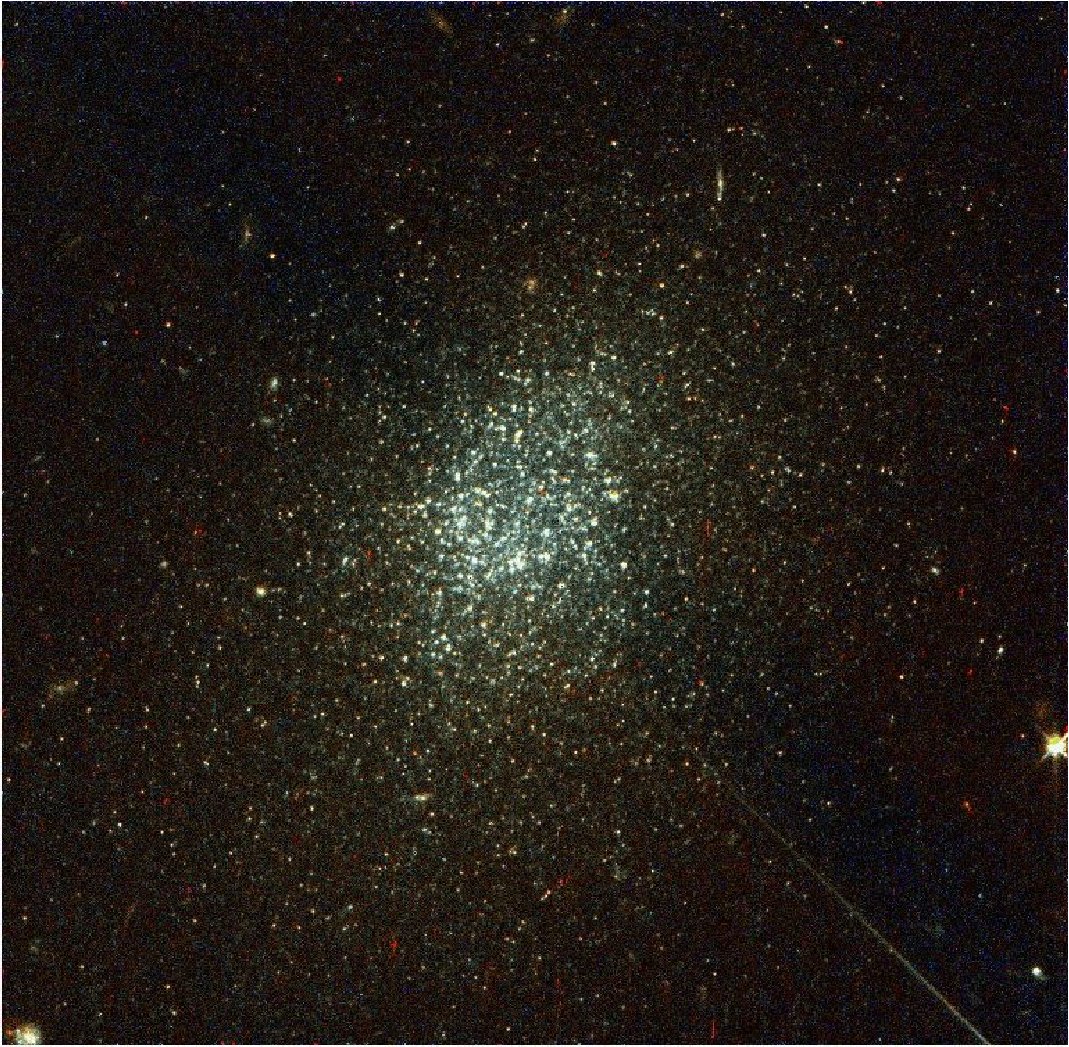}
\caption{Color images of the BCDs d0958+66 and d1028+70 observed with the WFPC2 camera.  
Color images are produced in the same manner as the ACS color images.
WFC chip 2 images are displayed, with size 1.3 arcmin on a side.  
\label{wfpccol}}
\end{centering}
\end{figure}

Imaging was obtained in both
$F814W$ and $F606W$ bands for each of our 22 candidates.   Although Johnson-Cousins $V$ and $I-$band filters
are traditionally used to obtain TRGB distances, the $F814W$ filter transforms well to
Cousins $I$, while the $F606W$ provides adequate color separation and
is chosen over the $F555W$ for its greater throughput.  
An initial set of 14 candidates was observed in Cycle 16 (GO-11126).  Due to the failure of the ACS, the program was  
carried out using the lower sensitivity WFPC2 camera. 
A second set of 8 candidates
were observed with the repaired ACS/WFC (GO-11584) in Cycle 17.  One candidate was re-observed
with ACS in order to probe the stellar population to fainter magnitudes. 
Regardless of instrument, each candidate was observed in both bands in a 
single orbit. This meant two exposures with combined total integration times of 
900-1000s in $F814W$ and 1000s in 
$F606W$ for WFPC2 observations, and $\sim 1250$s in $F814W$ and $F606W$ for ACS/WFC
observations.  See Table \ref{tabobs} for a summary of observations.  Thumbnail
images of each candidate are shown in Figures \ref{wfpc2bw1} - \ref{acsbw}.   
Color images are shown for a subset of these in Figures \ref{acscol} - \ref{wfpccol}.
Color figures for all can be found in the Extragalactic Distance Database (EDD), at
http://edd.ifa.hawaii.edu.

\subsection{Stellar Photometry}\label{phot}
                                                                                
We started with bias corrected and flat-fielded data produced by the STScI pipeline.
Photometry was then performed using the PSF-fitting stellar photometry packages from A. Dolphin.
HSTPHOT \citep{hstphot00} was originally run on the WFPC2 data, while the ACS module of
DOLPHOT was used for the ACS data.  Pre-processing with these routines
included bad column and hot pixel masking, cosmic ray rejection, and sky determination,
as described in the manuals.  Sources are then simultaneously detected and
photometered in both $F814W$ and $F606W$ images, using PSF fitting with model PSFs from 
TinyTim \footnote{http://www.stsci.edu/software/tinytim}.  Aperture corrections were
made based on stars in uncrowded regions.  Charge-transfer inefficiency and zeropoint
corrections are also applied, and measured magnitudes are provided as both flight
system instrumental VEGAMAG magnitudes and as corresponding Johnson-Cousins 
transformed apparent magnitudes.  CTE corrections and zeropoints are taken from 
\citet{dolphin00} for WFPC2 data, while DOLPHOT makes use of the newest
\citet{chia12} CTE corrections and revised magnitude zeropoints from the STScI webpages.

A detection threshold of 3.0$\sigma$ 
was applied.  This produced $F814W$ and $F606W$ matched catalogs with
output including $\chi^2$ for the fit, signal-to-noise, sharpness, roundness, position angle,
crowding, object type (distinguishing between star, faint, and elongated/extended objects),
and magnitudes. 
We then culled this output to reject anything that was not considered a point source 
with sufficiently high S/N.  Final object lists contained only those objects 
classified as stars with 
$\chi^2 < 2.5$, (sharpness)$^2 < 0.09$, S/N $> 5$ in both bands,  and quality
flag 0.  Final lists consisted of $\sim 10000$ and typically less 
than 1000 good stars in ACS and WFPC2 fields, respectively.

\begin{figure}[t]
\begin{centering}
\includegraphics[angle=270, totalheight=5.0in]{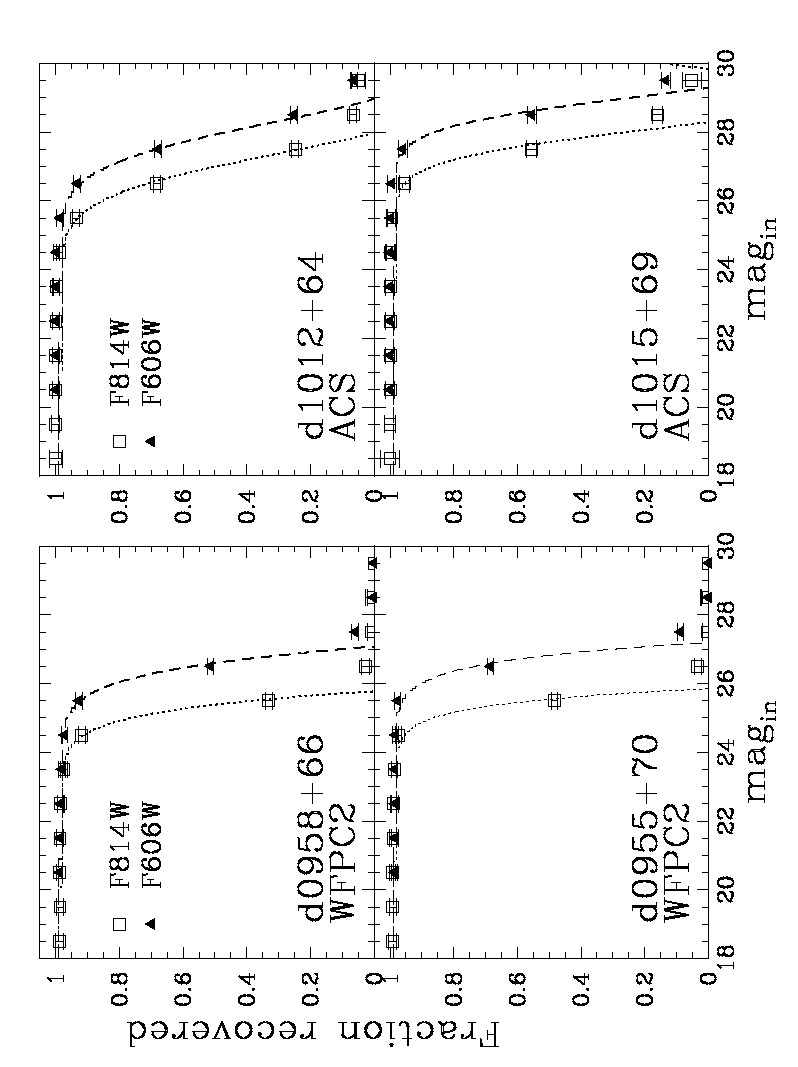}
\caption{Stellar detection completeness in our WFPC2 and ACS imaging from false star tests with HSTPHOT.  These examples
show the completeness in $F814W$ and $F606W$ bands for both crowded (d0958+66, d1012+64) and uncrowded 
fields (d0955+70, d1015+69).  Detections reach about 2 mag deeper in the ACS data.  Additionally, detections 
in uncrowded fields go slightly deeper than for crowded fields in both WFPC2 and ACS imaging.  
As shown for the ACS examples, the $\sim 50$\% 
completeness is about a half magnitude deeper for the uncrowded field.
Error bars are from Poisson statistics.
\label{WFPCcomp}}
\end{centering}
\end{figure}

Using artificial stars, we tested the detection completeness and photometric uncertainties of
HSTPHOT/DOLPHOT.  Between 100,000 - 200,000 sources were added using the Dolphin routines to 
WFPC2 and ACS imaging, one at a time
so as to not introduce additional crowding.  HSTPHOT/DOLPHOT were then run in the same way to recover
these artificial sources, including implementing the same rejection parameters that were
applied to the real data.   Detection completeness is shown in Figure \ref{WFPCcomp}. 
For WFPC2 data, we find an $F814W$ 50\% completeness limit of $\sim 25.5\pm0.2$.  The crowding 
of each field appears to affect the recovery rate, with crowded fields reaching about $\sim0.2$ 
mag shallower than this and sparser fields probing $\sim 0.2$ mag deeper. $F606W$ data go 
1.2 mag deeper than the $F814W$ data.  For ACS data, the difference in depth due to 
crowding is as large as $\sim0.8$ mag between dense and uncrowded fields.  We find a 50\% 
completeness limit of $\sim 27.3\pm0.4$ in $F814W$.  $F606W$ data reach about 1 mag deeper.  
At the tip of the RGB at $F814W \sim 23.8$, we expect to be 100\% complete in 
all datasets.  For WFPC2 data, the 50\% completeness limit is less than 2 mag below the tip.  
However, even in crowded fields, we expect completeness to be $> 90$\% one magnitude below the tip. 

\begin{figure}[t]
\begin{centering}
\includegraphics[angle=0,totalheight=3.in]{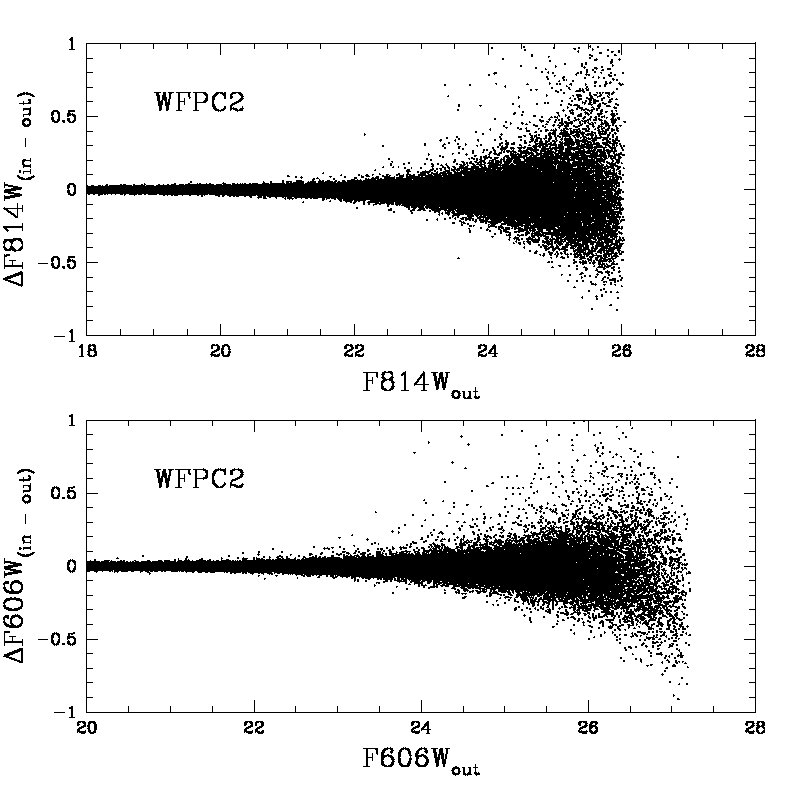}
\includegraphics[angle=0,totalheight=3.in]{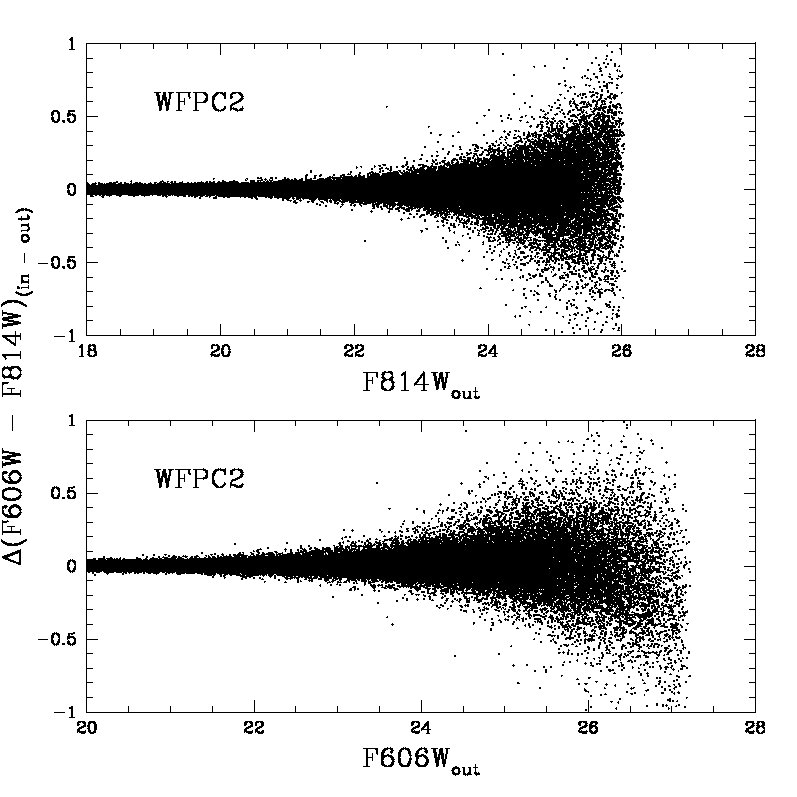}
\includegraphics[angle=0,totalheight=3.in]{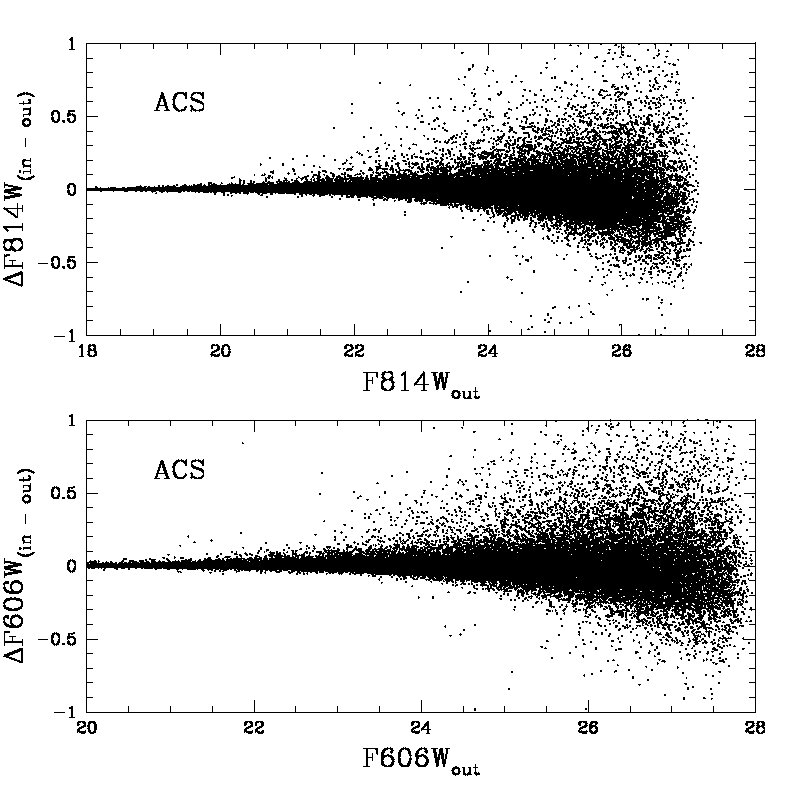}
\includegraphics[angle=0,totalheight=3.in]{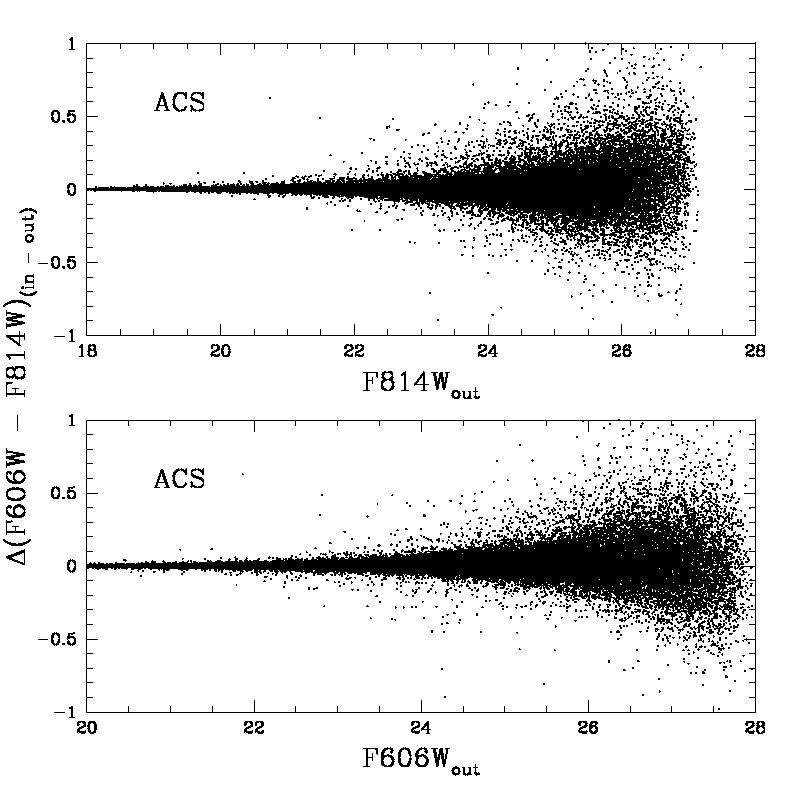}
\caption{Magnitude and color measurement errors from false star tests using HSTPHOT (WFPC2) and DOLPHOT (ACS).  
These examples show the results for galaxy d1028+70 observed with WFPC2 and d1012+64 observed with ACS, 
where the false stars shown were added to the same chip as the location of real galaxies.
Over 100,000 and 200,000 stars were added to
each of the WFPC2 and ACS images, respectively, of which over 60,000 were recovered as good stars within 
3 R$_e$ of the real galaxy centers.
\label{WFPCerror}}
\end{centering}
\end{figure}

Uncertainties in the photometry are also investigated using the artificial star tests.  Results
are displayed in Figure \ref{WFPCerror}. 
The scatter at faint magnitudes becomes quite large. At $F814W = 23.5$, close to the expected 
TRGB magnitude for M81 group members, the $1\sigma$ uncertainty is $\sim0.055$ mag for ACS 
and 0.11 mag for WFPC2 data.  The uncertainty is slightly higher, by
$\sim0.03$ mag, for the most crowded fields in each case.  For individual stars at the TRGB, this 
translates to an uncertainty in distance of 100 and 200 kpc for ACS and WFPC2 data, respectively.  
The second turnaround radius for the M81 group is 230 kpc, within which we expect all galaxies to be
group members.   Since the TRGB measurement is based on the ensemble of stars at the tip of the RGB,
the uncertainty decreases with increasing number of stars populating the upper TRGB.  
\citet{mmrt06} find that the maximum likelihood method for determining the TRGB which we implement
should produce reasonable results for RGBs consisting of $> 50$ stars in the top magnitude
before the tip.
Thus, the photometric uncertainties are within the range needed to
unambiguously establish group membership, at least for galaxies with an RGB population 
size sufficient for determining the location of the RGB tip.  

\begin{figure}[t]
\begin{centering}
\includegraphics[angle=0, totalheight=5.0in]{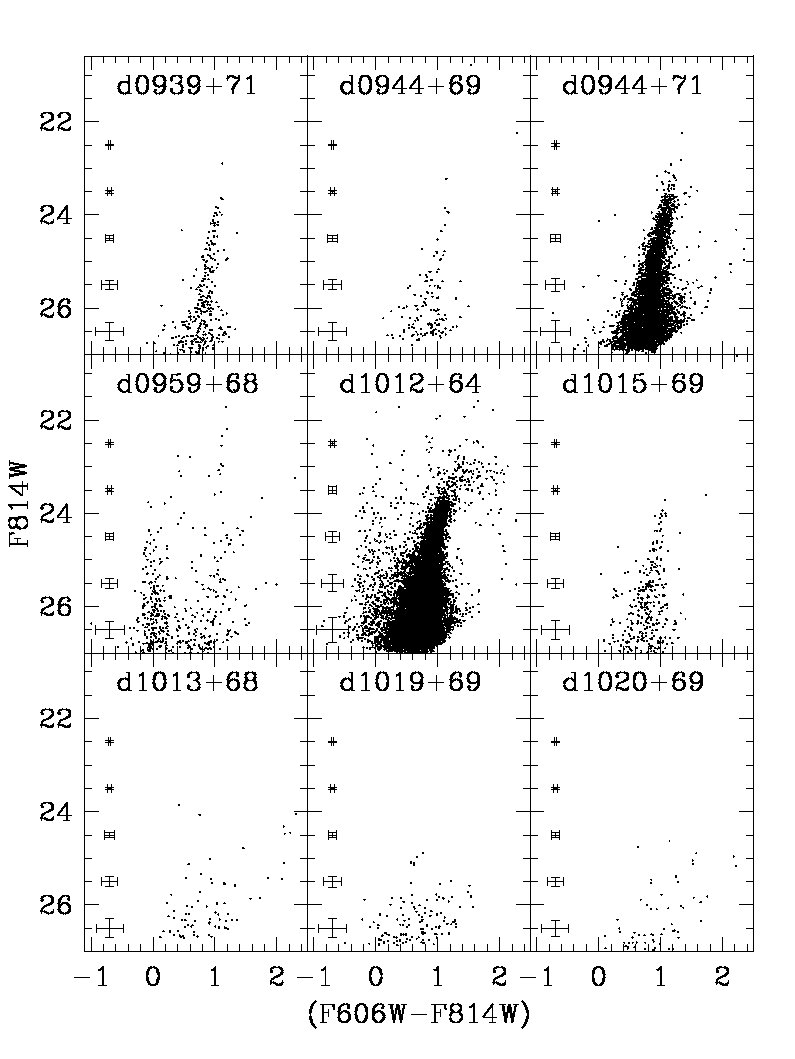}
\caption{CMDs within a $50^{\prime\prime} \times 50^{\prime\prime}$ region centered on each of the 9 
candidate M81 dwarf galaxies imaged with ACS.  Error bars denote photometric uncertainties.  
RGBs are evident in at least 5 of these panels, at ($F606W - F814W) \sim 0.8$, and may be present
in d0959+68.
\label{ACScmd}}
\end{centering}
\end{figure}

\begin{figure}[t]
\begin{centering}
\includegraphics[angle=0, totalheight=5.0in]{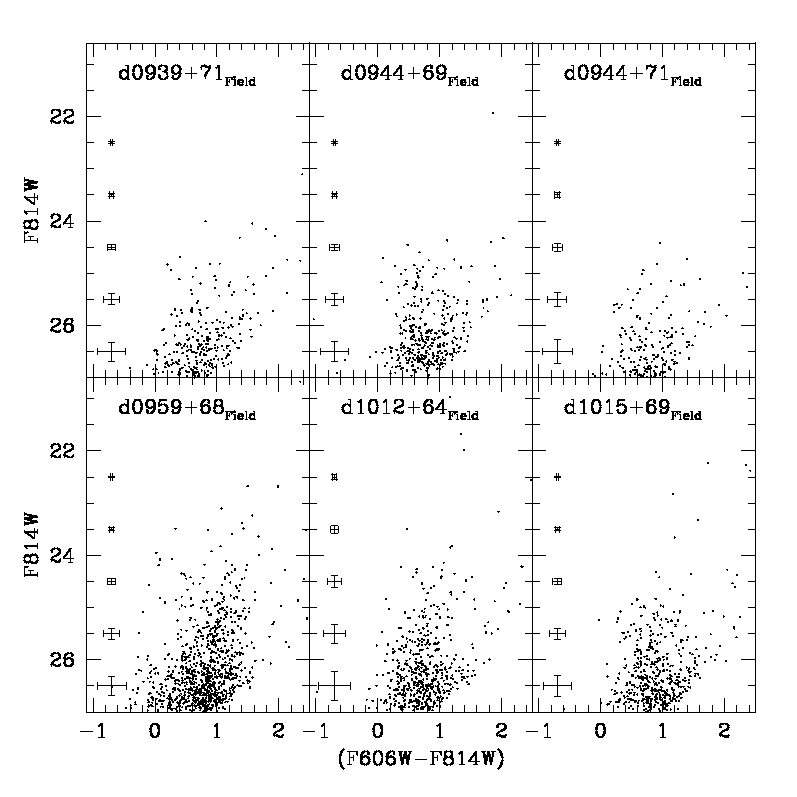}
\caption{CMDs for stars in ACS/WFC 1 for 6 pointings. All candidates were centered in WFC 2. Error 
bars denote photometric uncertainties.  Hints of an RGB may be present d0959+68,
indicating that the galaxy extends into this chip.
\label{ACSfldcmd}}
\end{centering}
\end{figure}

\begin{figure}[t]
\begin{centering}
\includegraphics[angle=0, totalheight=5.0in]{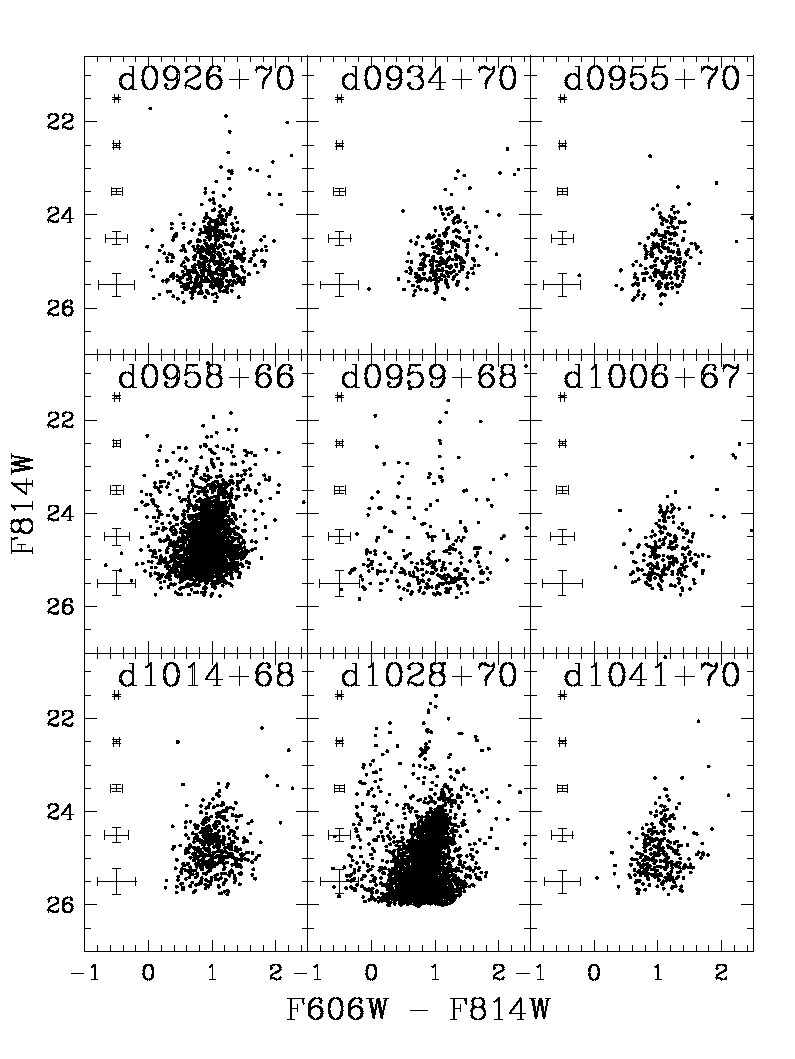}
\caption{CMDs for 9 candidate M81 dwarf galaxies imaged with WFPC2.  Error bars denote photometric uncertainties. Data from chip 
3 are plotted only.  In 8 cases, a RGB is evident as a wedged-shaped structure at $(F606W - F814W) \sim 0.8$. 
In one case, d0959+68, no RGB is apparent, but we expect this object to be a group member for reasons described in 
the text.
\label{WFPCcmd}}
\end{centering}
\end{figure}

\begin{figure}[t]
\begin{centering}
\includegraphics[angle=0, totalheight=5.0in]{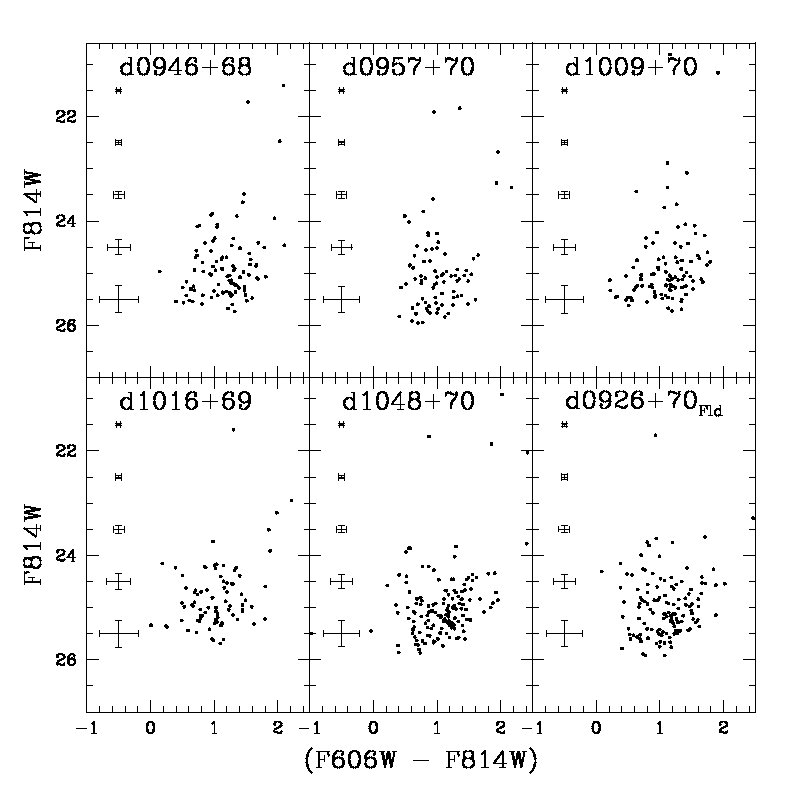}
\caption{CMDs for 5 additional candidate M81 dwarf galaxies imaged with WFPC2.  Error bars denote
photometric uncertainties.  There are no obvious RGBs.  The bottom right panel shows the CMD from
WFPC2 detector 2 for the d0926+70 pointing.  All candidates are centered in detector 3.
\label{WFPCfldcmd}}
\end{centering}
\end{figure}

CMDs for all candidates are shown in Figures \ref{ACScmd} - \ref{WFPCfldcmd}.  Error bars denote 
$1\sigma$ uncertainties in magnitude and color as determined from the artifical star tests.  
The most prominent feature in many of these CMDs is the red giant branch at $F606W - F814W \sim 0.8$,
the presence of which is confirmation of the galaxian nature of these objects. The RGB is visible
as a long, narrow strip in ACS data, seen in at least 5 of the top 6 panels of Figure \ref{ACScmd}.
For d0959+68 the presence of an RGB is more questionable.
Since all candidates were centered in the WFC chip 2, we show the CMDs for stars in the bottom detector for
several cases in Figure \ref{ACSfldcmd}.  Hints of an RGB in d0959+68 may indicate that
the target extends onto the other CCD.
The RGB has a wedge-shaped appearance in candidates shown in Figure \ref{WFPCcmd} observed with WFPC2. 
The broadness at the detection limit is primarily due to large photometric errors.  No
such feature is apparent in the remaining 5 candidates observed with WFPC2 (Figure \ref{WFPCfldcmd}).
For comparison we also include in this plot the stellar detections for the d0926+70 observation taken
from a different chip away from where the target was centered and which exhibits fewer stellar detections. 
We have no independent
control fields, but as many of our galaxies are spatially quite small, we can use part of some of the 
images as an approximation of the expected foreground stellar contamination.  

In a few of these CMDs (d0959+68, d1012+64, d0958+66, and d1028+70), a main sequence (MS) is visible 
as a vertical strip around
$F606W - F814W \sim -0.2$. Blue loop stars are also likely present at bright magnitudes. These stars
indicate the existence of a young stellar population.   Asymptotic
giant branch (AGB) stars, representative of intermediate age populations, are also apparent in
some of these CMDs above and redward of the RGB.  These populations will be discussed further in 
Section \ref{stellpop}.

\subsection{TRGB measurements}\label{methods}

Red giant stars have ceased hydrogen burning in their cores while continuing
to burn hydrogen in outer shells. This shell burning causes the outer envelope of the star to expand
and become redder and brighter while dumping material on a degenerate core, thereby increasing the core 
mass and temperature.  Because of the degeneracy, the star cannot compensate for the increased core
temperature and, when the core reaches the required temperature, at a constant core mass and 
therefore a predictable luminosity, the onset of helium burning occurs as a runaway thermonuclear reaction.
The degeneracy is lifted before the star explodes, and the star then spends the
next stage of its life on the lower luminosity horizontal branch.
Thus, the tip of the RGB, immediately before a star undergoes the helium flash, has
a near constant bolometric luminosity and can therefore be used as a standard candle.  
Although this luminosity is strongly dependent
on the underlying stellar population metallicity and, to a much lesser extent,
the population age, it is least sensitive to these evolutionary variables in the $I-$band \citep{luca07}.  

Accurate TRGB distance measurements can therefore be made with good calibration for the 
$I-$band tip luminosity. 
For this, we use the work of \citet{luca07}. They define a zeropoint calibration of the TRGB, accurate
to 1\% statistical uncertainty, as a function of the stellar population color to account for 
variation due to metallicity and age.  In addition, the
zeropoint is provided in the HST flight system for $F606W$ and $F814W$ for both ACS and WFPC2 cameras.
They find 
\begin{eqnarray*}
M^{ACS}_{F814W} = -4.06 + 0.20 [(F606W - F814W) - 1.23] \\
M^{WFPC2}_{F814W} = -4.01 + 0.15 [(F606W - F814W) - 1.12]
\end{eqnarray*}

\noindent
The measurement of the tip of the RGB can therefore be made directly within the flight system
magnitudes as measured by DOLPHOT/HSTPHOT, eliminating additional uncertainty from magnitude
transformations into the Johnson-Cousins system.

In order to determine the precise location of the apparent magnitude of the tip we use the
maximum likelihood method of \citet{mmrt06}, as described in \citet{brad09}.  This has the advantage
over edge detection algorithms in that no binning or smoothing is required, although a functional
form must be assumed for the stellar luminosity function.  We use the TRGBTOOL program written
by D. Makarov which assumes a simple power-law with a cut-off faintward of the TRGB and a second
power-law with a different slope brightward. Default values for the slopes are provided although
these can be varied as necessary to fit the data.  The discontinuity in the fit 
determines the TRGB location.  The procedure 
incorporates the results of the artificial star tests discussed above to account for completeness, 
crowding, and
photometric errors and provides realistic uncertainties in the tip measurements.

We run TRGBTOOL on HSTPHOT/DOLPHOT photometry for all the candidates.  In most cases where
we believe a red giant branch is visible in the CMD, the automated tip finder locates a tip magnitude close
to the expected value for M81 Group members.  In several cases, eg. d0934+70 and d0959+68, the RGB is so 
poorly populated that
uncertainties in the tip measurement are very large. 
In addition, for d0959+68 the tip measurement obtained from the tip finding software
may be affected by AGB stars.  The few cases with well populated RGBs have small TRGB measurement errors
and well constrained distances.

\section{Results}\label{res}

\subsection{Confirmation of 14 new M81 companions}\label{members}

\begin{figure}[t]
\begin{centering}
\includegraphics[angle=0, totalheight=3.0in]{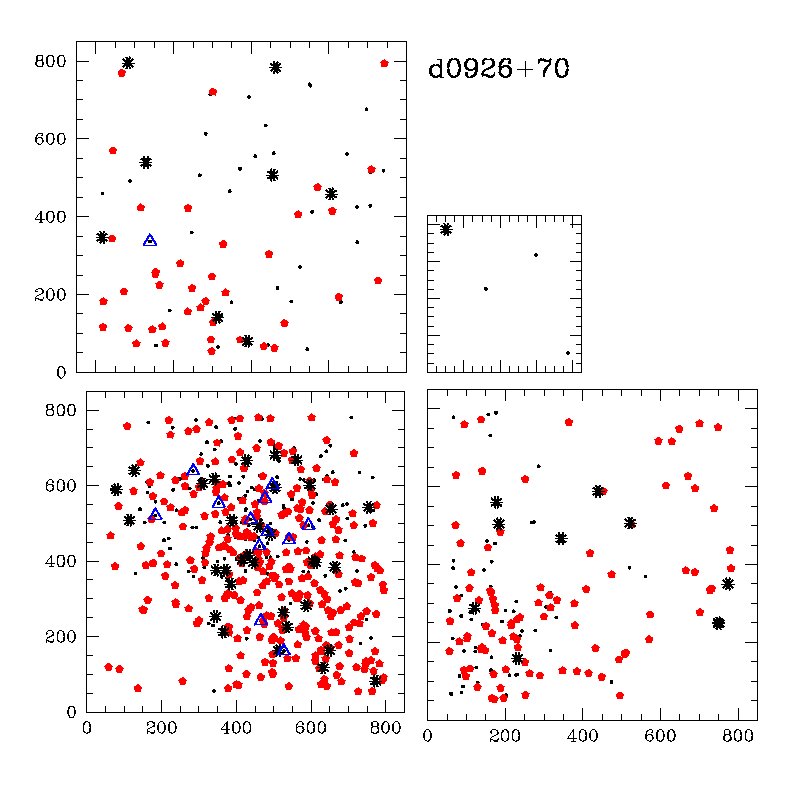}
\includegraphics[angle=0, totalheight=3.0in]{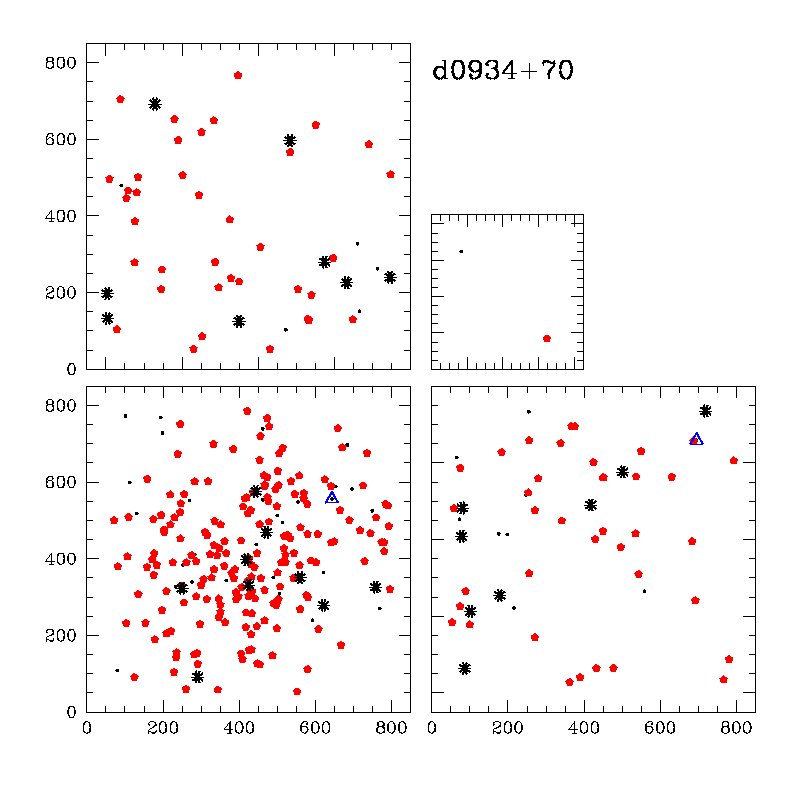}
\includegraphics[angle=0, totalheight=3.0in]{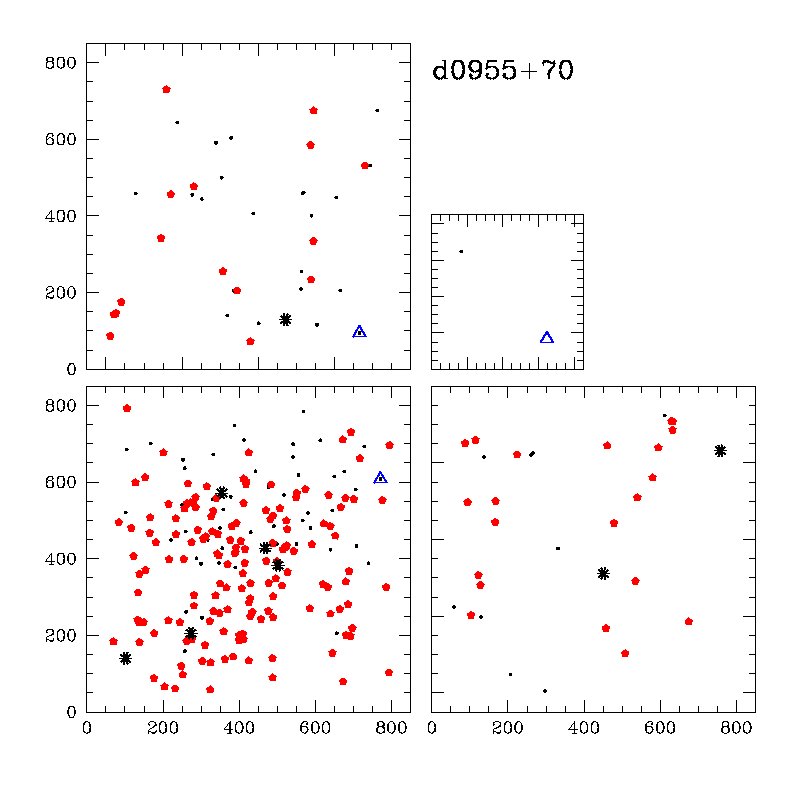}
\includegraphics[angle=0, totalheight=3.0in]{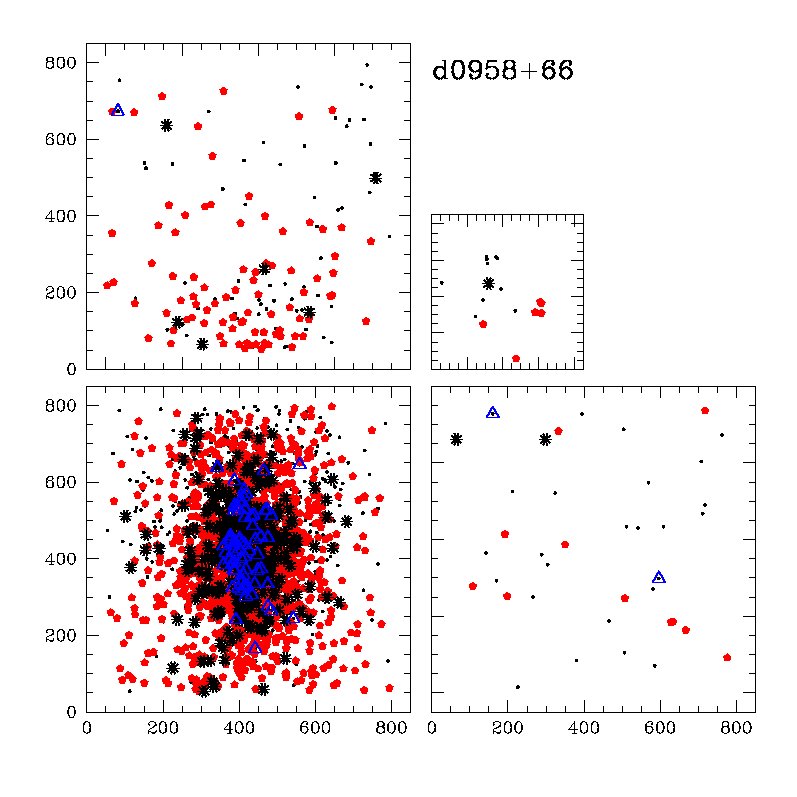}
\caption{Locations of detected stars in the WFPC2 fields.  Black dots denote all 'good' stars recovered
by HSTPHOT.  Red pentagons are stars within the range of the RGB: $F814W_{TRGB} < F814W < 25$ and
$0.4 < F606W-F814W < 1.4$, black asterisks represent possible AGB stars having $F814W < 23.4$ and
$F606W-F814W > 0.6$, and blue open triangles denote potential main sequence and blue loop stars
with $F814W < 25$ and $F606W-F814W < 0.2$.  
\label{Locs1}}
\end{centering}
\end{figure}

\begin{figure}[t]
\begin{centering}
\includegraphics[angle=0, totalheight=3.0in]{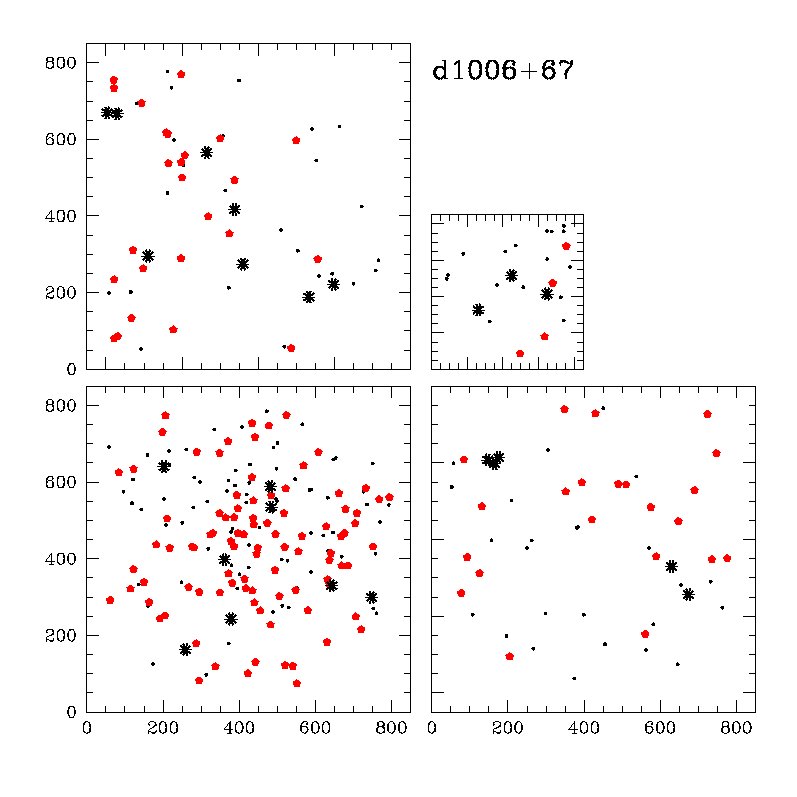}
\includegraphics[angle=0, totalheight=3.0in]{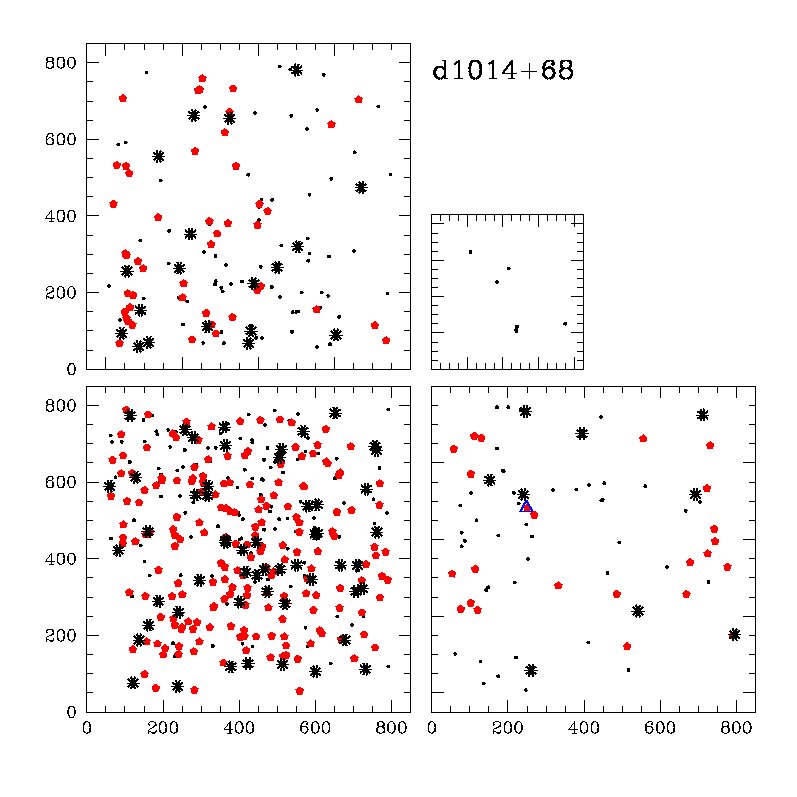}
\includegraphics[angle=0, totalheight=3.0in]{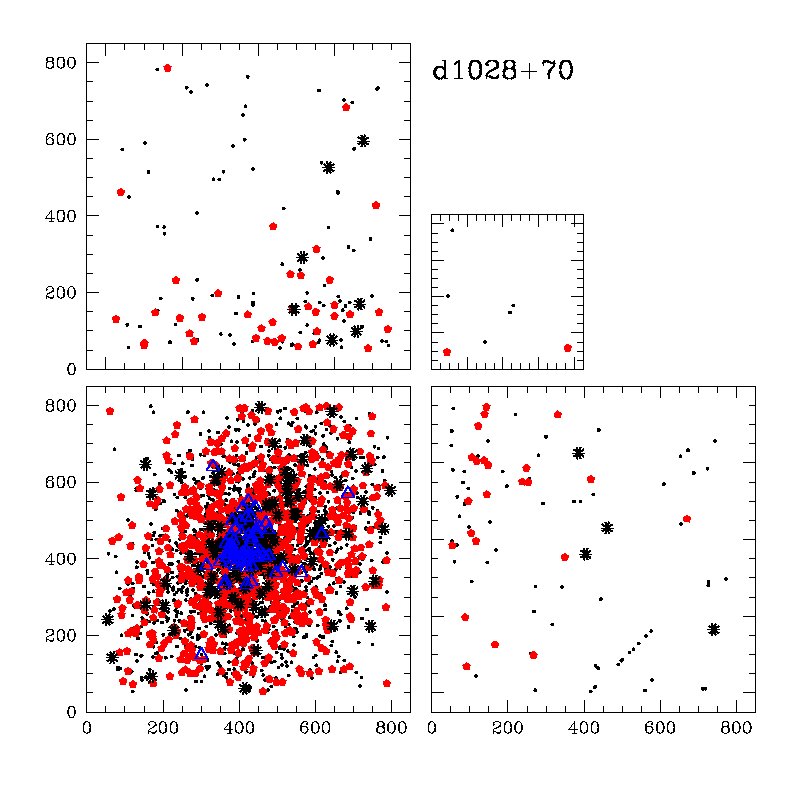}
\includegraphics[angle=0, totalheight=3.0in]{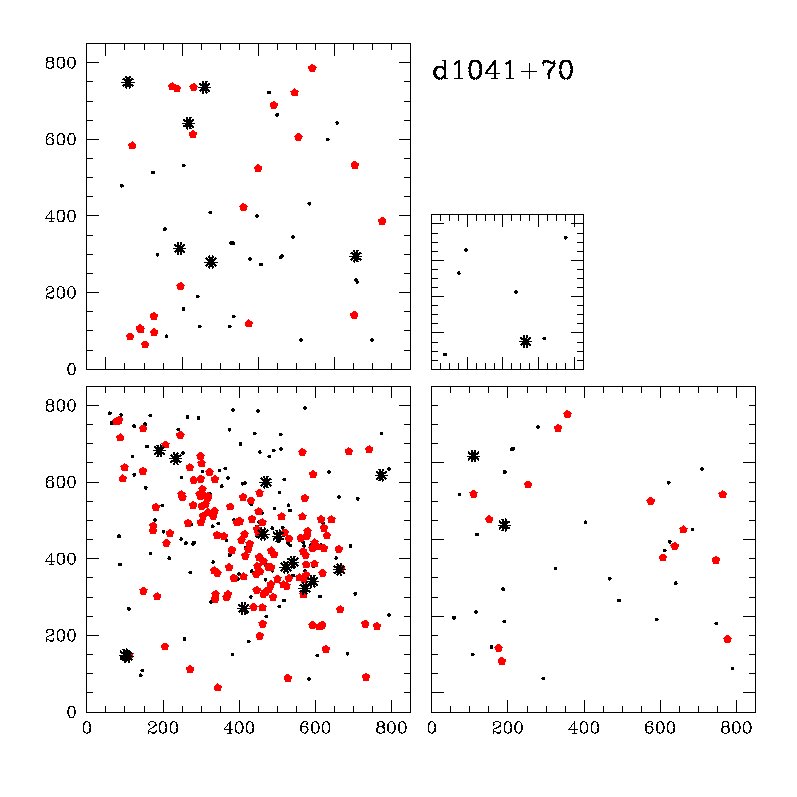}
\caption{Locations of detected stars in the WFPC2 fields.   Symbols as in Fig. \ref{Locs1}.
\label{Locs2}}
\end{centering}
\end{figure}

\begin{figure}[t]
\begin{centering}
\plottwo{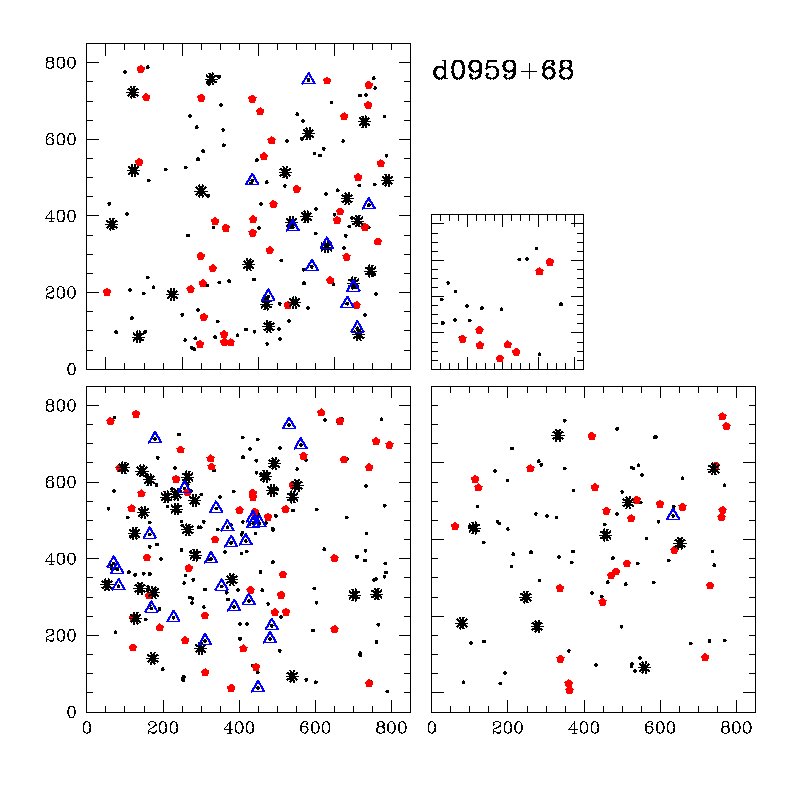}{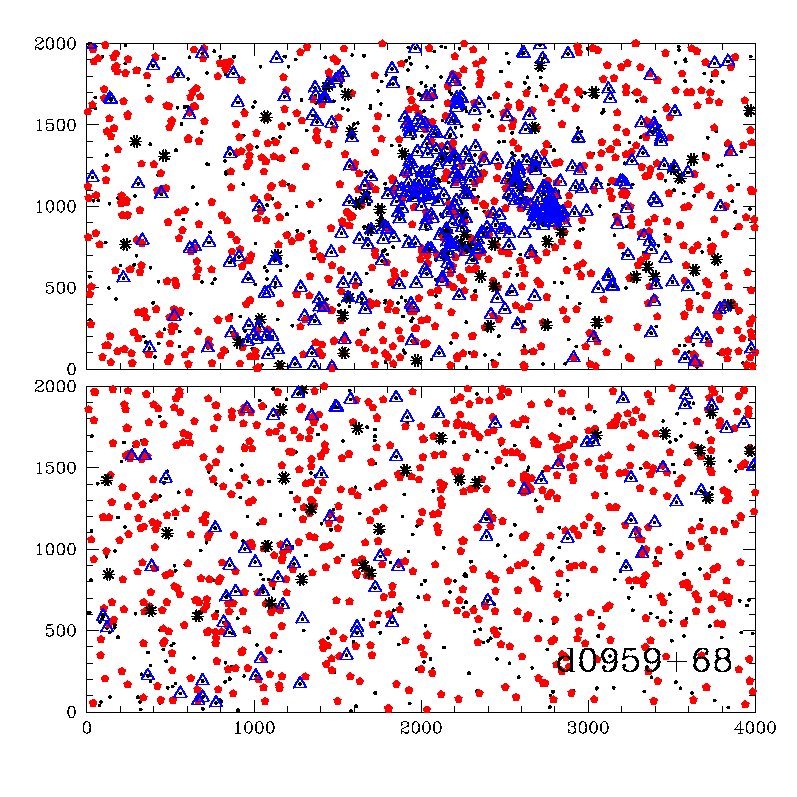}
\caption{Locations of d0959+68 stars observed with both WFPC2 and ACS cameras.  
Symbols as in Fig. \ref{Locs1} with limiting depth of $F814W < 26.5$ for ACS RGB and main
sequence stars.
\label{Locs3}}
\end{centering}
\end{figure}

\begin{figure}[t]
\begin{centering}
\includegraphics[angle=0, totalheight=3.0in]{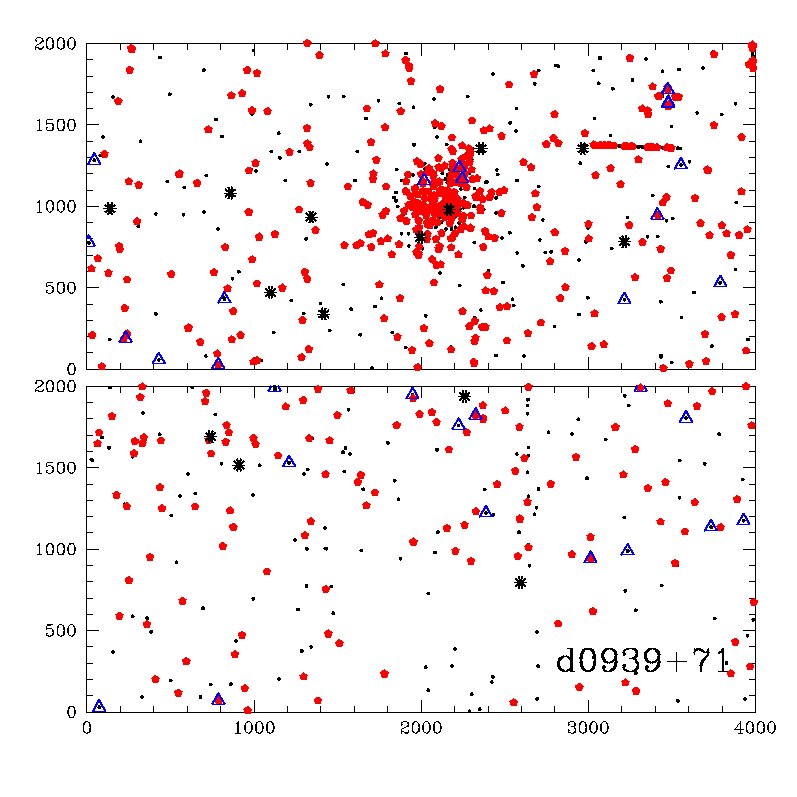}
\includegraphics[angle=0, totalheight=3.0in]{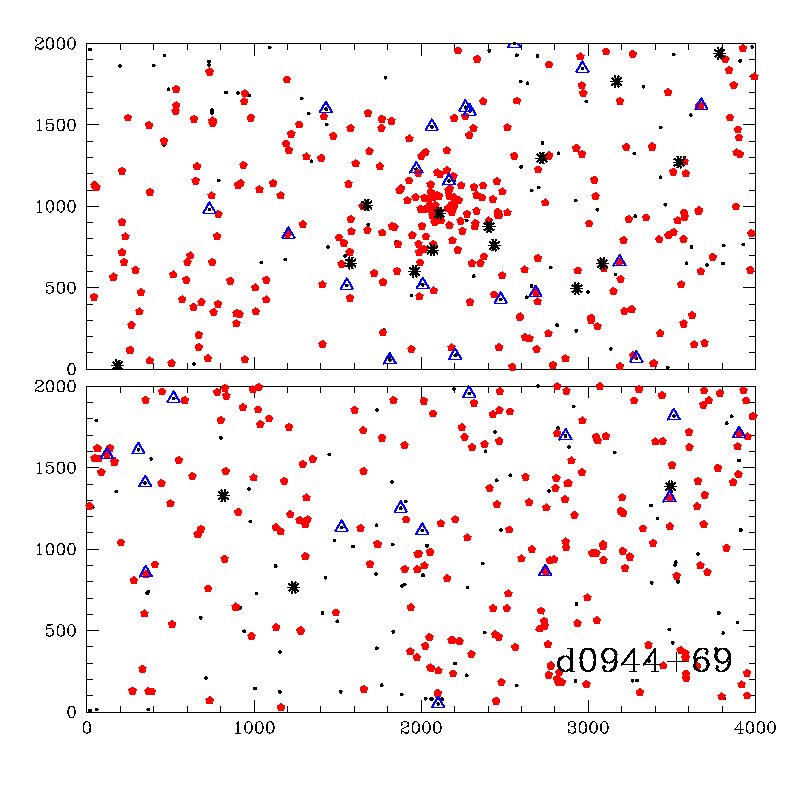}
\includegraphics[angle=0, totalheight=3.0in]{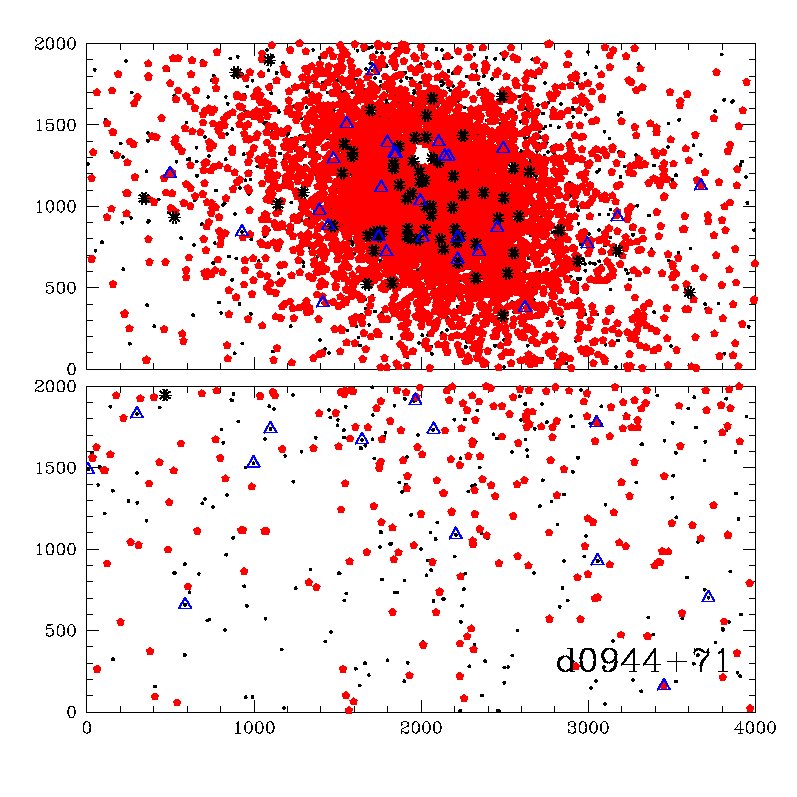}
\includegraphics[angle=0, totalheight=3.0in]{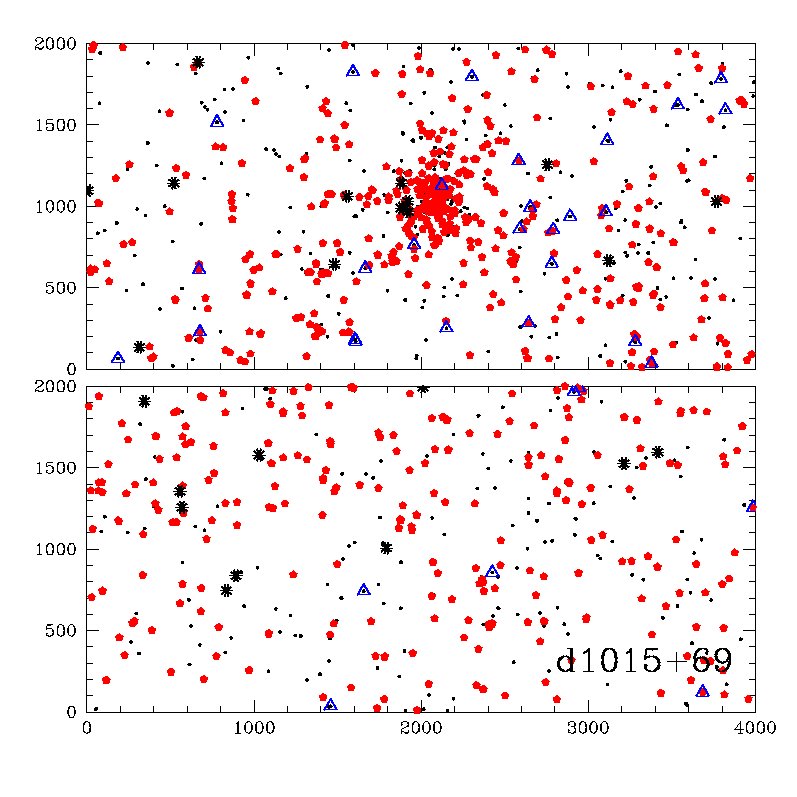}
\caption{Locations of detected stars in the ACS fields.  
Red pentagons are stars within the range of the RGB: $F814W_{TRGB} < F814W < 26.5$ and
$0.4 < F606W-F814W < 1.4$, black asterisks represent possible AGB stars having $F814W < 23.4$ and
$F606W-F814W > 0.6$, and blue open triangles denote potential main sequence and blue loop stars
with $F814W < 26.5$ and $F606W-F814W < 0.2$. 
\label{Locs4}}
\end{centering}
\end{figure}

The spatial distribution of detected stars in WFPC2 and ACS images for 
14 candidates are displayed in Figures \ref{Locs1} - \ref{Locs5}.
Symbols are coded by region of the CMD.  Black points denote all good stellar
detections. Red pentagons are stars which fall within the range
of the RGB: $F814W_{TRGB} < F814W < 25$ and $0.4 < F606W-F814W < 1.4$. Black asterisks
represent possible intermediate age AGB stars having $F814W < F814W_{TRGB}$ and $F606W-F814W > 0.6$. Blue
open triangles denote potential young main sequence and blue loop stars
with $F814W < 25$ and $F606W-F814W < 0.2$.

Although the stellar detections have been culled for only 'good' stars, some artifacts
are still present in the ACS imaging. 
All 14 objects shown here exhibit obvious concentrations of stars at the targeted location of
the candidate, further proof that these are real galaxies.
WFPC2 images not only have shallower magnitude limits than ACS, but cover smaller regions
on the sky, with a field of view of only
$\sim 150 \times 150$ arcsec as compared to the larger $202 \times 202$ field
of ACS.  Although concentrations are obvious in WFPC2 chip 3, the stellar populations
spill out onto the other chips and outside the field of view.  The
smaller dwarfs observed with ACS are largely contained within the imaging;
larger galaxies extend beyond the ACS images.

Many of these objects consist predominently of RGB stars with
perhaps a handful of stars in the AGB range.  These two stellar types have
similar distributions.  A few objects exhibit evidence for a younger stellar population
with upper main sequence and blue loop stars.   These are found as concentrations in 
the core for some cases and scattered throughout the object in others.  A wide range
of galaxy sizes and stellar population densities is noted.

In Figures \ref{trgb1} - \ref{trgb9} we show the extinction corrected CMDs of all expected
members.  Extinction in the ACS and WFPC2 bandpasses was determined using 
the reddening values from the dust maps of \citet{sfd98} with the recalibration for extinction in 
different bandpasses by
\citet{schlafly}, as provided in their Table 6 assuming a \citet{fitz} reddening law with R$_V = 3.1$.   
The measured location of the TRGB is indicated with a broken solid
line while dashed lines denote the uncertainty in the tip location.
Table \ref{tabX} provides TRGB $F814W$ magnitude measurements, along with the extinction
corrected values, associated
tip absolute magnitudes calculated as a function of TRGB color, and the resultant distance moduli and
calculated distances.  In addition, we correct our previous estimates of absolute $r^{\prime}$ magnitude
(see Paper 1) for these new distance measurements.  Uncertainties in extinction are not taken into account,
so total uncertainty in distance measurements is expected to be larger.

The largest uncertainty in TRGB measurement is for object d0934+70.
This galaxy appeared to be strongly affected by galactic cirrus in MegaCam imaging,
and is in a region with high extinction according to the dust maps of \citet{sfd98}.
The RGB is poorly populated, leading to the large uncertainty in TRGB
measurement. The RGB is also strangely shaped, displaying what appears to be color/metallicity
shifts along the extension of the RGB.  This could perhaps be caused by the low number
statistics and large photometric errors, or by high and variable extinction spatially
across this object.  The maximum likelihood best fit distance is 3.0 Mpc, but an extinction 
over-correction would produce an underestimate in the measured distance.  We believe this may be
the case.  The large $1\sigma$ uncertainties place the distance between 2.8 - 4.1 Mpc.  
In comparison to the CMDs generated from 'field' stars in
regions on the detectors away from the targetted objects, there are significantly
more stars associated with this object and these are arranged largely along the expected 
location for a RGB.    
Because of this, and coupled with the object's spatial location in the group, 
we consider it an M81 group member, albeit with an uncertain distance measurement.  

One object, d0959+68, was observed with both WFPC2 and ACS.  We do obtain a TRGB measurement for
this object from ACS data, with corresponding distance 4.2 Mpc. However, we consider this measurement 
uncertain due to the likely contamination of the RGB region with AGB stars.  The majority 
of stars associated with this object are blue, young, main sequence and blue loop stars. 
Since this object lies right along the tidal
stream between M81 and NGC 3077, and given its young population and lack of a built-up
RGB, we consider this object to be a tidal dwarf formed along this bridge of 
material likely having a distance similar to that of these 2 giants, $\sim 3.7 - 3.9$ Mpc.

Other objects, d0926+70, d0955+70, d1006+67, d1014+68, d1041+70, d0939+70,
d0944+69, d1015+69, have uncertainties in distance ranging 200 kpc to 1 Mpc.
The TRGB measurements themselves place all these objects near to M81 at 3.69 Mpc.
The large uncertainties are due to the nature of these objects. They all
appear to be small dwarfs with very small stellar populations. With such
poorly populated RGBs, the uncertainties in the distance measurements will
be large. We expect these dwarf galaxies are all associated with 
the M81 group.

Finally, four objects have small uncertainties in the tip magnitude due to the much
more populated RGBs.  Three of these were considered to be blue compact dwarf
galaxies (BCDs) in our first paper and are now established to be associated with the
M81 group.  These are small objects with reasonably large and concentrated stellar
populations consisting of both old stars in a well developed RGB and
younger main sequence, blue loop, and AGB stars.

Thus, based on the distance measurements from the CMDs along with supporting evidence from
concentrations of resolved stars at the targeted locations of the candidates, we confirm a 
total of 14 new M81 companion dwarf satellites. 
Of the original 22 candidates,
12 were expected to be members based on the degree of resolution of the stellar population
in $r^{\prime}$ MegaCam imaging.  Of these, we find that 11 are bonafide
members of the M81 group.  Of 6 candidates which we thought were galaxian in nature,
but were uncertain members due to a lower degree of resolution, we find that 3 are
indeed M81 group members while 3 are non-members and likely more distant dwarf galaxies.
Of the final 4 candidates considered to be possible artifacts or distant galaxy clusters,
none prove to be group  members.
Nine of these confirmed M81 Group members exhibit predominantly RGB sequence stars in
their CMDs displaying little or no evidence for any AGB, Blue Loop, or Main Sequence stars. These are the
dSph galaxies.  A tenth object may also contain a small number of younger stars. 
The remaining 4 dwarfs all exhibit more significant younger populations in their
CMDs.  We display the projected and 3-D distribution of 22 previously known and
14 new M81 group members within the region of our survey in Figure \ref{3d1}.  
Table \ref{M81sep} lists the projected and physical separations of each group galaxy from M81.
CMDs, spatial distributions, and color figures for
each of these galaxies can also be found in the Extragalactic Distance Database (EDD), at
http://edd.ifa.hawaii.edu, by selecting the catalog CMDs/TRGB.

\subsection{Spatial Distribution}\label{spatial}

Figure \ref{corelb} shows a projection of the three-dimensional distribution of all known galaxies
within a cube centered on M81 and extending 600 kpc on the cardinal axes.  The plot is
in supergalactic cartesian coordinates and the orientation is chosen so measured distances (and
their errors) are in the line-of-sight.  This orientation allows a projection of the
\citet{sfd98} dust map to be superimposed as a reminder that a band of cirrus
crosses the region.  Revised \citep{schlafly} selective to total extinction
E(B-V) values reach 0.22 at the location of one of our galaxies, although values are
below 0.1 for 85\% of cases, with a median of 0.06.

The gas deficient systems are given emphasis in this plot, with the new dSph/non-star forming galaxies
colored red and previously known gas deficient systems colored orange.  Normal giant and
dwarf galaxies with both young and old stars and detected HI \citep{begum, sambit} are identified
by open symbols, with the 3 newly discovered cases given emphasis.  Five tidal dwarfs
(young stars but no convincing evidence of old stars) are identified by green triangles.
The measured TRGB distances for the tidal dwarfs are considered quite unreliable since
the RGB stars in these fields may not be associated with the targets.  All the tidal
dwarfs are in close proximity to M81 in projection and we assume the M81 distance of
3.69 Mpc for them.

The projection chosen with Figure \ref{corelb} is minimally influenced by errors in distance.  We
note an apparent flattening of the distribution of gas deficient systems to the
supergalactic equatorial plane.  M81 itself lies just 40 kpc off this plane.  The
significance of this configuration can be evaluated from the histograms in Figure \ref{cumrad}.
The galaxy KKH57 lies outside the CFHT survey region so will not be considered further.
Of the 20 gas deficient galaxies in the survey region only d1041+70 deviates
substantially from the supergalactic mid-plane (or more precisely, a plane parallel to
the supergalactic equator displaced 40 kpc to pass through M81).  The rms deviation in
supergalactic latitude (SGB) with respect to M81 is $\pm61$ kpc for 19 galaxies excluding
d1041+70.  By contrast the rms deviation in supergalactic longitude (SGL) with
respect to M81 is $\pm123$ kpc for the same galaxies.  The difference between SGB and SGL
scatter has a 95\% probability of significance; that is, there is suggestive but not
conclusive evidence for a physical flattening.  The tentative postulation is that this sample
on a scale of a few hundred kpc shares flattening about the same pole as the 50 times
larger structure that it is embedded in, the Local Sheet \citep{tully08}.

For the same 19 galaxies the scatter in distance is $\pm231$ kpc.  This value is the
convolution of true scatter and measurement errors.  If the true scatter in the
line-of-sight is the same as the scatter in SGL then measurement errors at the level of
5\% are implied.  This estimate of errors is in accordance with our expectation.  This
level of error in distances limits our ability to define the properties of a putative
plane in the distribution of the M81 dwarfs.  It can be supposed that such a plane is
tilted to our line-of-sight.  However it would be necessary to have distances accurate
to 1\% to evaluate such a scenario, well beyond our present capability.

It is evident from Figure \ref{corelb} that the distribution of galaxies with gas and young
populations does not manifest the flattening seen by the gas poor systems.  Discounting
M82 and NGC 3077 which are directly engaged in an encounter with M81, the later type
systems are at greater projected radii from M81 than the early types.  This difference
is illustrated in Figure \ref{cumrad}.  It is a familiar pattern \citep{grebel97,tt08}.
The median and mean projected separations from M81 for early types within the CFHT
survey region are 122 and 133 kpc (149 and 162 kpc statistically deprojected) while the
median and mean for late types are 158 and 169 kpc (193 and 206 kpc deprojected).
Beyond the CFHT survey region all known galaxies associated with the M81 Group
except KKH57 are late types.
In the common picture, minor galaxies that have made multiple crossings within a group
halo have lost their gas through tidal disruption, stripping, or starvation.  The
outer extent of their distribution roughly coincides with the radius of `second
turnaround' \citep{tully10} which is a direct function of the mass of the parent halo.
The gas-bearing systems, by contrast, are likely more recent arrivals.  They may be on
first infall or they may have high orbital angular momentum that has kept them from
harms way.

The apparent flattening of the distribution of early types, although not overwhelmingly
convincing, is given credence by similar phenomenae seen more locally.  The most
remarkable case to date has been discussed by \citet{ibata13}.  
Half of the small
companions to M31 lie in a very thin ($\pm13$ kpc) disk with coherent kinematic properties.
This disk is askew from the supergalactic plane but oriented such that the Milky Way
lies in the same plane.  Meanwhile most of the early type companions to the Milky Way
appear to lie in a plane \citep{lb76,Metz08,Pawl12}.
This plane is essentially orthogonal to the supergalactic plane although the long axis
is aligned with the long axis of the Local Sheet \citep{tully08}; ie, toward the
Virgo Cluster.  The past histories of these planes and two other apparent planes
of Local Group satellites are discussed by \citet{st13}.
In review, in the few instances around nearby major galaxies where we
have information, in every case there is evidence that gas poor companions lie in
flattened distributions.

\subsection{Structural properties of the new members}\label{structpar}

To determine centroids of the new objects, we used the geometric mean of the
distribution of the resolved RGB stars (or main sequence stars in the case of d0959+68)
in the galaxy field using the $F814W$ ACS/WFPC2 data.  Ellipticity
and position angle around these centroids were then derived from
second moments \citep{sh97}.  

To obtain structural parameter measurements for each galaxy observed with ACS,
we perform stellar surface density profile fitting. 
Stellar number counts were extracted in successive elliptical apertures around
each galaxy out to a radius of 40 arcsec.  This was truncated if counts exhibited
a significant increase in slope, chosen emperically to be $>$ 30\%, attributed to contamination 
by an excess of field stars.
The central measurement was also discarded. Aperture sizes for each galaxy depended on galaxy stellar
density and were optimized to contain statistically significant counts over the background
while also maintaining enough apertures to perform the fitting.  Foreground contamination 
was estimated in these same sized apertures using stellar detections in the other ACS chip, and a mean foreground
stellar count density was determined for each galaxy.  This was then subtracted
from the galaxy counts.  For larger galaxies, the stellar component
may extend over the entire ACS field of view, and thus this background
correction may be a slight over-correction. The effect of this would be to produce size
measurements that are too small, and indeed for larger galaxies, these size measurements
are almost certainly underestimates.  Smaller galaxies are less likely 
to extend very far onto the other chip.  However, these galaxies are more likely to suffer from 
statistical variance affecting the low number counts.

We then fit both the differential 
count density with a Sersic function and the curve of growth with a cumulative
Sersic function.  The generalized Sersic function is defined as

\begin{equation}
 I(r) = I_{\circ} \ e^{-(r/r_{\circ})^{1/n}}
\end{equation}

\noindent
where $n$ is the Sersic parameter.  For $n = 1$, this reduces to
an exponential profile which usually provides a good fit for
disk and dwarf galaxies, and for $n = 4.0$ becomes a de Vaucouleurs profile.
The cumulative Sersic function is then

\begin{equation}
I(r) = 2\pi\sigma_{\circ}r_{\circ}^{2}n \cdot \gamma[2n,(r/r_{\circ})^{1/n}]
\end{equation}

\noindent
where $\gamma$[a,x] is the Incomplete Gamma function,
\begin{equation}
\int_{0}^{x}exp(-t)t^{a-1}dt 
\end{equation}
\citep{j95}. 
\noindent

The fitting of this nonlinear function to the data was done using a
Levenberg-Marquard algorithm \citep{press92} which performs a
$\chi^{2}$ minimization that implements an
inverse-Hessian method far from the minimum and switches to
a steepest decent method as the minimum is approached.
In Figures \ref{prof1} - \ref{prof6}, we
display both curve of growth and count density profile fits.  
Parameters were extracted from the best of these two fits
for the six galaxies imaged with ACS. WFPC2 fields are so small that even
the smaller galaxies fill the field and it is not possible to obtain a good
estimate of the foreground contamination.  For these galaxies, we therefore defer
to the MegaCam size measurements from Paper 1.

From these fits we directly obtain scale length and profile type
(Sersic index n),
where scale length in the generalized Sersic function is directly
related to the half-light or effective radius through

\begin{equation}
R_{\circ}^{1/n} = R_{e}^{1/n} / (2.3026 b_{n})
\end{equation}

\noindent
with the approximation b$_{n}$ = 0.868 \ n - 0.142
\citep{ccd93}.
\noindent
This assumes that the mean
brightness of the stars does not vary as a function of radius.
It further assumes that we can use counts in the opposing chip as an 
approximation for the foreground correction.  For the larger galaxies,
where we may be over-correcting for this affect, we find that
if no background correction is applied
d1012+64 and d0944+71 would be 1 and 2\% larger, respectively.
Results are listed in Table \ref{tabS}.  In Figures \ref{prof1} - \ref{prof6},
we also display contour plots of these galaxies on which we overlay ellipses
having shapes and sizes corresponding to the measured PA, ellipticity, and
effective radius for each galaxy.

As we found with the CFHT imaging, the majority of the galaxies are best fit 
with Sersic index $n < 1.0$.  It was determined that the
low n fits were primarily driven by large cores, typical
of dwarf galaxies, whereas, in the outer radii, the surface brightness
profile tends to drop exponentially. 

For d0959+68, we fit only the blue stellar component which we take to be the 
tidal dwarf (see Section \ref{d0959}).
However, these young stars are found
in star forming clumps, and the stellar surface density is not well fit with a 
Sersic profile. This is true in the case of the measurement made with CFHT imaging
as well.  

To estimate uncertainties in the size measurement, we repeat the process, 
using different regions of the images for the foreground correction. 
The $1 \sigma$ uncertainties quoted in the table come from the variance in size 
measurements from a total of 20 profile fits.  As can be seen in Table \ref{tabS}, the uncertainties
are large for the fainter dwarfs, reaching as high as 12.5\% in apparent size for the smallest dwarf 
d0944+69 (or 33\% for physical size when including distance uncertainties).
The largest and best populated objects observed with ACS, d0944+71 and
d1012+64, have much smaller 6\% uncertainties in physical size.

Differences from CFHT structural measurements can be attributed to large uncertainties in
measurements for faint, low surface brightness objects due to the shallow depth of
CFHT/MegaCam imaging, contamination 
by non-member stars in the CFHT imaging data, uncertainties in the foreground correction
for ACS data and small size of the HST fields, and the different methods used. In
the case of CFHT imaging data, we used the low surface brightness light profile,
just visible above the sky in most cases, while for HST imaging, we use the distribution 
of individual stars with colors and brightnesses corresponding to the RGB. 
Shape measurements are also expected
to differ since measurements from CFHT imaging were based on the light distribution 
and will be strongly affected by brighter sources including foreground stars and 
luminous blue main sequence stars, while the HST measurements are based on the RGB star
counts (with some amount of foreground contamination), each weighted equally.

In Figure \ref{remr}, we plot M$_{r^{\prime}}$ vs log Re for the newly confirmed members with late
and early types distinguished by color, along with previously known M81 group members.  Points
are based on the MegaCam imaging for which we have size and magnitude measurements for all M81 members
within the survey region.  The new BCDs are seen as the high surface brightness objects at 
magnitudes similar to previously known galaxies while the majority of the new detections lie
at magnitudes fainter than M$_{r^{\prime}} > -10$.  A surface
brightness limit $< 27$ mag arcsec$^{-2}$ for all new detections is apparent.  
Also shown in the plot are size measurements based on the ACS imaging for
5 galaxies. These are shown as crosses at the corresponding $r^{\prime}$ magnitude.  The points
are seen to shift slightly although insignificantly in most cases.  Only for our smallest dwarf 
d0944+69 is the shift noticably large.  The two dashed vertical lines indicate the region of
parameter space where few star clusters and galaxies tend to be found. The CFHT/MegaCam imaging measurement places
d0944+69 within this gap while the ACS measurement inflates it to well within the typical size range
for galaxies.  

\subsection{Total magnitudes of the new members}\label{magpar}

Measuring the brightness of these galaxies from the HST data based on the diffuse
resolved stellar populations is not straightforward.  For WFPC2 data, only $\sim2$ 
mag are visible below the TRGB, with incompleteness and increased photometric errors 
setting in before the detection limit.
ACS data are not much improved for this purpose, reaching only $\sim3$ mag below the TRGB.  In addition, the 
WFPC2 fields are too small to fully contain
the full spatial extent of the galaxies, and foreground contamination is not, in many cases,
easily determined.  This is true for the larger galaxies imaged with ACS as well. 
Nevertheless, we attempt to determine absolute $F814W$-band (I) magnitudes for these 
galaxies using the following method.  For dSphs with pure old stellar populations,
we use Padova model stellar luminosity functions (LFs) with age 12.5 Gyr and metallicity determined
from the color of the RGB based on isochrones.  Details on the models used for the luminosity
functions and isochrones are provided in Section \ref{stellpop}.
We assume these objects have single stellar 
populations and that the models provide a good match to the data.  We then fit the observed RGB 
luminosity function with a power law. The RGB slope is typically $\sim0.3$ but we allow it to range 
between 0.25 - 0.35 although the models permit a slightly smaller range, $0.3\pm0.025$.
Integrating over this fitted LF from the TRGB down to the base of the RGB, we obtain
an estimate of the total magnitude of the RGB population.  Using the model LF, we then 
determine the fraction
of total flux that is in main sequence and horizontal branch vs. RGB stars and apply this 
correction to the RGB magnitude.  Finally, we make a correction for the fraction of the
galaxy that falls off the detector.  This method relies on being able to determine the
slope of the RGB LF, but sparse RGBs extending only 1.5 - 2 magnitudes below the tip in the case of
WFPC2 data, and contamination from foreground stars cause poor fits.  For WFPC2 data,
we therefore simply assume a slope of 0.3, as suggested by the models.  The largest uncertainties 
in the total magnitude come from the uncertainties in power law slope and area correction and together are
typically 0.4 mag for ACS data and 0.6 mag for observations with WFPC2.  The effect of using different
metallicity and age models for an assumed old ($> 4$Gyr) single stellar population is less than
$\pm0.1$ mag. 

For objects with multiple or younger populations (see Section \ref{stellpop}) we additionally include 
these estimated contributions to the total magnitude.  For d0959+68, we only include the contribution for
young main sequence stars under the assumption that this component constitutes the
tidal dwarf and scale a model LF having age 30 Myr with Z = 0.008 to
the observed flux brightward of $M_{F814W} = -2$ to estimate the total flux. 
Objects with obvious multiple components 
are more challenging and we only provide very rough estimates for these by assuming the composite of 
2-3 discrete populations.  For d0926+70 we use a model with
single old 12.5 Gyr stellar population with Z = 0.001, under the assumption that any younger
population present will contribute negligibly to the total magnitude.  For the BCDs d1012+64 and
d1028+70, we assume the populations are a composite of an old 12.5 Gyr stellar population with
Z = 0.0015 and 0.001, respectively, and a young 40 Myr population with Z = 0.008 and 0.003,
respectively.  In these cases, the contribution from the young component is only 0.1 mag.  
The intermediate aged population of d0958+66 cannot be ignored.  We therefore 
factor in missing flux below the detection limit for an assumed 3 discrete populations:
12.5 Gyr with Z = 0.0008, 1.5 Gyr with Z = 0.002, and 60 Myr with Z = 0.004.  The intermediate
age model is scaled in a similar way as is done for the young population using the total flux of visible
AGB stars brightward of the RGB.  The young and intermediate stars here contribute 0.3 mag to the total
based on these models with a model-dependent uncertainty of less than 0.1 mag. 
These estimated total $F814W$ magnitudes are listed in Table \ref{tabX}.

\subsection{Stellar populations of the new members}\label{stellpop}

We investigate the stellar populations and evolutionary histories of these confirmed
dwarf galaxies.
Attempts were made to perform stellar evolution synthesis on the most populated
objects observed with ACS. However, with the bright limiting depth of the observations
providing access to only the brightest RGB and AGB and handful of
upper blue main sequence stars, along with the degeneracy between age and metallicity at the
level of the RGB, uncertainties proved to be very large.  
Because of this we prefer to keep the discussion more general until deeper data are acquired.
We therefore instead use isochrones, overlaying model isochrones on the CMDs for each galaxy
in order to determine the plausible range in ages and metallicities of the stellar populations. 
Padova isochrones
\citep{marigo08,gir08} are used, which are transformed directly on
the ACS/WFC and WFPC2 filter systems.  From the web interface form, we
chose bolometric corrections from \citet{loidl}, circumstellar dust calculations 
as in \citet{groenewegen}, and a Chabrier lognormal IMF \citep{chab01}. We also
choose detailed AGB tracks from \citet{mg07}. 
Padova isochrones were chosen over other models largely due to the fact that
they include stellar isochrones down to very young ages, and several of
our galaxies show evidence for young main sequence populations. Comparisons
using Dartmouth \citep{dotter08} and Basti \citep{basti} isochrones found these to
be systematically more metal poor than the Padova isochrones.
For internal consistency, we therefore only include Padova isochrones in this paper.

For old stellar populations, the color of the RGB is more greatly affected by metallicity
than age.   
In Figures \ref{trgb1} - \ref{trgb9} we investigate the old stellar population
by overlaying isochrones extending up to the RGB phase for a constant 
12.5 Gyr age, having metallicities Z = 0.0001, 0.001, 0.002, 0.004, 0.01.   For
many of the galaxies, the CMDs are well described by a pure old stellar population
without any indication of intermediate age AGB stars or younger main sequence
and blue supergiant stars.  The large spread in the RGB sequence due to photometric
uncertainties makes it difficult to constrain the metallicities. For
WFPC2 observed galaxies, because of photometric uncertainties the old RGB stars can 
be described by a wide range of
metallicities typically between $0.0001 < Z < 0.002$.  Since the photometric errors for the ACS data 
are much smaller, it is possible to put tighter constraints on the metallicity of the
old stellar population.  We find for these galaxies that the majority of RGB stars tend to
fall between $0.0001 < Z < 0.001$ or between $0.001 < Z < 0.002$.

We display the RGB metallicity Z vs total $r^{\prime}$ magnitude for our sample
of new M81 dwarfs in Figure \ref{zmr}.   The bars along the Z axis are not standard uncertainties, but
rather the metallicity of the isochrones having assumed age 12.5 Gyr and colors within $\pm 1\sigma$ of the 
mean RGB color at $F814W = 24.5$.  The photometric
uncertainties are expected to produce most of the observed spread in these RGBs, although in
a couple cases a width slightly larger than can be accounted for by photometric 
uncertainties (d1012+64 and d0959+68) indicates possible intrinsic broadening
and a real metallicity spread.  
The metallicity ranges are provided in Table \ref{Tstellpop}.
For M81 galaxies not part of our HST
follow-up, we use ACS data from the ANGST survey \citep{angst} taken in similar passbands and
reaching about 1 mag deeper.  Using the same method we estimate the allowed metallicity range
of the RGB for these galaxies.  Early and late type galaxies are denoted by color, and as has been noted
previously \citep[e.g.][]{mm98}, separate relations are found for galaxies that are still forming stars and for 
those that are not, offset in the sense that late type galaxies are more metal poor at
a given luminosity.  
Given the large uncertanties, the differences are not signficant in our data. 

To assess ages, we can again use isochrones to investigate the existence of young, intermediate, and old
populations, and can also compare the relative fractions of different populations.
The older stellar populations ($> 1$ Gyr) visible in these CMDs are formed by RGB and AGB
stars. Since AGB stars evolve faster than RGB stars, the ratio of the 2 populations 
provides information about the ages of the older stars.  Following \citet{brad11}, we
take the ratio of RGB and AGB stars as stars 2 mag below and 2 mag above the TRGB, respectively, 
which gives a crude estimate of the ratio of intermediate to old stars.  We use the same color cuts
for the ACS data that they use to extract counts from simulated CMDs for a range of
metallicities and ages and compare our results to their simulations for single 
stellar populations as shown in Figure \ref{agbrgb}.   
For our WFPC2 data, due to the larger uncertainties, we broaden the color constraints defining the
RGB by a factor 1.5, consistent with the increased magnitude errors.   Because the model
values were extracted using older \citet{b94} and \citet{Girardi00} evolutionary tracks while we use more recent
models to estimate metallicities, and as these are strictly model estimates, the reader
is cautioned against directly reading off ages from the plot based on the measured population ratio and metallicity. 
However, we expect the trends to be valid and thus the ratio to be useful for estimating the relative
ages of our galaxies.  Of course, these results also assume a single age for
the old stellar population, whereas some of these objects may have hosted continual
or at least multiple bursts of star formation. 

\subsubsection{BCDs: d1012+64, d0958+66, d1028+70}\label{d1012}

Three of the new objects were considered to be blue compact dwarf galaxies (BCDs) in paper 1:
d0958+66, d1012+64, and d1028+70.  All three are compact in size with effective radii 220-250 pc and
contain significant
numbers of blue stars. For the most part these galaxies are found towards the periphery of the 2nd turnaround
radius of the group, as can be seen in the distribution maps in Figure \ref{3d1}, although
d0958+66 does lie closer to the core region with a physical separation from M81 of only 180 kpc.  

One of these galaxies, d1012+64, has deeper imaging observed with ACS, while the other two
were observed with WFPC2 and only reach $\sim 1.5$mag below the TRGB.  
Originally detected in the Kiso Survey for UV excess galaxies, d0958+66 was previously classified as a 
spiral galaxy while d1012+64 is also cataloged as UGC 5497 and was considered a "diffuse"
object.  Distances found here confirm their M81 group dwarf galaxy status. 

The CMDs shown here corroborate our original classification of BCD for these objects.
All three host
significant young populations in addition to well developed RGBs. Main sequence ($F606W-F814W < 0$),
blue loop, and red supergiant stars are visible indicating very recent or ongoing star 
formation.  We find 24, 56, and 40 stars with colors bluer than $F606W-F814W < 0.6$ and magnitudes 
brighter than the TRGB, indicative of a young population, for d1012+64, d0958+66, and d1028+70, respectively, 
whereas all dSph and transition type objects exhibit no such stars in the CMD with the exception of d1014+68
which has 2.  Isochrones with a range of ages and 
metallicities are overlaid in Figs. \ref{trgb1} - \ref{trgb2}.

We estimate the metallicity of the old population using the overlaid isochrones
assuming the RGB is defined by old stars with age $\sim 12.5$ Gyr.
For d1012+64 we find that isochrones with Z $= 0.001 - 0.0015$ match the RGB well.  However, 
lower ages or metallicities in the range $0.0004 < Z < 0.0018$ are possible. 
For a younger assumed 7 Gyr age, tracks with $0.001 < Z < 0.004$ would also match the RGB.
The metallicities for the other two BCDs are even less well constrained due to photometric
measurement errors and the resulting broadness of the RGB.  However, we find a 
metallicity $\sim 0.001$ ($0.05 Z_{\odot}$) for both of these with an allowed
range of $0.0001 < Z < 0.003$ assuming 12.5 Gyr ages, slightly more metal poor than d1012+64.  
\citet{sambit} estimate metallicities for these galaxies using a L-Z relation for BCDs, 
finding Z $\sim 0.05$ Z$_{\odot}$ for all three.
Metallicities are therefore fully consistent with that
expected for their luminosities. As d1012+64 is the brightest of this sample, it is therefore not
surprising that it also appears to have the highest metallicity.
The young population in d1012+64 is best described with isochrones having Z $= 0.004 - 0.008$ while
for the other two BCDs, more metal poor isochrones with Z $= 0.002 - 0.004$ yield better matches to
the data.  
In Paper 1, from longslit spectroscopy and a comparison of [NII] $\lambda 6584$ and H$\alpha$ lines,
we found an upper limit of $Z < 0.0057$ for the ionized gas in regions around young stars
in d0958+66 and d1028+70, again consistent with these results.  

The width of the RGB provides information about the number and extent of star formation
episodes.  In particular, the width of the RGB for
d1012+64 is visibly larger than for most other objects observed with ACS/WFC (see Figure \ref{ACScmd})
but photometric uncertainties here are also larger due to effects of greater crowding.  The 
measured intrinsic width is slightly broader than expected from the magnitude errors as determined
from false star tests, but not significantly
so.  In the other 2 cases, the width of the RGB is consistent with the WFPC2
photometric uncertainties. This therefore provides no conclusive proof for multiple stellar 
populations with intermediate and old ages. However,
in such cases where old RGB, intermediate aged AGB, and young MS stars are all present, clearly
more than one period of star formation has occured.

In all three cases isochrones for young populations with ages 20-200 Myr provide acceptable 
matches to the bluer points in the data. Even younger ages may be possible for d1012+64 and d1028+70.
For both d1012+64 and d1028+70, star formation likely occurred at least as recently as 60 Myr ago,
while for d0958+66, the CMD exhibits only a weak scatter
of such very blue stars. The majority of young stars in this object fall redward of
$F606W - F814W = 0$, and may indicate that the most recent episode of star formation took place
$\ge 60$ Myr ago.  For d0958+66 and d1028+70, longslit spectra show post starburst
signatures with strong Balmer absorption.  In addition, H$\alpha$ emission is found in both galaxies 
(See Paper 1).

From interferometric radio observations with the Giant Meterwave Radio Telescope (GMRT),
\citet{sambit} detected HI in all 3 of these BCDs.  HI masses were determined to be
$\sim 1.7 - 3.5 \times 10^6$ M$_{\odot}$ for these, with d1012+64 having the lowest mass and
d1028+70 having the highest.  Based on H$\alpha$ and FUV observations, they also
determined star formation rates $\sim 10^{-4}$ M$_{\odot}$ yr$^{-1}$, making these some of the lowest mass
star forming galaxies known.
H$\alpha$ fluxes, indicative of instantaneous star formation, were found to be very low and
capable of being produced by a single O star.  The stronger FUV fluxes they measure from GALEX data
trace recent star formation
within the past $\sim10^8$ years consistent with the results based on the CMDs.
For d1012+64, they also find evidence for possible HI outflows.

The location of red supergiant stars in the CMD are strongly affected by metallicity, thereby
constraining the metallicities of the young populations in these galaxies.
For d1012+64, we compare isochrones to helium burning red supergiant stars just blueward of 
$F606W - F814W = 1$ and find that isochrones with
metallicity Z $> 0.004$ describe the data well.   For d0958+66,
we find $0.003 < Z < 0.008$ with a good match to the data for Z = 0.004, while for
d1028+70, $0.002 < Z < 0.004$ best describe the data. 
As was found from the color of the RGB, d1012+64 appears to be the most metal rich
of the BCDs.  

A significant AGB component also visible in these plots above the RGB may be evidence for an 
intermediate
aged population and may partially contaminate the RGB sequence of old stars.
From visual comparison of the BCD CMDs, the largest AGB population is seen in d0958+66.
AGB tracks with ages between 500 Myr and 4 Gyr approximately bracket the AGB stars in 
all 3 objects assuming metallicities of $0.001 < Z < 0.003$. 

\begin{figure}[t]
\begin{centering}
\includegraphics[angle=0, totalheight=3.0in]{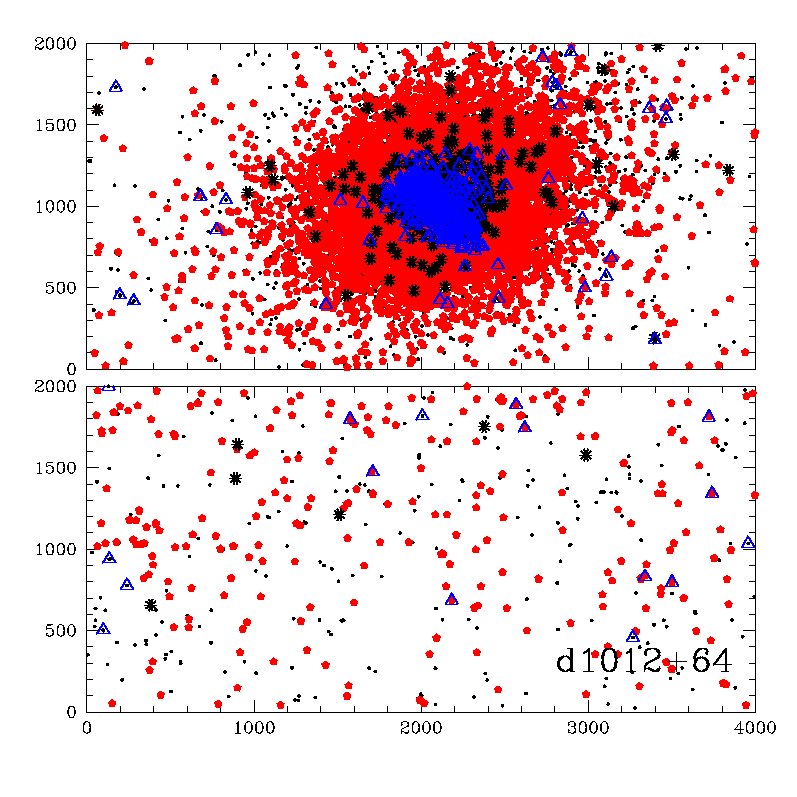}
\caption{Locations of d1012+64 stars in the ACS fields.   Symbols as in Fig. \ref{Locs4}.
\label{Locs5}}
\end{centering}
\end{figure}

\begin{figure}[t]
\begin{centering}
\includegraphics[angle=0, totalheight=3.1in]{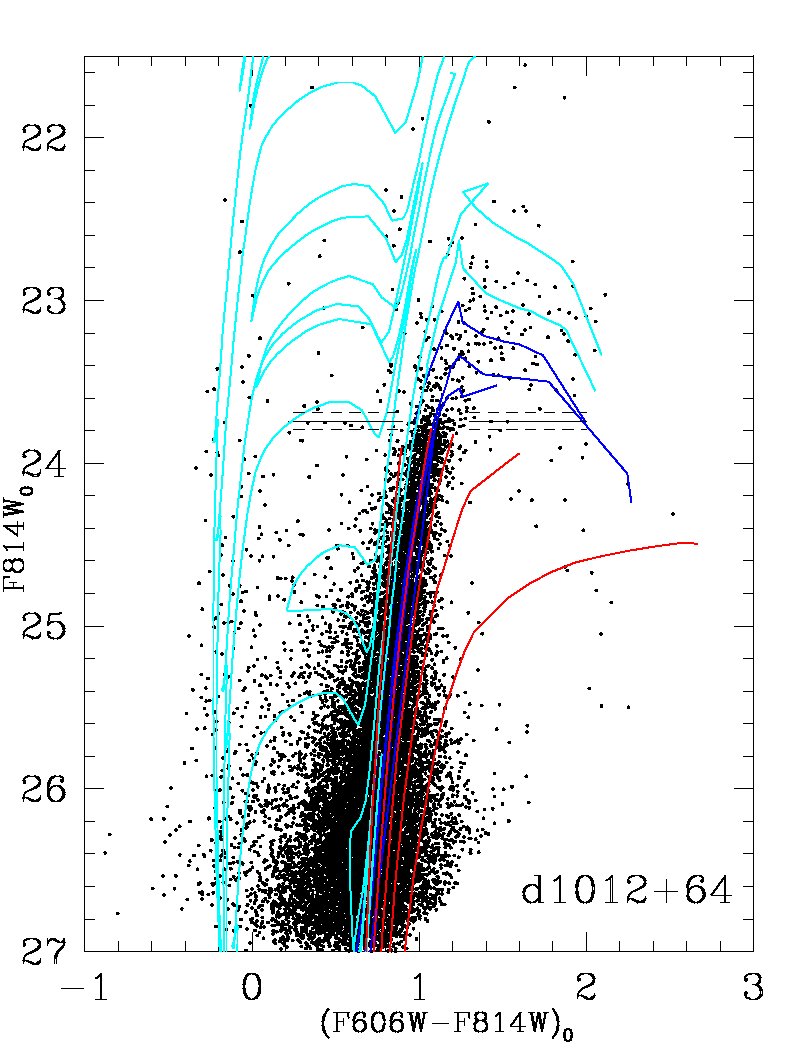}
\caption{CMD for the BCD, d1012+64, observed with ACS.  The TRGB is indicated by a horizontal 
broken solid line with $1\sigma$ uncertainties represented by dashed lines.
Stellar detections from the full ACS
WFC chip 2 are shown. Stellar isochrones from Padova models \citep{marigo08} are provided
for constant age 12.5 Gyr and, from left to right, metallicities Z =
0.0001, 0.001, 0.002, 0.004, 0.01 (red, up to the TRGB). Tracks for (left to right) 1, 2, and 4 Gyr
intermediate age AGB stars are coded in blue with Z = 0.003, 0.003, and 0.002, respectively.
Isochrones for young stellar populations are shown in cyan for (left to right) ages 20, 30, 60, 
and 80 Myr with Z = 0.008, and 200, 500, and 750 Myr with Z = 0.004.
\label{trgb1}}
\end{centering}
\end{figure}

\begin{figure}[t]
\begin{centering}
\includegraphics[angle=0, totalheight=3.1in]{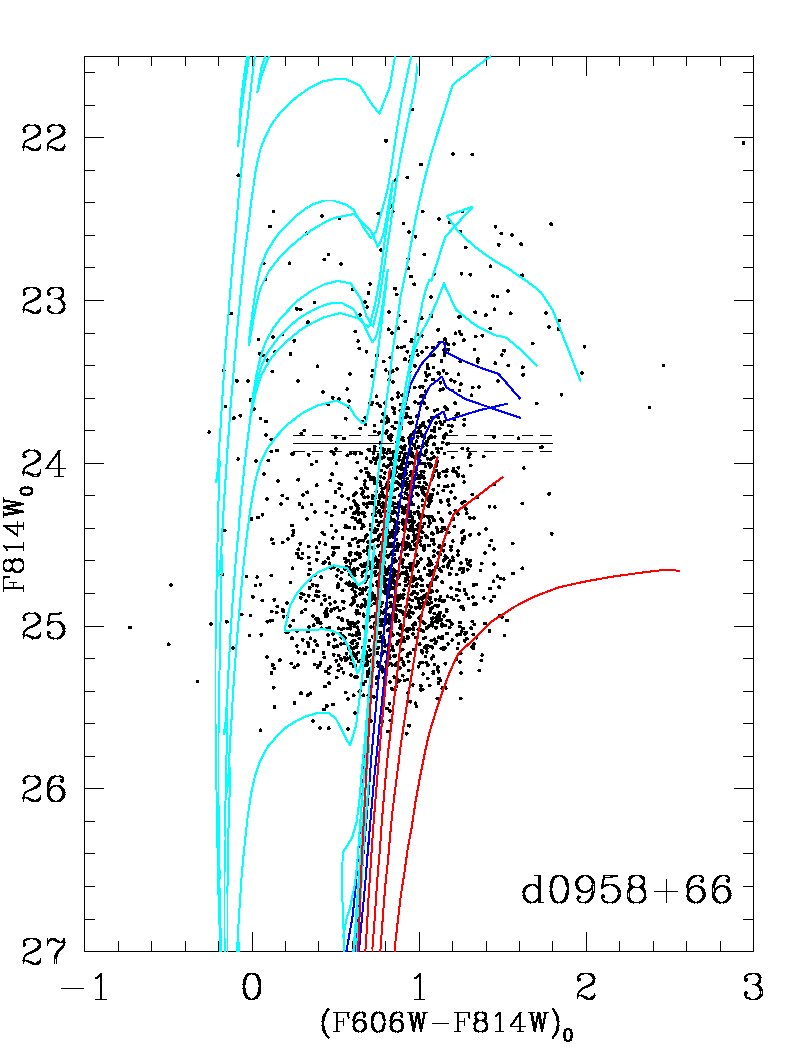}
\includegraphics[angle=0, totalheight=3.1in]{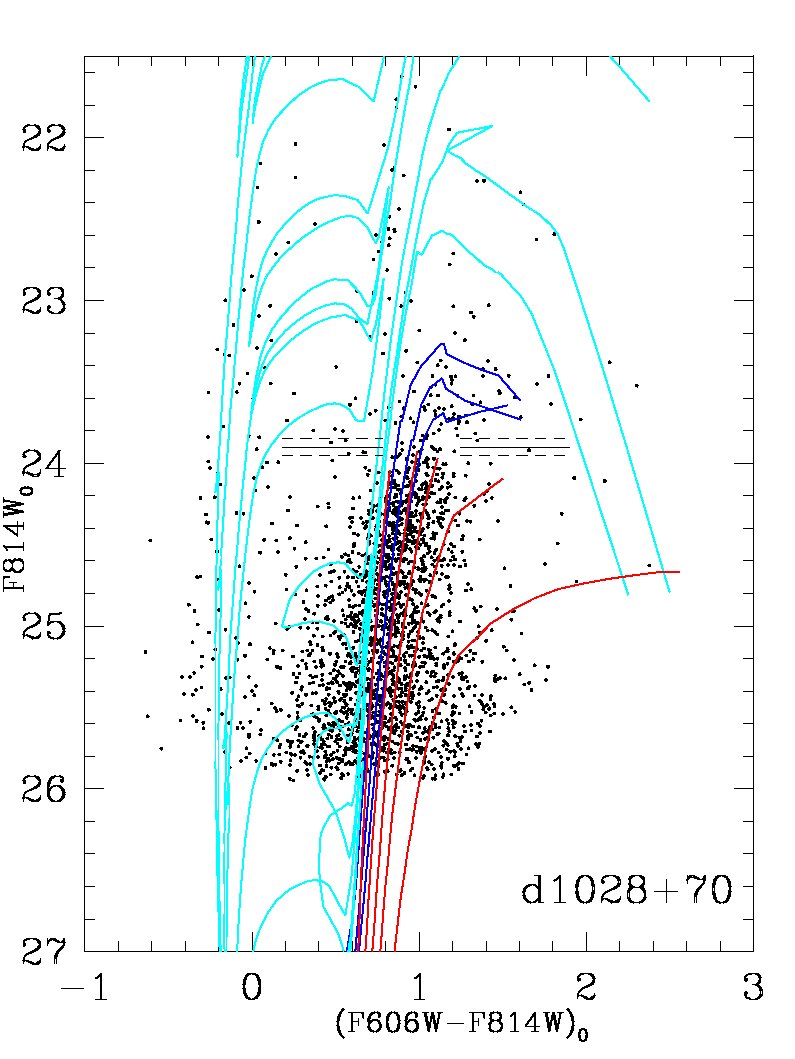}
\caption{CMDs for two M81 BCDs observed with WFPC2 in chip 3.
The TRGB is indicated by a horizontal broken solid line with $1\sigma$ uncertainties
represented by dashed lines. For both galaxies, red isochrones correspond to old 12.5 Gyr RGB stars 
(left to right: Z = 0.0001, 0.001, 0.002, 0.004, 0.01) while blue isochrones represent intermediate age 
AGB sequences for (left to right) ages 1, 2, and 4 Gyr with Z = 0.002.   Young stellar 
population isochrones are color coded cyan. For d0958+66, these correspond to (left to right) 
ages 20, 30, 60, 80, 200, and 500 Myr with Z = 0.004, and 750 Myr with Z = 0.003 while
for d1028+70 these are displayed for ages 20, 30, 60, 80, and 200 Myr with Z = 0.003, and ages
350 and 500 Myr with Z = 0.002.
\label{trgb2}}
\end{centering}
\end{figure}

\begin{figure}[t]
\begin{centering}
\includegraphics[angle=0, totalheight=3.1in]{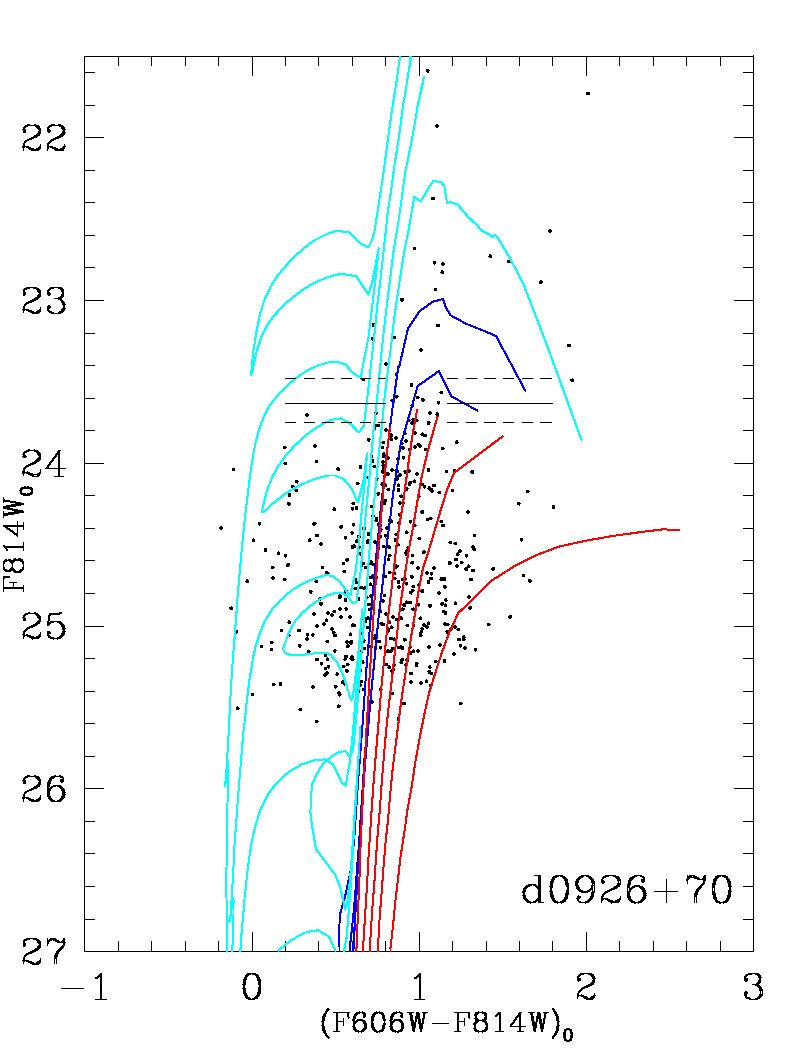}
\caption{CMD for a transition dI/dSph, d0926+70 observed with WFPC2.  Data
from chip 3 are displayed.  The broken solid line denotes the best fit TRGB,
the dashed lines represent the $1\sigma$ uncertainties in the RGB tip measurement.
Red isochrones correspond to old 12.5 Gyr RGB stars (left to right: 
Z = 0.0001, 0.001, 0.002, 0.004, 0.01), blue tracks correspond to intermediate
age AGB stars (1 Gyr with Z = 0.0015 and 4 Gyr with Z = 0.001), and cyan 
represent young populations for (left to lower right): 80 Myr with Z = 0.002 and 
150, 270, and 500 Myr with Z = 0.0015. 
\label{trgb3}}
\end{centering}
\end{figure}

\begin{figure}[t]
\begin{centering}
\includegraphics[angle=0, totalheight=3.1in]{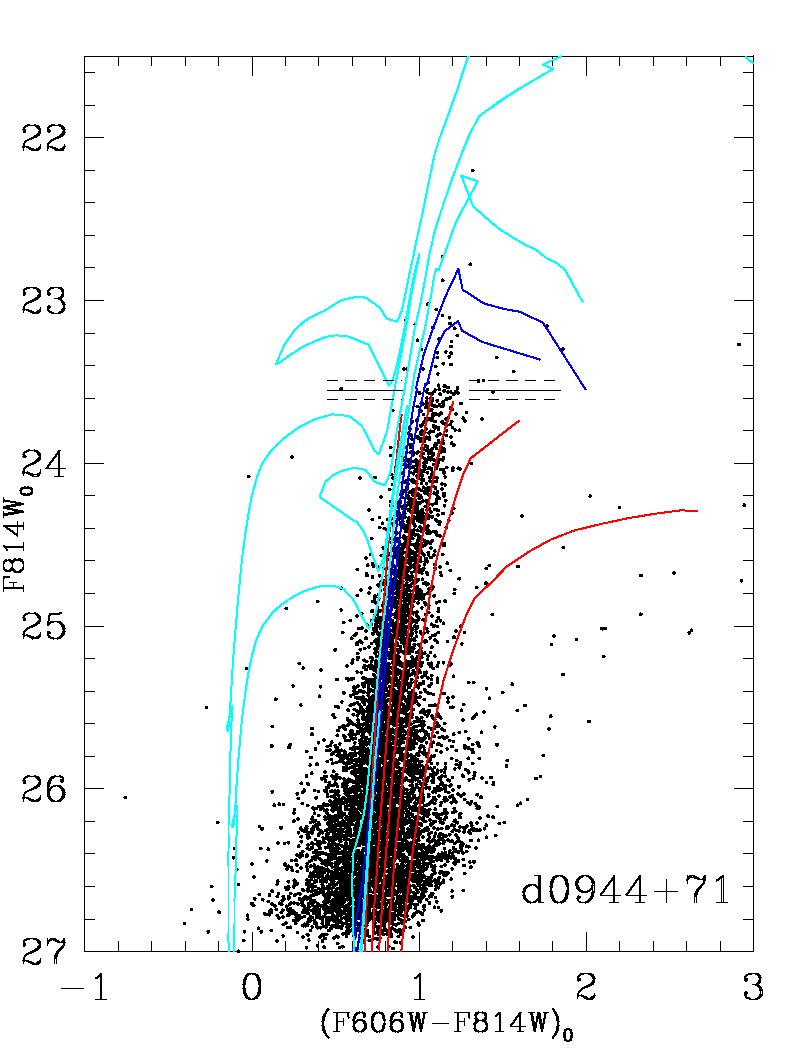}
\caption{CMD and associated TRGB for dwarf spheroidal d0944+71 
observed with ACS.  Stellar detections from the full ACS
WFC 2 detector are shown. Stellar isochrones from Padova models \citep{marigo08} are provided
for constant age 12.5 Gyr and, from left to right, metallicities Z =
0.0001, 0.001, 0.002, 0.004, 0.01 (red, up through the RGB phase) and for ages 1 and 2 Gyr
with Z = 0.003 and 0.002, respectively (blue, AGB phase). 
Cyan sequences represent young stellar populations with ages 90 and 150 Myr 
with Z = 0.01 and 600 Myr with Z = 0.004 (left to right).
\label{trgb4}}
\end{centering}
\end{figure}

\begin{figure}[t]
\begin{centering}
\includegraphics[angle=0, totalheight=3.1in]{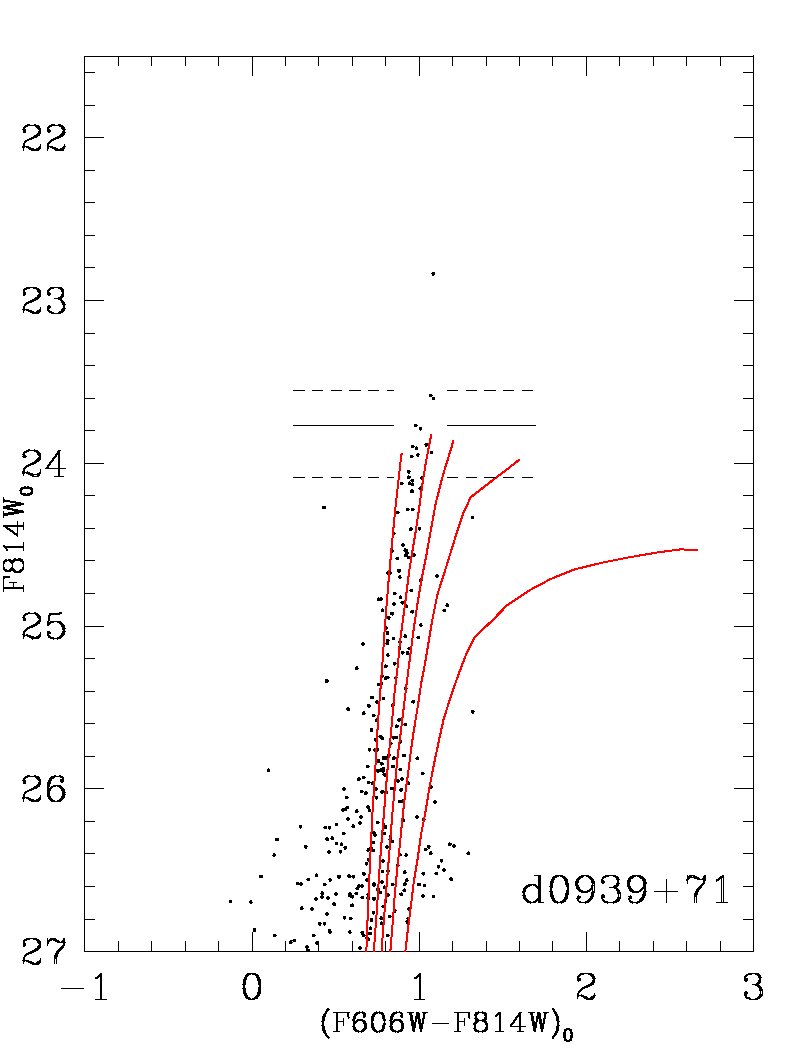}
\includegraphics[angle=0, totalheight=3.1in]{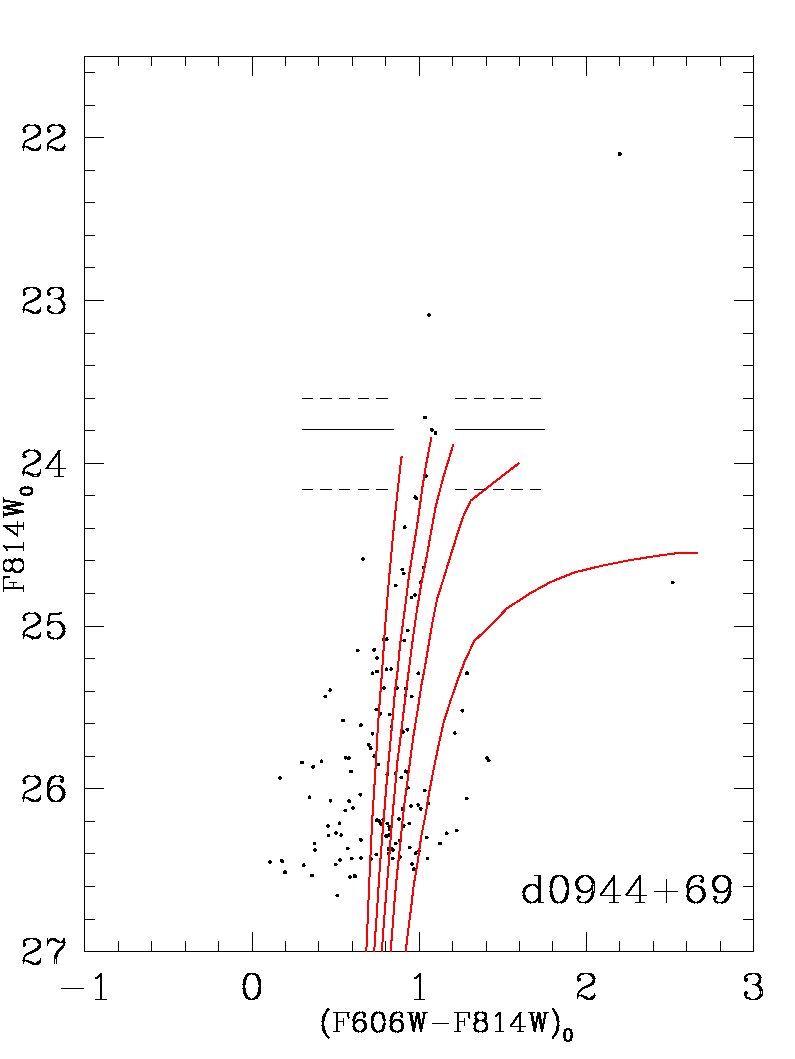}
\caption{CMDs and associated best fit TRGBs (solid broken line) for 2 
dSphs observed with ACS. Stellar detections within a 1 arcmin$^2$ region
centered on the target are displayed.  $1\sigma$ uncertainties are denoted
by dashed lines. Isochrones correspond to old 12.5 Gyr RGB stars (left to right:
Z = 0.0001, 0.001, 0.002, 0.004, 0.01).
\label{trgb5}}
\end{centering}
\end{figure}

From the ratio of AGB/RGB stars, we investigate the ages of the old stellar population
under the assumption of an initial dominant star formation episode.
For BCD d1012+64, we find a ratio of AGB/ RGB stars in the core region of
0.06 while over the full extent of the chip the ratio is slightly lower, at 0.05.  
From Figure \ref{agbrgb},
we find that this corresponds very roughly to an age of around 5-8 Gyr in the core for $-1.5 < [Fe/H] < -1$
(based on the color of the RGB)
and 8-10 Gyr when including the galaxy outskirts. The radial spread in age over the galaxy is not 
surprising since we know that very recent star formation has only occured in the central core 
based on the location of the blue stars in this galaxy, as seen in Figure \ref{Locs5}.   
For BCD d1028+70, we find a similar but slightly higher ratio of 0.067 which, with metallicity
between $-1.78 < [Fe/H] < -1.0$, corresponds to an age of about 5-9 Gyr for the old
population, while for BCD d0958+66, we obtain a much higher ratio of 0.19, suggesting a 
major period of star formation more recently than 3 Gyr ago.

Spatial distributions (Figures \ref{Locs1}, \ref{Locs2}, \ref{Locs5}) of the different stellar 
populations can also be used to probe
the stellar evolution. For these 3 galaxies with significant populations of RGB, AGB, and
MS/blue He-burning stars, we find that AGB stars generally follow approximately the same distribution 
as RGB stars although are more concentrated towards the core in d1012+64 while the
young stellar component for all three is found to be concentrated
towards the core in each.  In the case of d0958+66, the concentration
appears to be extended along the major axis of the galaxy, while in d1028+70, the
blue stars are centrally concentrated.  For d1012+64, the blue stars are clustered
in the center of the galaxy, but are elongated along the minor axis of
the galaxy.  This can be seen in a contour plot, Figure \ref{prof1}, where black contours correspond
to the RGB component while blue contours represent the younger stars.  \citet{sambit}
display FUV contours for these objects which, as one would expect, coincide with our
blue population contours and for d1012+64 are also elongated along the minor axis and
concentrated near the core of the galaxy.  

To investigate whether star formation episodes may have
propagated through the galaxy over long timescales, we compared CMDs for each BCD as a function 
of distance from the center, along both the minor and major axes.  We find no significant changes
in the location of the RGB, and therefore in metallicity or age, further from the core region.  The primary differences are
a slightly broader RGB for stars near the core (due in part to larger photometric errors
where crowding is greater and to the presence of multiple stellar populations), and the gradual 
disappearance of younger stars further
from the core consistent with the picture of a constricting star formation region confined
to the galaxy cores.  
Recent star formation limited to the central core for all 3 galaxies
suggests the HI gas is only dense enough for star formation in the deepest part of the potential
well of these small galaxies.

\subsubsection{dI/dSph Transition: d0926+70}\label{d0926}

This galaxy was considered a dwarf irregular in our original discovery paper. 
In the WFPC2 imaging, it is found to contain several blue stars clustered near the 
galaxy centroid, although perhaps slightly
off-center (Figure \ref{Locs1}). This is difficult to determine as the object is not entirely
imaged within the WFPC2 field, and the shape of the galaxy appears to be slightly extended and
irregular as well.

Plotting
stellar detections on the CMD, Figure \ref{trgb3}, we find a handful of stars in this galaxy have
$(F606W - F814W)_\circ \sim 0.0$.  There are also a number of stars brightward of
the RGB.  We include tracks for young stellar populations having age and metallicities
80 - 500 Myr with Z = 0.0015 - 0.002, and for intermediate aged populations
with ages and metallicies 1 Gyr with Z = 0.0015 and 4 Gyr with Z = 0.001.

Based on the isochrones we find some evidence for
a young population of young blue and red helium burning stars with ages $> 100$ Myr.   
It is not clear whether the
stars brightward of the RGB truly form an intermediate aged population or
whether they are also part of a fairly young stellar component.  RGB stars
are consistent with an old metal poor population having $0.0001 < Z < 0.0023$.  
However, the AGB/RGB ratio is rather high, at 0.08, corresponding to an age
of $\sim 4-7$ Gyr.  If the dominant population has an age as young as the
fraction of AGB stars would suggest, the RGB and most stars brightward of the RGB could 
potentially be accounted for by an intermediate aged population with Z $\sim 0.001$.  
All stellar detections are well represented with metal poor isochrones.  The AGB         
population is much better described using metal poor isochrones than slightly more
chemically enriched ones and even the young component can be represented with 
metal poor isochrones.  There is no evidence for significant metal enrichment in this galaxy.

\citet{sambit} observed d0926+70 with the GMRT and did not detect
HI to a $3\sigma$ upper limit of $0.31 \times 10^{6}$ M$_{\odot}$.  It is plausible that
the HI mass in this galaxy is simply lower than this, as the star bursting brighter BCDs
contain as little HI as $1.7 \times 10^6$ M$_{\odot}$.  Given the excess of 
blue stars towards the core of this object, the irregular shape of the galaxy, and the 
roughly determined young age for the blue stellar component, we consider
this object to be a dwarf 
transition object with properties intermediate between dIs and dSphs. Such objects known in the
Local Group include e.g. Phoenix, LGS3, and Leo T.   With an effective radius of 240 pc and brightness
M$_R = -9.7$, this is one of the smallest galaxies with recent star formation known.  
As would be expected for a late type galaxy, this object does not lie in the core of the group.
In the lower panel of Figure \ref{3d1}, it can be seen as the object well separated from
other galaxies at supergalactic cartesian coordinates (2.7, 2.1, 0.006), and with a distance from M81 of 
345 kpc.

\subsubsection{dSph: d0944+71}\label{d0944ds}

The one dwarf spheroidal with well built up RGB is d0944+71, observed with ACS.
Although a few AGB stars are present brightward of the TRGB, the stellar population of 
d0944+71 is dominated by RGB stars (Figure \ref{trgb4}).  The RGB itself has a slightly
higher metallicity than the other ACS observed dSphs, with the sequence
bracketed largely by $0.001 < Z < 0.002$ isochrones. 
No obvious main sequence is present in the CMD to our limiting
magnitude, although we find several very blue stars associated with this galaxy.  
These exhibit a slight excess and concentration in the core of this galaxy over the 
scatter of blue stars seen in other ACS fields (see Figure \ref{Locs4}), indicating that they are associated 
with the galaxy. However, from the CMD we find that these are clustered at the faintest magnitudes 
where photometric uncertainties are largest.  If these are young stars, 
any low level residual star formation is constrained by isochrones to have occured $\geq 150$ Myr
ago.   AGB stars are sparse, and we find an AGB/RGB ratio of $\sim 0.02$, indicative of a very old 
population.  The few AGB stars present can therefore be accounted for with a strictly old population.  
The picture from the CMD then is of 
a dominant, and perhaps exclusively, old population.  Blue stars are attributed to either
foreground contamination, photometric errors, or very low level star formation at more recent epochs.  
Another non-detection in HI, \citet{sambit} find a 
$3\sigma$ upper limit of $0.31 \times 10^{6}$ M$_{\odot}$.

In our original paper, we had considered this
galaxy to be a possible dwarf irregular due to the large stellar population
which included a few visibly brighter stars.  Given the lack of a significant young component to the
stellar population present in the CMD, we now consider this object to be a
dSph even though it does host a scattering of potentially young stars.   Interestingly, this object
lies at large distance from the core of the group, $\sim335$ kpc distant from M81.  It is
easily seen on the lower panel of Figure \ref{3d1} well separated from other large galaxies
at supergalactic coordinates (2.6, 2.1, 0.1).  This large distance may explain the very
low level of star formation this galaxy may have undergone perhaps as recently as 150 Myr ago.  

\subsubsection{Classical and Ultra-faint dSphs}\label{d0939}

\begin{figure}[t]
\begin{centering}
\includegraphics[angle=0, totalheight=3.1in]{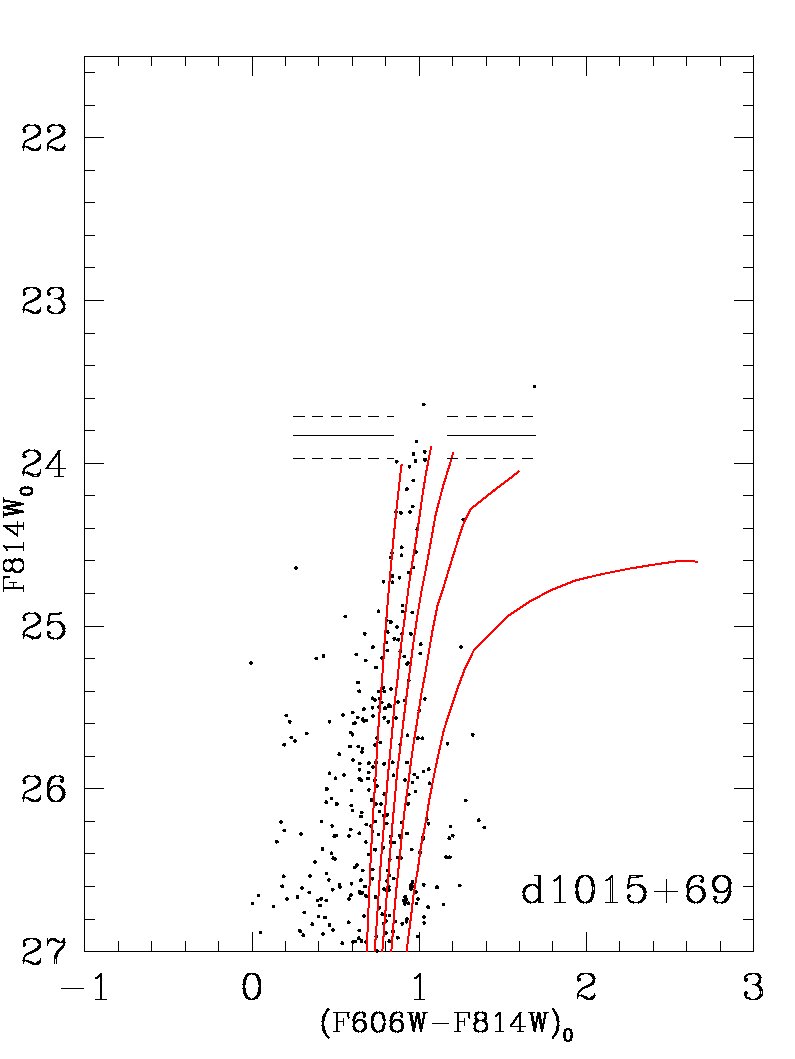}
\includegraphics[angle=0, totalheight=3.1in]{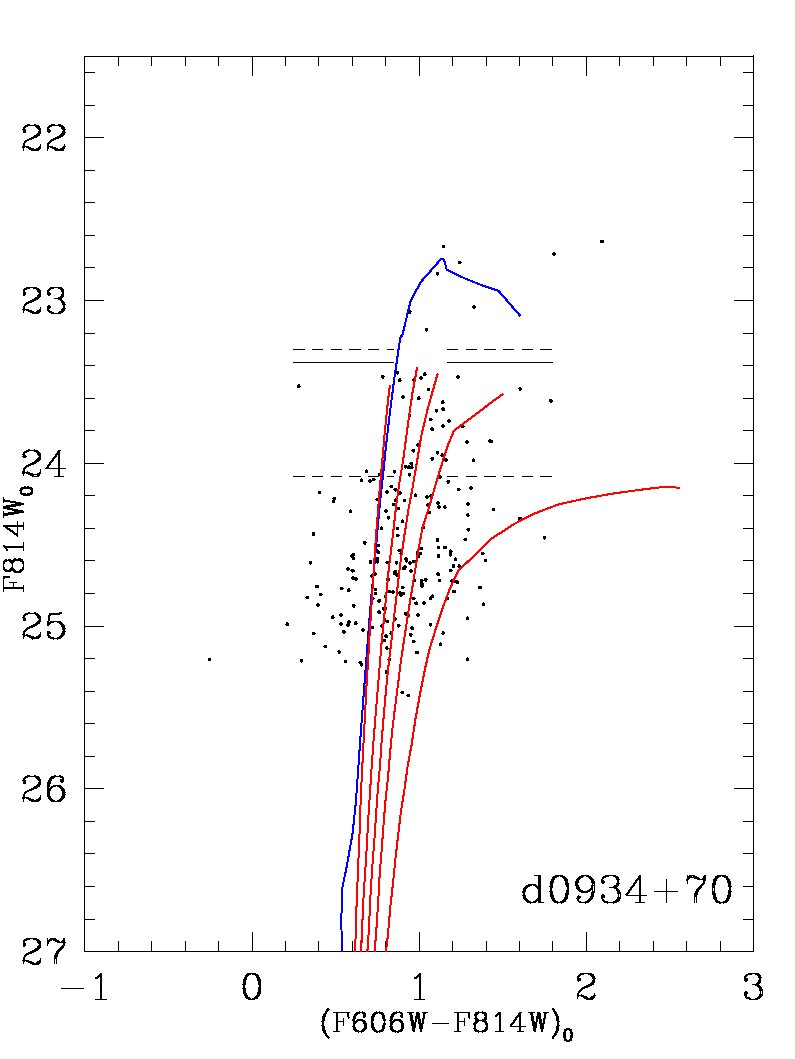}
\caption{CMDs and best fit TRGB denoted (solid broken line) for another
2 dwarf spheroidals. d1015+69 and d0934+70  were observed with ACS and 
WFPC2, respectively.  For d1015+69, stellar detections from a 1 arcmin$^2$ region are plotted.
For d0934+70, we show only stellar detections from WFPC2 chip 2.
Overlaid are stellar isochrones 
for constant age 12.5 Gyr and, from left to right, metallicities Z =
0.0001, 0.001, 0.002, 0.004, 0.01 (up to the RGB phase).  
For d0934+70, we additionaly include an
AGB isochrone (blue) for an intermediate aged 1 Gyr population with Z = 0.002. 
\label{trgb6}}
\end{centering}
\end{figure}

\begin{figure}[t]
\begin{centering}
\includegraphics[angle=0, totalheight=3.1in]{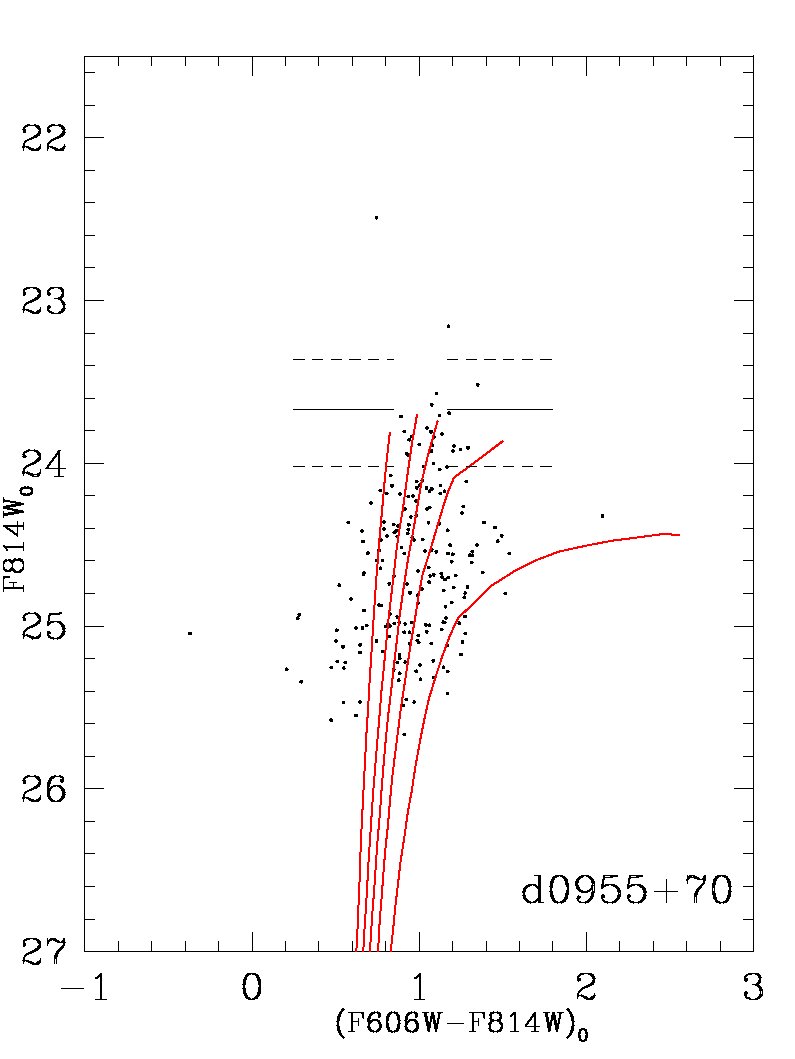}
\includegraphics[angle=0, totalheight=3.1in]{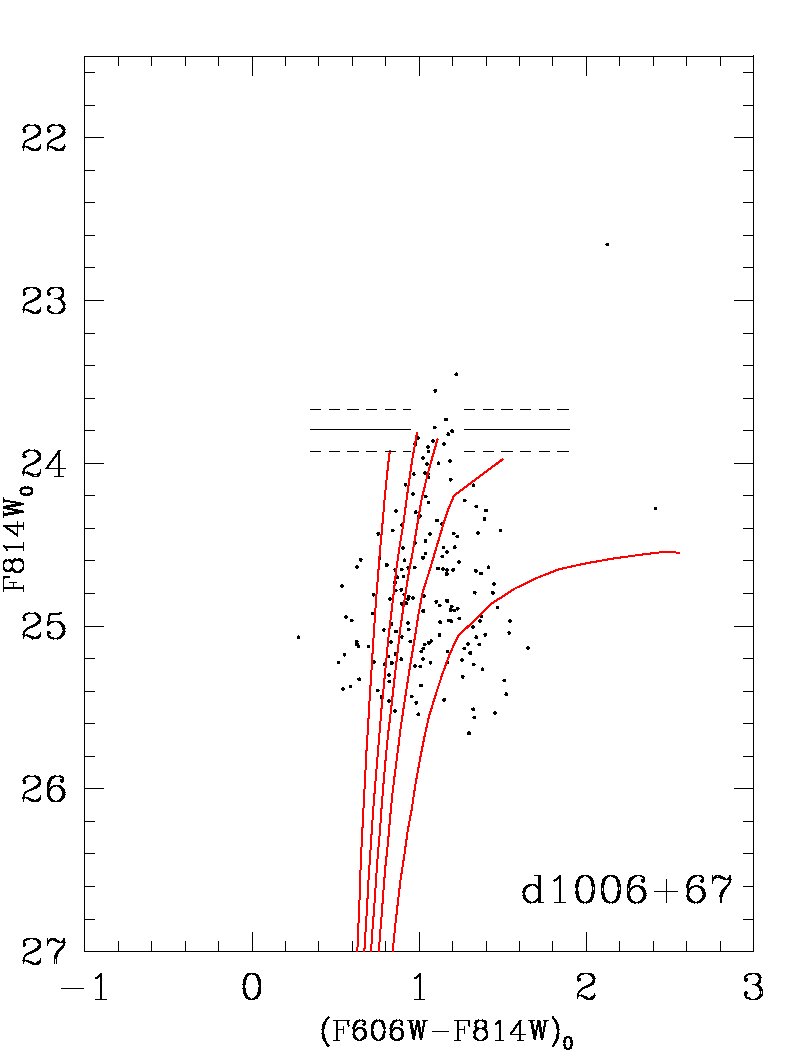}
\caption{CMDs for 2 dSphs observed with WFPC2.  Data
from chip 3 are displayed.  The broken solid line denotes the best fit TRGB,
the dashed lines represent the $1\sigma$ uncertainties in the RGB tip measurement.
Isochrones correspond to old 12.5 Gyr RGB stars (left to right:
Z = 0.0001, 0.001, 0.002, 0.004, 0.01).
\label{trgb7}}
\end{centering}
\end{figure}

\begin{figure}[t]
\begin{centering}
\includegraphics[angle=0, totalheight=3.1in]{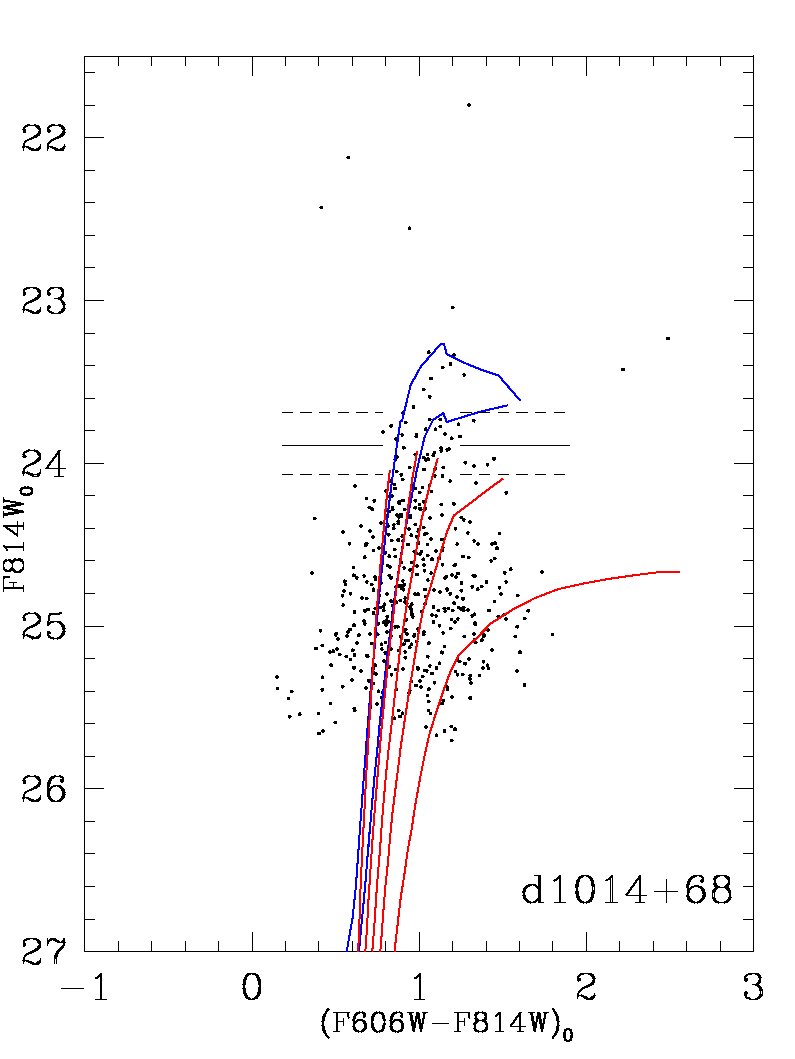}
\includegraphics[angle=0, totalheight=3.1in]{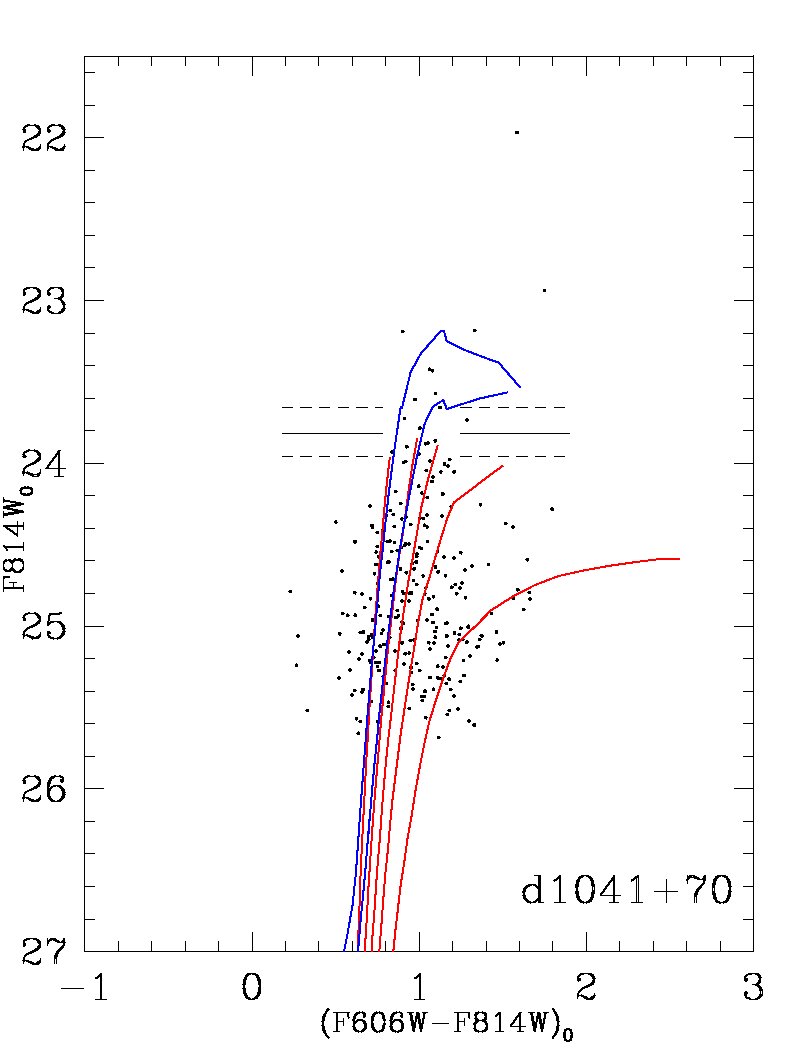}
\caption{CMDs with TRGBs shown for another 2 dSphs with
data from WFPC2 in chip 3. Isochrones correspond to old 12.5 Gyr RGB stars (Red, left to right:
Z = 0.0001, 0.001, 0.002, 0.004, 0.01).  
Two intermediate aged AGB tracks of 1 and 4 Gyr with Z = 0.002 are also shown (left to right).  
\label{trgb8}}
\end{centering}
\end{figure}

\begin{figure}[t]
\begin{centering}
\includegraphics[angle=0, totalheight=3.1in]{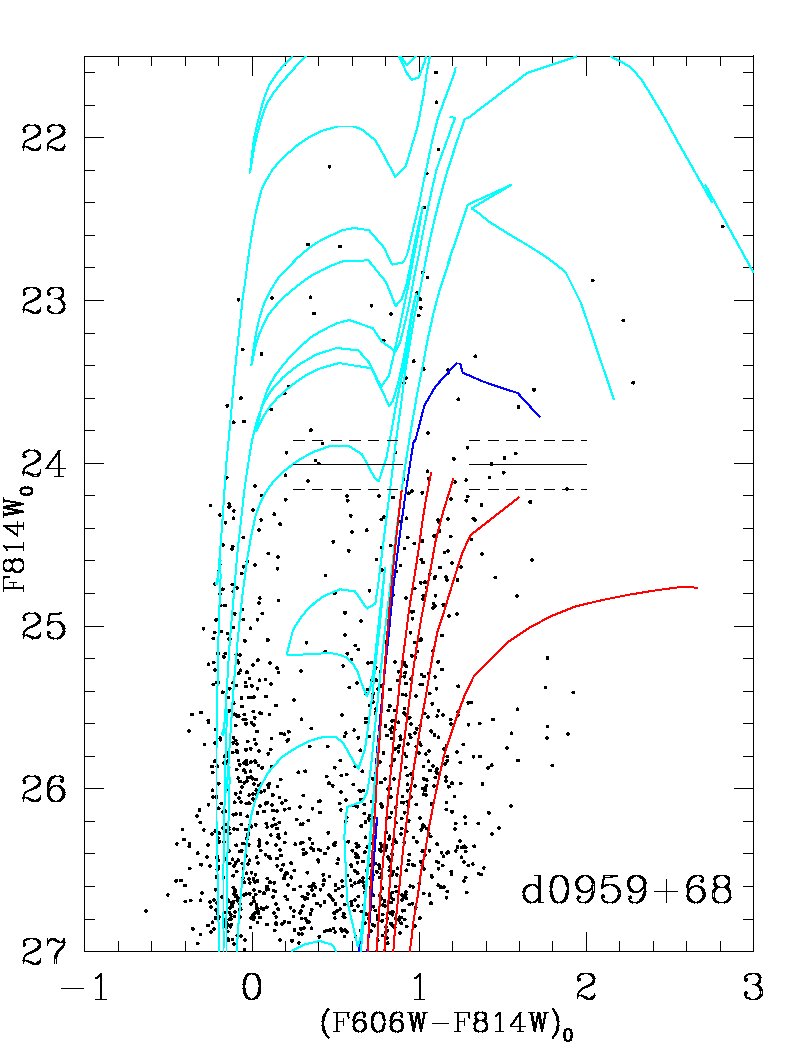}
\caption{CMD and best fit TRGB denoted (solid broken line) for tidal dwarf 
galaxy d0959+68.  Stellar detections from the full ACS
WFC chip 2 are shown. Stellar isochrones 
are shown for constant age 12.5 Gyr and, from left to right, metallicities Z =
0.0001, 0.001, 0.002, 0.004, 0.01 (red, up to the TRGB), and for a single 4 Gyr
intermediate aged AGB population with Z = 0.002 (blue).
Isochrones for young stellar populations are displayed for
(left to right, cyan): 30, 60, and 80 Myr with Z = 0.008, and 200 and 400 Myr with Z = 0.004.
\label{trgb9}}
\end{centering}
\end{figure}

For ACS observed galaxies, we find pure old stellar populations in the very small 
galaxies d0939+71,
d0944+69, and d1015+69. These have poorly populated, but still reasonably well defined RGBs.  The
only stars present in the CMDs (Figures \ref{trgb5}-\ref{trgb6}) are those which
could be described by an old red giant population having low metallicity.
For d0939+71 and d1015+69, 12.5 Gyr isochrones suggest an RGB metallicity 
between $0.0001 < Z < 0.001$, consistent with an old, single age/metallicity population.  
The RGB for d0944+69 is very poorly populated, but may have a slightly higher
metallicity than these other two galaxies.  The isochrones which best bound this 
RGB have metallicities between $0.0001 < Z < 0.002$.  This is also our smallest
and faintest dwarf at $M_{r^{\prime}} = -6.8$ with size R$_e = 90$ pc (as estimated from 
$r^{\prime}$ MegaCam data) or R$_e = 130$ pc and $M_{I} = -9.1$ (as measured from $F814W$ ACS data),
likely within the
range of the ultra-faint dwarf galaxies ($M_V > -8.0$) being discovered within the Local Group.
With no evidence for a younger stellar population, we consider these 3 galaxies to be
non-starforming, old dwarf spheroidals.  Object d1015+69 was considered a possible dI in our
previous paper but is now established to be an early type dSph.

Galaxies observed with WFPC2 having pure old stellar
populations as found from the CMDs (Figure \ref{trgb7}) include d0955+70 and d1006+67.
These are likely to be metal poor, although the metallicities are poorly constrained
due to large photometric uncertainties.  Besides the shallow depth of the imaging, the
RGBs are poorly populated since these galaxies host very small stellar
populations. Three other galaxies observed with WFPC2 (d1041+70, d1014+68, and d0934+70,
Figures \ref{trgb8} and \ref{trgb6}) also display predominantly old
stellar populations although may host a few stars above the tip
of the RGB.  We find that a 1 Gyr, $Z=0.002$ isochrone could explain the handful
of AGB stars in d0934+70 coinciding with this track.
For d1041+70 and d1014+68, AGB sequences
for ages 1 and 4 Gyr with Z = 0.002 approximately bound the few AGB stars in these
galaxies. Although there may have been some small amount of star formation as recently as 
$\sim 1$ Gyr ago, there is no evidence for significant intermediate (or
young) populations in any of these aforementioned galaxies.  Thus, we
also consider these 5 galaxies to be dwarf spheroidals.

Object d1041+70  was considered a possible dI in our
previous paper but is now determined to be an early type dSph without any recent
star formation.  However, the pear shape of this galaxy observed in
MegaCam imaging is noted again here (Figure \ref{Locs2}).  This object is spatially extended along
one axis with very few associated stars outside of the main concentration. It appears to be tidally
disturbed. Interestingly this object lies far from M81 and the core of the group.  This can be seen 
as the faint early 
type galaxy near the BCDs at positive SGZ in the 3-D maps (Figure \ref{3d1}), and in particular nearest to d1028+70.
The distance of d1041+70 from M81 is 265 kpc while the separation from d1028+70 is 122 kpc.  We 
find that 
the major axis of the galaxy (as projected on the sky) does not, however, point toward either of these
galaxies, being $\sim 65$ and 80 deg off from M81 and d1028+70, respectively, in 3-D space or 25 and 32 deg
off in 2-D space.  

\subsubsection{Tidal dwarf: d0959+68}\label{d0959}

The final group member is d0959+68.  The CMD for this object (Figure \ref{trgb9})
exhibits similar numbers of blue and red stars including a main sequence, red and blue helium
burning stars, an RGB, and possibly a small population of AGB stars.  
A young population is clearly 
present and, based on isochrones, may include stars younger than 60 Myr. 
\citet{kkkm} find evidence for star formation within $\sim 10$ Myr
through H$_\alpha$ emission and a derived star formation rate of log(SFR [M$_\odot$ yr$^{-1}$]) $= -3.77$.
As this object has a sparse RGB, we conclude
that it has not had time to build up a significant RGB.  
Stars no older than 4 Gyr could make up the bulk of the RGB.

We propose that the RGB that is present more likely consists of 
older stars in the M81 - NGC 3077 bridge which formed originally
within one of these galaxies, while the young population is related to recently
induced star formation attributed to the interaction between these galaxies.
This object was imaged with both WFPC2 and ACS. Both fields are shown in Figure \ref{Locs3}
for comparison.   The deeper imaging of ACS is
apparent with the much greater number of stellar detections.  The ACS field exhibits
a uniform distribution of red stars, which are more numerous than seen
in other ACS fields away from the dwarf galaxy cores, whereas at the location
of the targetted object we find a strong concentration of blue stars. 
While it is possible that the blue stars represent a clump of younger stars within a more
extended object, we believe that the
very uniform and broad spread of RGB stars indicates that these originate
from the tidal stream itself.
The blue stars, having then formed recently within the HI stream, 
constitute the tidal dwarf.  It is not known whether there is a large enough mass
associated with this object for it to remain a separate (long-lived) bound entity.
The TRGB distance for this object of $4.2\pm0.3$ Mpc may be more 
indicative of the distance to the tidal stream stars than to the tidal dwarf itself.

Table \ref{Tstellpop} provides a summary of the stellar populations found for
each object.  This includes the approximate age and metallicity ranges for young, intermediate, or old
stellar populations.  

\section{Discussion}\label{disc}

Overall, we find a wide range of properties in these small, faint group members - 
from dSphs with very old stellar populations, a
dI/transition dwarf with intermediate age, 3 BCDs forming stars as
recently as 20-200 Myr ago, and a tidal dwarf with presumed pure young stellar 
population.

\begin{figure}[t]
\begin{centering}
\includegraphics[angle=0, totalheight=3.2in]{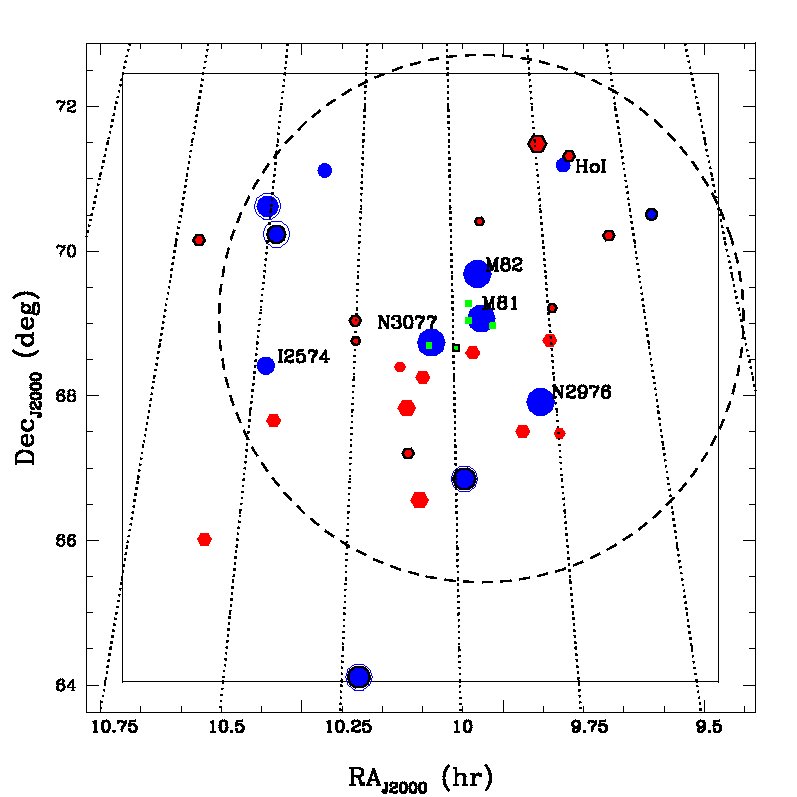}
\includegraphics[angle=0, totalheight=3.2in]{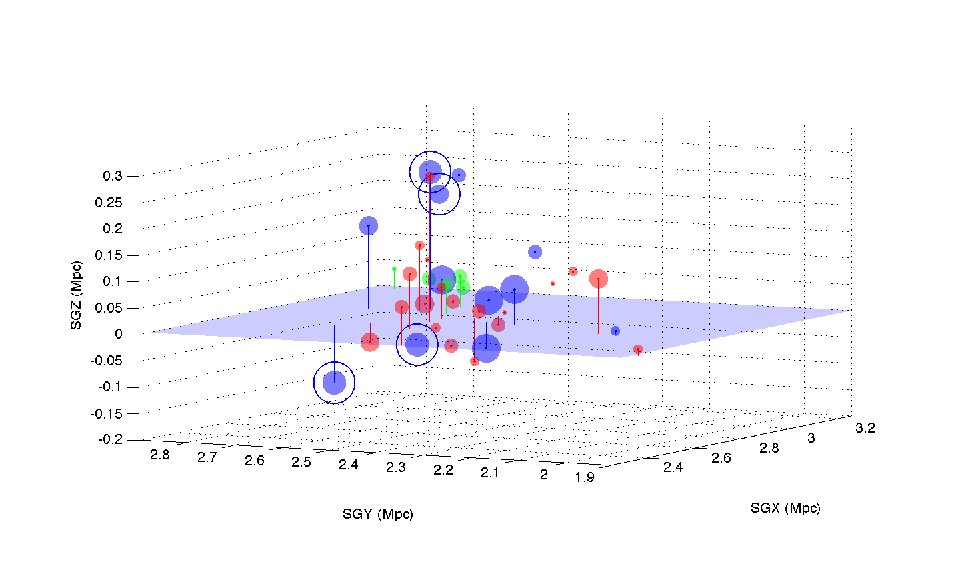}
\caption{Top: Projected distribution of all known M81 group members within this region of the sky.  The box indicates
the original survey region and the dashed circle the projection of the putative surface of 
second turnaround at 230 kpc from M81.  Late, early, and tidal dwarf galaxy types are denoted by symbol:
blue circles, red hexagons, and green squares, respectively.  Size indicates surface brightness 
with larger points shown for galaxies with brighter effective surface brightness.  BCDs are 
encircled and the 14 new detections are outlined. 
Bottom: 3-D distribution in supergalactic cartesian coordinates. Symbols, sizes, and colors as in the 2-D plot.
\label{3d1}}
\end{centering}
\end{figure}

\begin{figure}[t]
\begin{centering}
\includegraphics[angle=0, totalheight=4.8in]{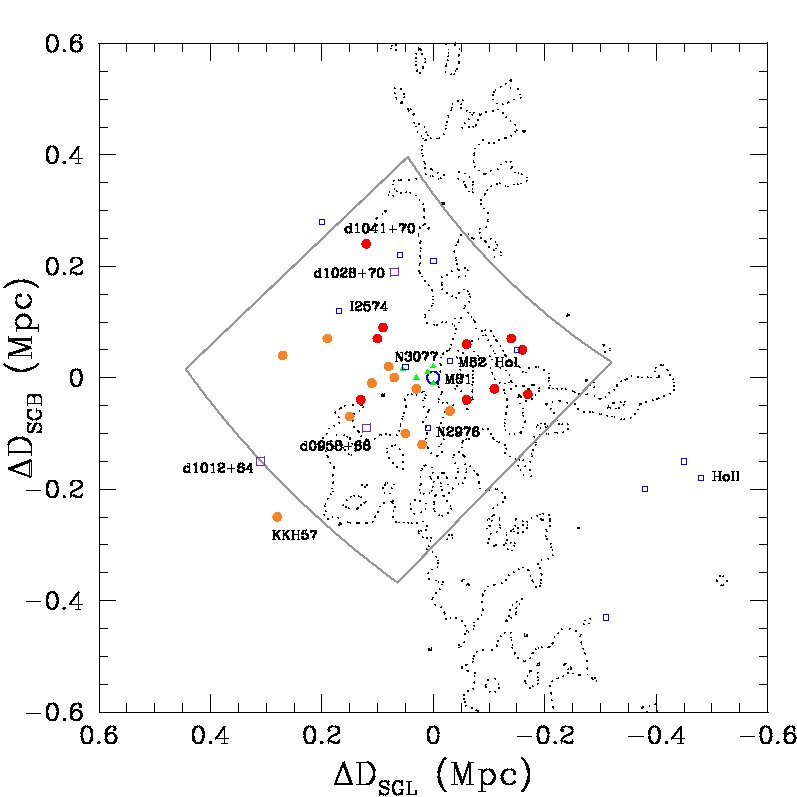}
\caption{Distribution of M81 galaxies within our survey region (gray box). 
Contours from the \citet{sfd98} dust maps are overlayed. Solid red (orange)
circles represent new (previously known) gas deficient galaxies while open symbols
denote gas rich galaxies.  Larger symbols are used for BCDs. 
\label{corelb}}
\end{centering}
\end{figure}

\begin{figure}[t]
\begin{centering}
\plottwo{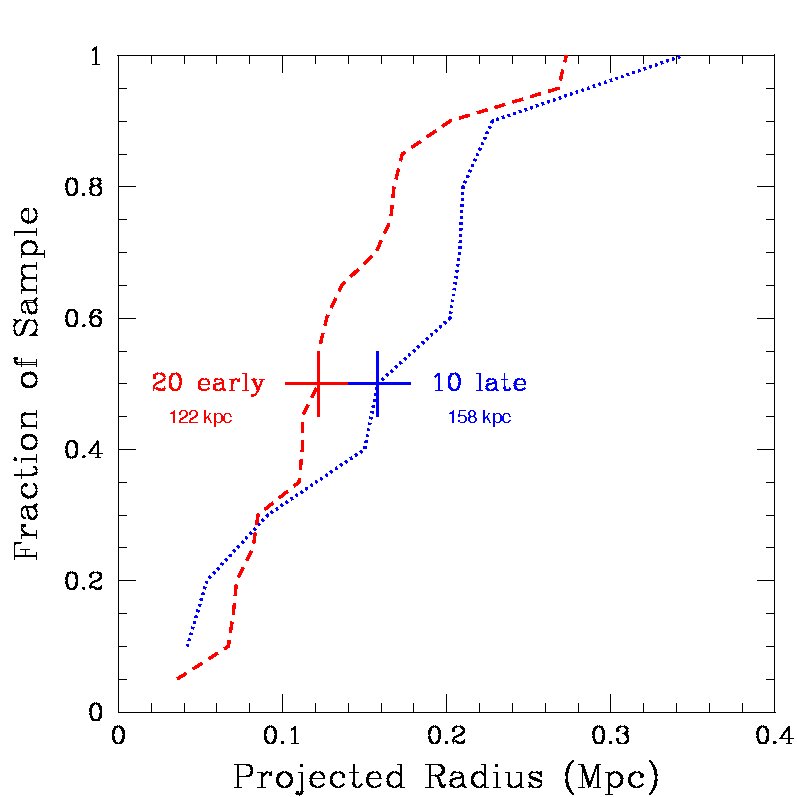}{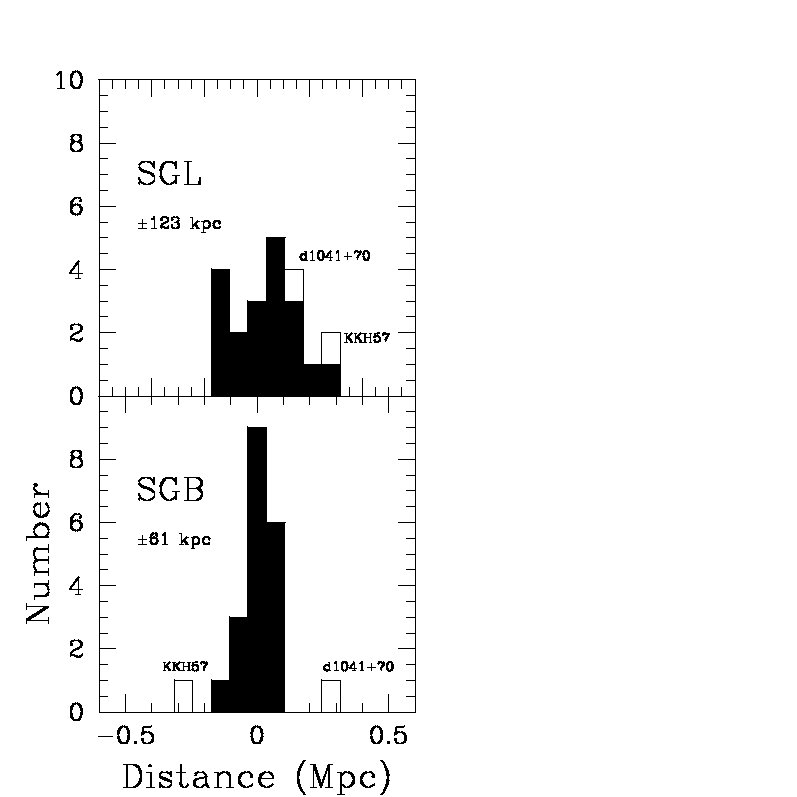}
\caption{
Left: Cumulative distribution of the projected separation of early
and late type galaxies from M81.
Right: Histogram showing the number counts of gas deficient group
members as a function of distance from M81 along supergalactic latitude and longitude.
\label{cumrad}}
\end{centering}
\end{figure}

\begin{figure}[t]
\begin{centering}
\includegraphics[angle=270, totalheight=2.8in]{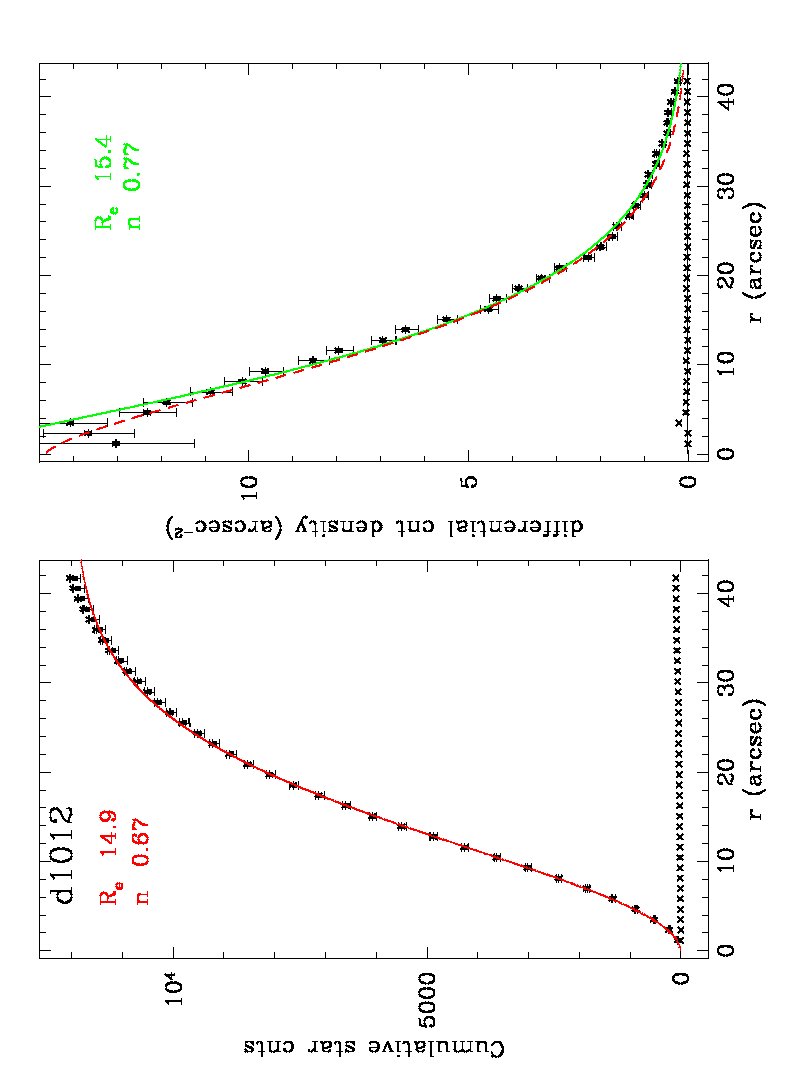}
\includegraphics[angle=270, totalheight=2.8in]{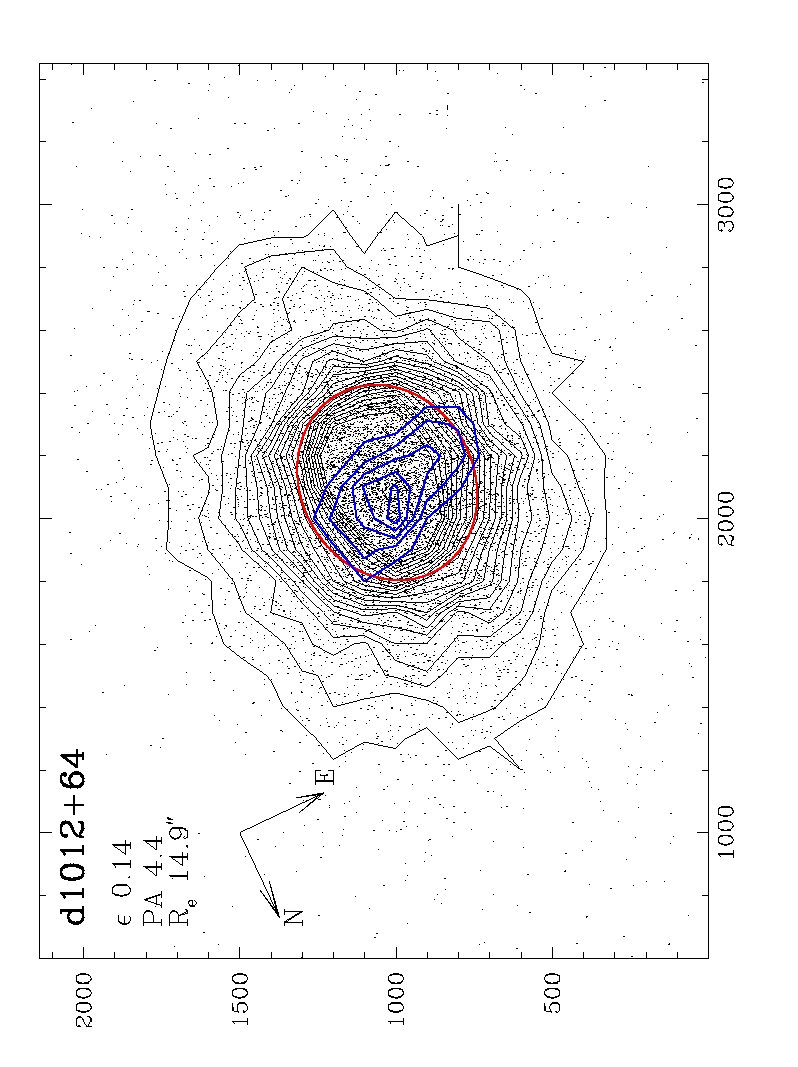}
\caption{Surface brightness profile fit with a Sersic function for d1012+64 star counts.
Both stellar count curve of growth and differential count density plots are shown with
best fits to the profiles overlaid.  Stellar counts in the vicinity of the dwarf galaxy are
shown as asterisks, foreground counts (as measured further from the object centroid, see text)
as crosses, and subtracted stellar counts corrected for foreground contamination
as points.  An average background level, shown by the solid black line,
is used for the differential count correction.  Solid curves are the best fit Sersic function
fits to the data.  The best fit for the curve of growth is also shown in the differential count density
plot as the dashed red line.  A stellar density contour plot is shown in the bottom panel.
Black contour lines indicate RGB stellar density with 0.32 star arcsec$^{-2}$ spacing, blue contours
represent main sequence stars.  The size of the red ellipse corresponds to the effective
radius.
\label{prof1}}
\end{centering}
\end{figure}

\begin{figure}[t]
\begin{centering}
\includegraphics[angle=270, totalheight=3.0in]{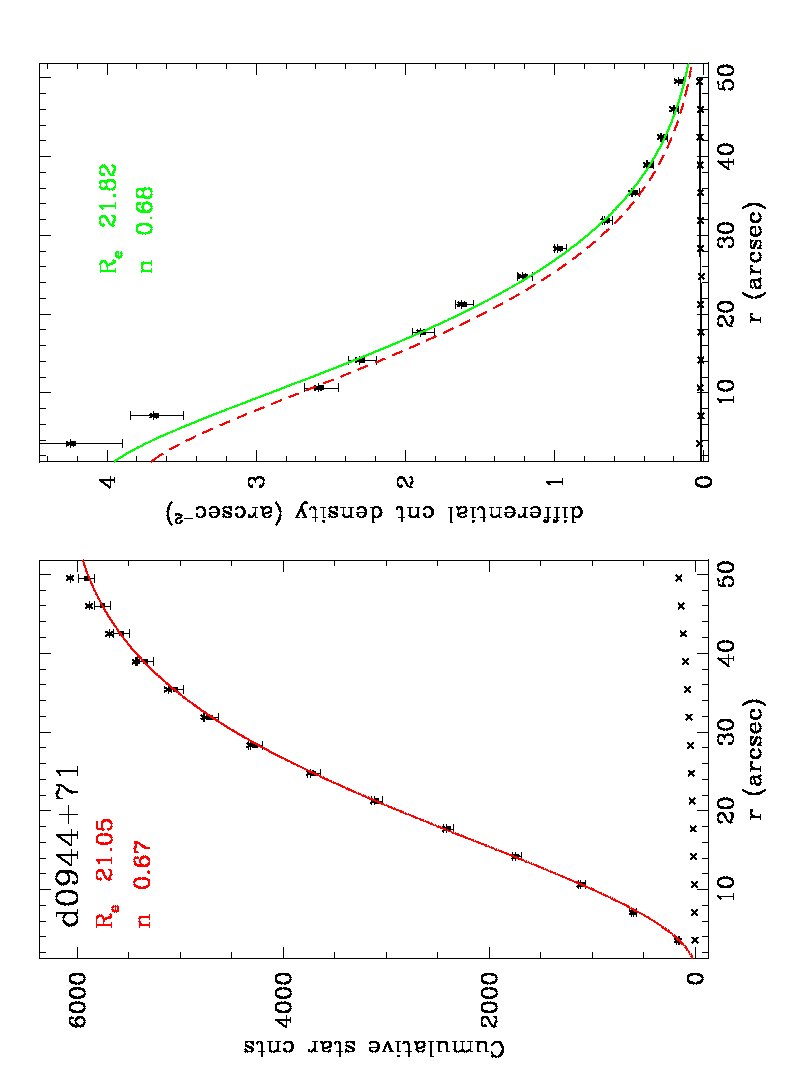}
\includegraphics[angle=270, totalheight=3.0in]{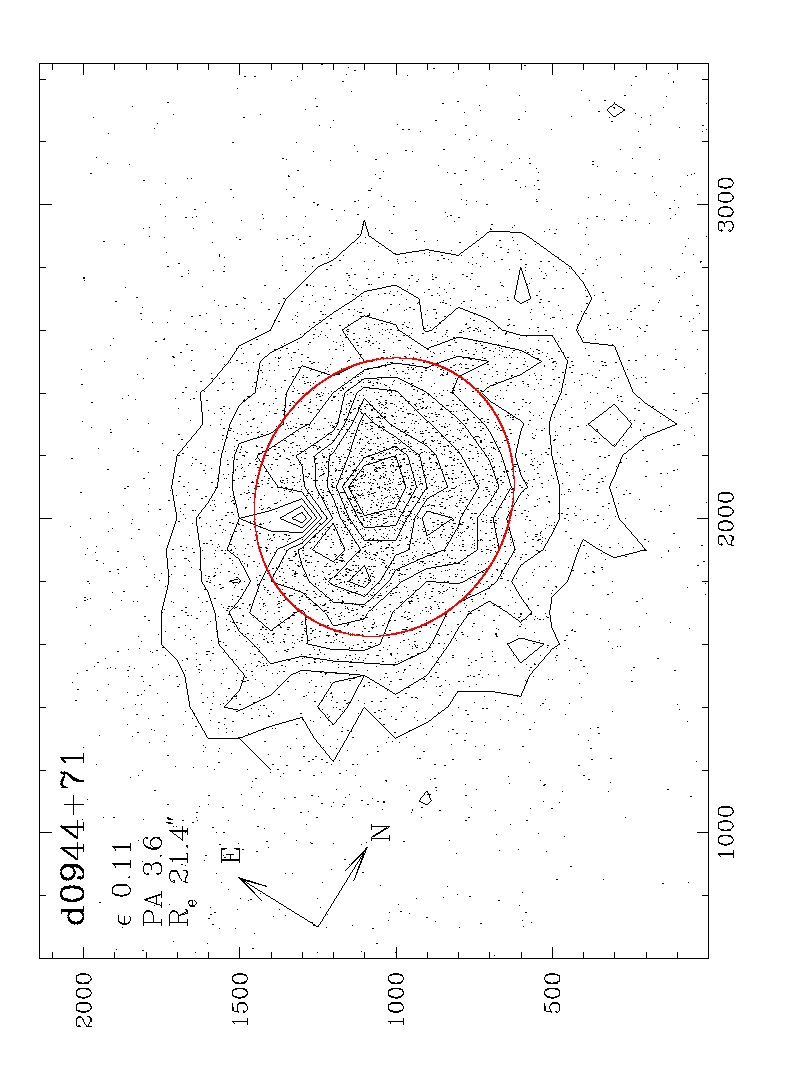}
\caption{Surface brightness profile fit with a Sersic function for d0944+71 star counts,
symbols as in Fig. \ref{prof1}.  Contour lines in the bottom panel indicate RGB stellar density
with 0.32 star arcsec$^{-2}$ spacing.
\label{prof2}}
\end{centering}
\end{figure}

\begin{figure}[t]
\begin{centering}
\includegraphics[angle=270, totalheight=3.0in]{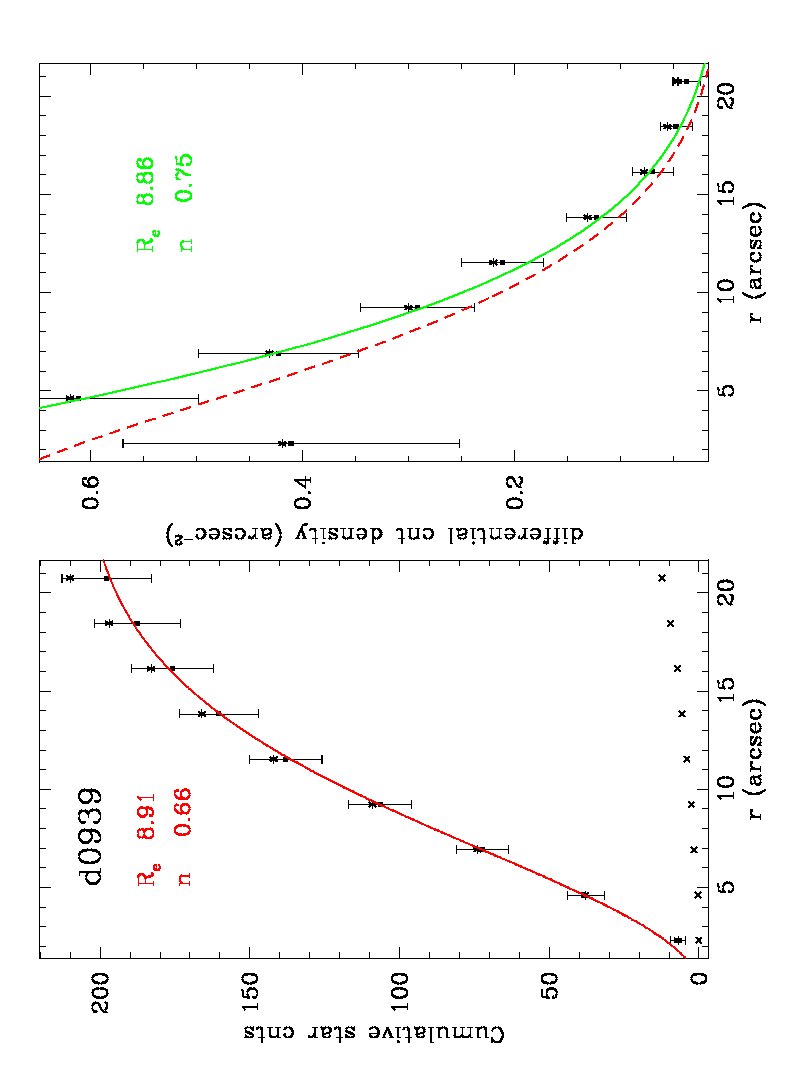}
\includegraphics[angle=270, totalheight=3.0in]{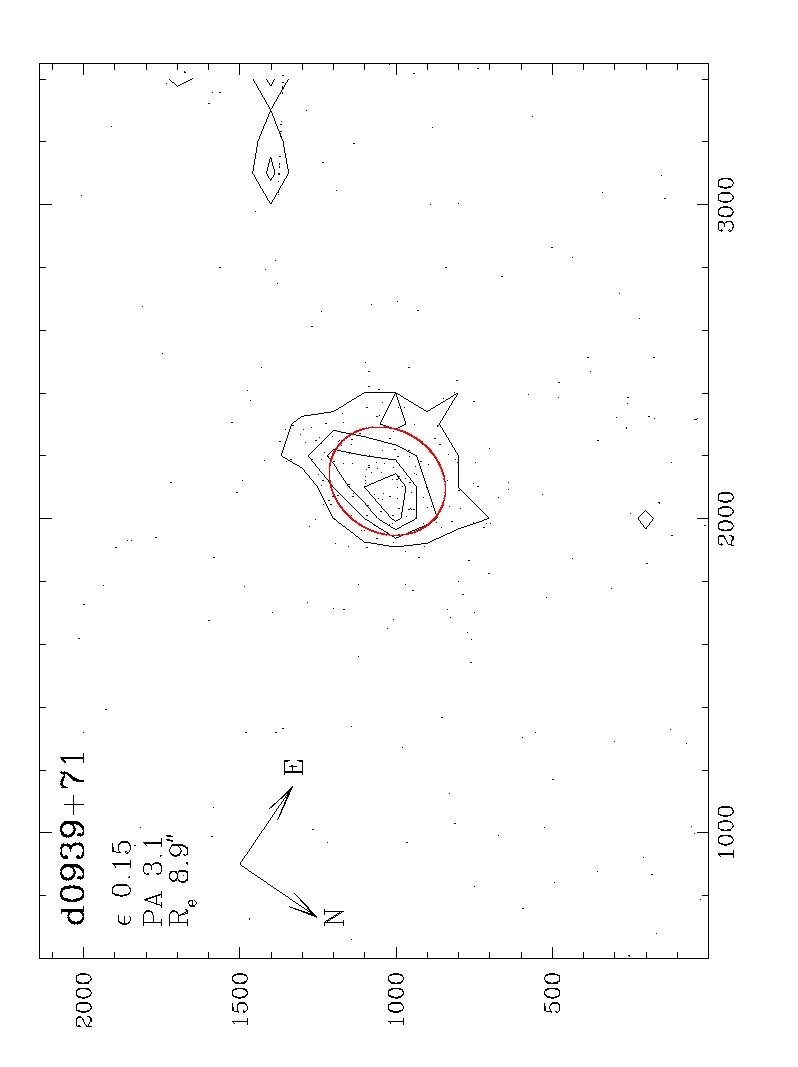}
\caption{Surface brightness profile fit with a Sersic function for d0939+71 star counts,
symbols as in Fig. \ref{prof1}.  Contour lines in the bottom panel indicate RGB stellar density
with 0.12 star arcsec$^{-2}$ spacing.
\label{prof3}}
\end{centering}
\end{figure}

\begin{figure}[t]
\begin{centering}
\includegraphics[angle=270, totalheight=3.0in]{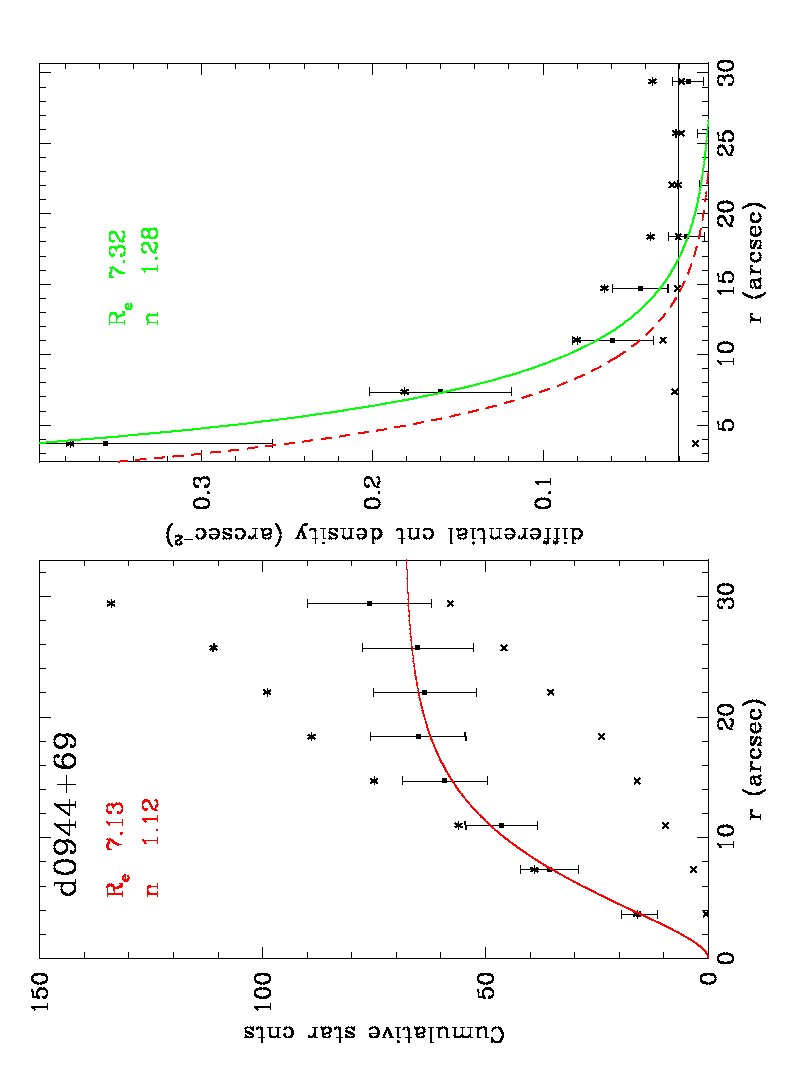}
\includegraphics[angle=270, totalheight=3.0in]{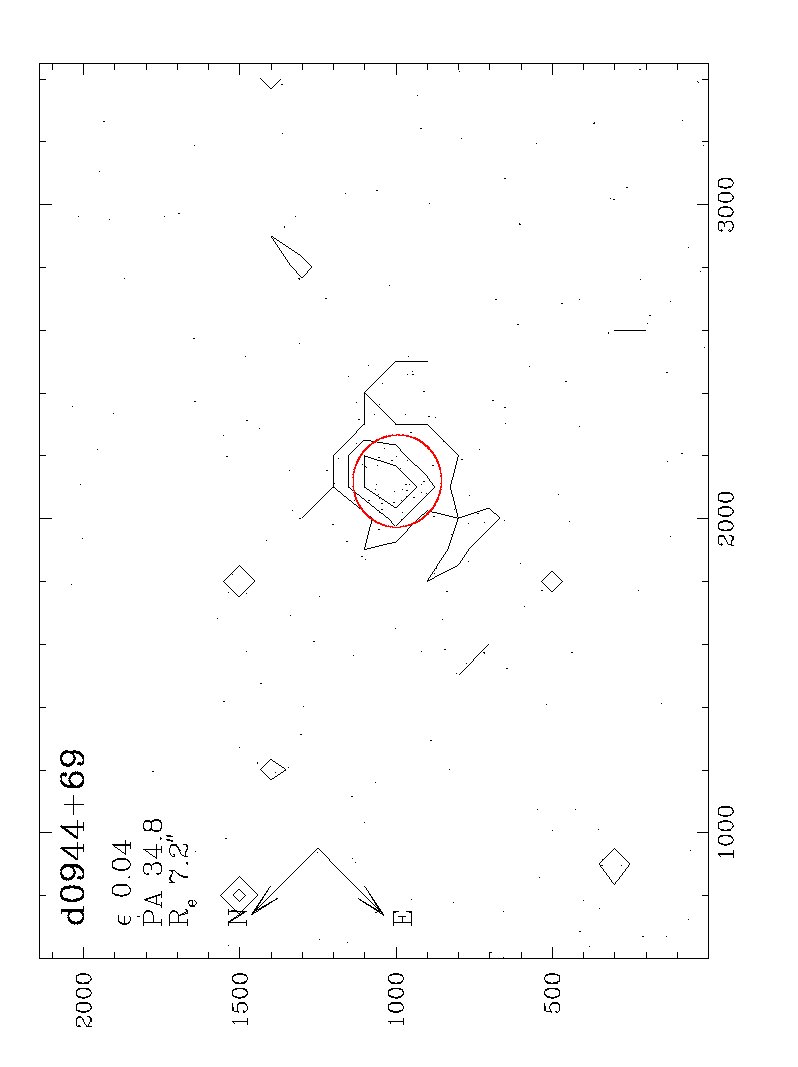}
\caption{Surface brightness profile fit with a Sersic function for d0944+69 star counts,
symbols as in Fig. \ref{prof1}.  Contour lines in the bottom panel indicate RGB stellar density
with 0.08 star arcsec$^{-2}$ spacing.
\label{prof4}}
\end{centering}
\end{figure}

\begin{figure}[t]
\begin{centering}
\includegraphics[angle=270, totalheight=3.0in]{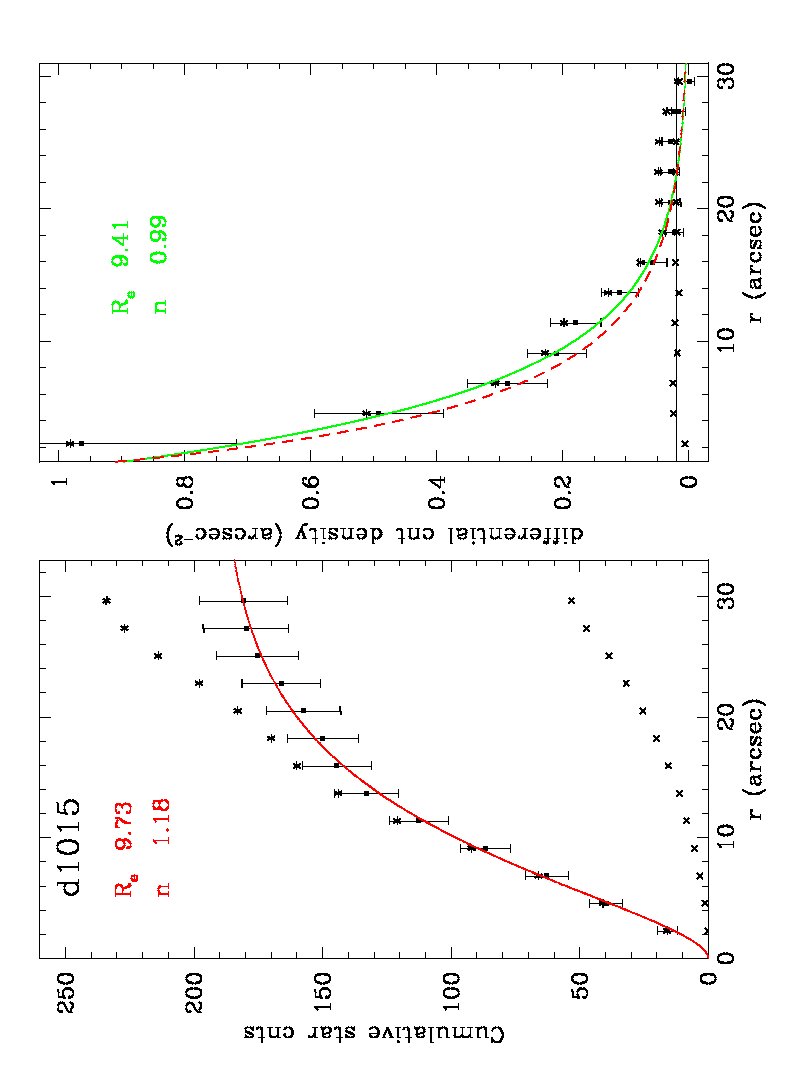}
\includegraphics[angle=270, totalheight=3.0in]{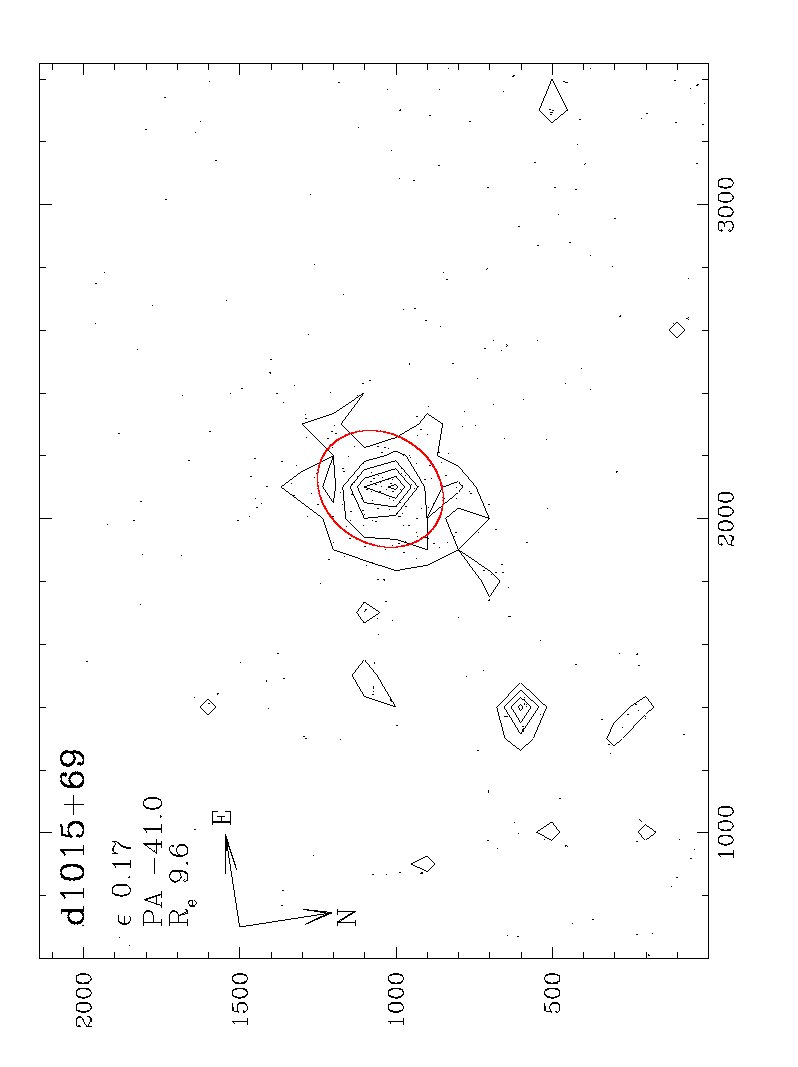}
\caption{Surface brightness profile fit with a Sersic function for d1015+69 star counts,
symbols as in Fig. \ref{prof1}.  Contour lines in the bottom panel indicate RGB stellar density
with 0.12 star arcsec$^{-2}$ spacing.
\label{prof5}}
\end{centering}
\end{figure}

\begin{figure}[t]
\begin{centering}
\includegraphics[angle=270, totalheight=3.0in]{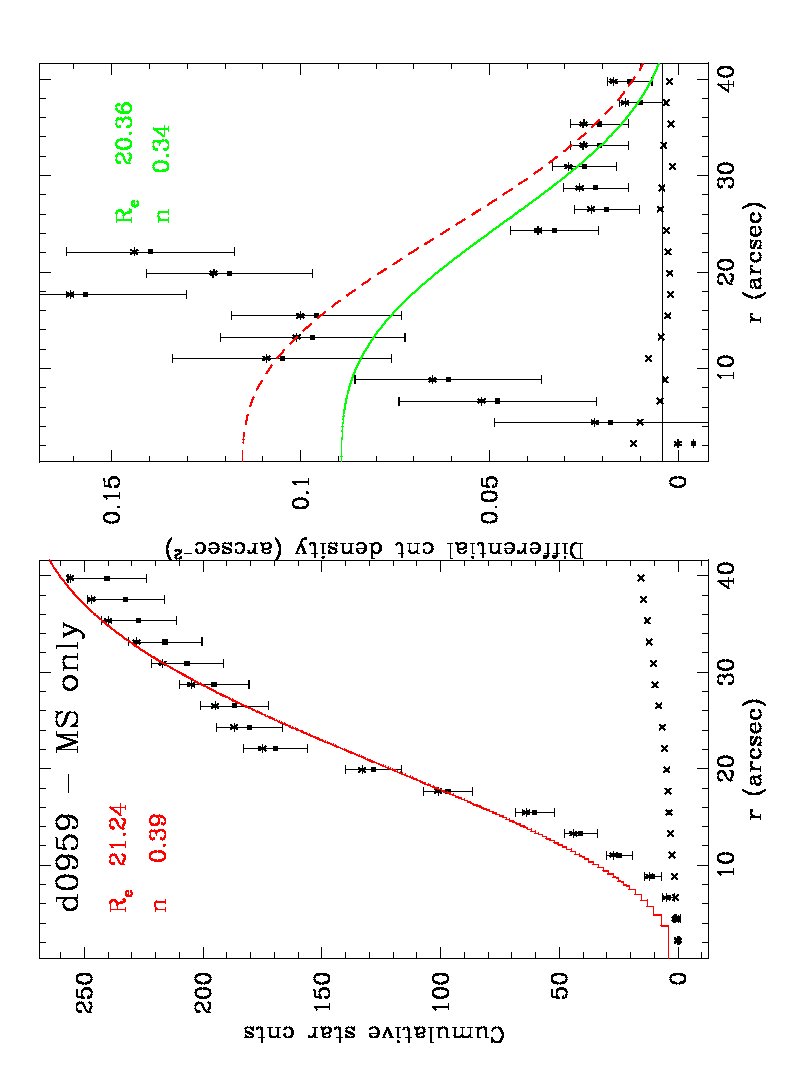}
\includegraphics[angle=270, totalheight=3.0in]{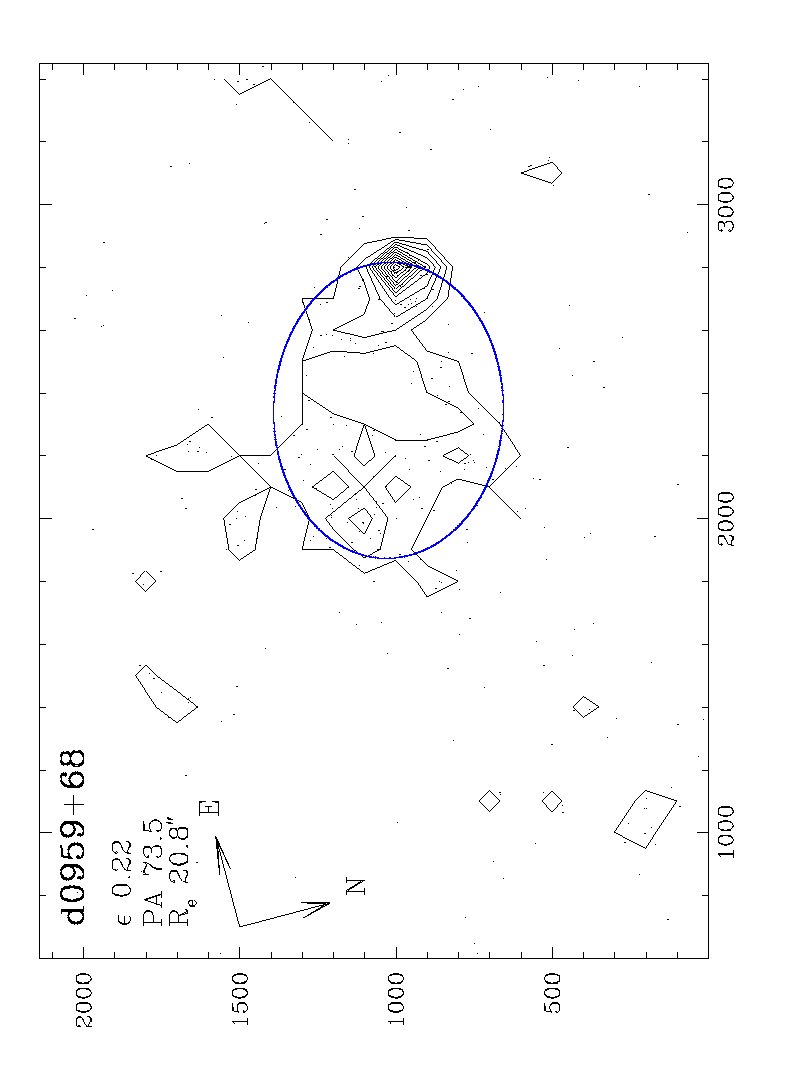}
\caption{Surface brightness profile fit with a Sersic function for d0959+68 star counts,
symbols as in Fig. \ref{prof1}.  Contour lines in the bottom panel indicate main sequence stellar density
with 0.08 star arcsec$^{-2}$ spacing.  The clumpy structure of separate star forming knots is
apparent.
\label{prof6}}
\end{centering}
\end{figure}

In the original surveyed region, there are now a total of 36 galaxies which belong to the 
M81 group. These include 4 BCDs (including DDO 82), 8 other late types (including the M81, M82,
NGC 3077, and NGC 2976 giants), 19 early type dwarfs, and 5 tidal dwarfs.
The projected distribution and 3-D plots in supergalactic cartesian coordinates are shown in 
Figure \ref{3d1}.  We see evidence for a morphology-density relation - late types dwarfs
(excluding tidal dwarfs) are 
found with a median physical separation from M81 1.9 times larger than that of early type dwarfs.
Early type dwarfs 
reside at a mean distance of 0.21 Mpc from M81 compared with 0.30 Mpc for
late type dwarfs.
Distinguishing between BCDs and dIs, we find mean distances
of 0.27 and 0.34 Mpc, respectively. Both of these mean distances are outside the second turnaround
radius of the M81 group, 230 kpc.  As BCDs are some of the least
clustered objects in the universe, and typically found in low density regions, 
this result is not surprising.

The core of the M81 group contains three closely interacting galaxies.  \citet{yun99} mapped out
HI tidal streams that connect the three giant galaxies M81, M82, and NGC 3077. From models, they
find the time of nearest approach of M82 and NGC 3077 to M81 to be $\sim 220-280$ Myr ago, coinciding with 
the ages of starbursts in all 3 galaxies and producing the observed HI streams. This core region is filled with
a number of star forming knots and tidal debris.  Amongst this debris and within the streams lie several
previously known or suspected tidal dwarfs including Arp's loop (and its brightest portion A0952+69),
Garland, HoIX, and BK3N \citep{K2002,mak02}. \citet{td08} also identified three tidal debris objects within
$\sim 20$ kpc of M81 likely to be short-lived structures, and more recently, two other small clumps of 
star formation were discovered lying along the HI bridge connecting M81 and NGC 3077 \citep{mi10,kkkm}.
We now include d0959+68 in this family.  This object was originally
discovered by \citet{durrell} in a survey for red giant stars around M81. In addition to a very low
surface brightness contribution of RGB stars, they identified a knot of young blue stars lying in projection on
the southeastern stream connecting M81 and NGC 3077, within a region of high HI density. From isochrone 
fitting to the CMD populated by these blue stars, they found evidence for star formation in the last
30-70 Myr. Similarly, we also find that isochrones with ages between 30-80 Myr and Z = 0.008 bracket much of the
main sequence.  
Red and blue supergiant stars found between the main sequence and RGB may indicate the presence of a population
as old as 200-400 Myr, around the timeframe of the expected formation period of the streams.

The Garland, which consists of several knots of star formation spanning $7^{\prime} \times 4^{\prime}$, lies
in projection near NGC 3077 while BK3N and A0952 lie near M81.  Attempts to measure the distances using the
TRGB found distances that more likely applied to the major galaxies themselves \citep{K2002}.
HolmIX, on the other hand, had no identifiable RGB population so distance estimates are based on photometry
of the brightest stars.  
In the 2D distribution of M81 cluster members 
(Figure \ref{3d1}), it is seen that the tidal dwarfs lie very near to the 3 giants involved in the 
tidal interaction, and in projection lie within the HI tidal streams (see Figure 31 in Paper 1). 
Estimated distances for these 
objects as found in the literature \citep{K2002} are used in the 3D plot in Figure \ref{3d1}.  
Given the large uncertainties in the distance measurements for these objects, the 3-D positions
displayed for the tidal dwarfs should be considered highly uncertain.  

The remaining star forming objects are found towards the periphery of the group.  One of these is 
the dI/dSph transition galaxy, d0926+70.  
This is one of the smallest known recently star forming dwarfs, surpassed only by Leo T and Leo P of the Local Group.  
Although slightly smaller and fainter, it could be said to resemble the Local Group dI/dSph transition
objects Phoenix and LGS 3.  A
study of the star formation history of these two Local Group galaxies \citep{hid11} found that a 
majority $50-90$\% of the
stars in these galaxies formed early in initial episodes of star formation during which there is 
no evidence of any chemical enrichment.  Since that first epoch of star formation, these galaxies 
have continued
to form stars at a very low rate until very recently, with chemical enrichment only occurring at very
recent times.  Similarly, d0926+70 appears to have had a 
dominant early period of star formation without significant chemical enrichment and exhibits evidence for
later and recent star formation at very low rates based on the presence of a handful of blue and red He burning
stars and possible low metallicity AGB stars.  The total HI gas mass measured for LGS 3 is 
$2.7 \times 10^5$ M$_{\odot}$ \citep{hid11}.  \citet{sambit} do not detect any HI in d0926+70 to an upper limit
of M$_{HI} = 3.1 \times 10^5$ M$_{\odot}$.  As d0926+70 is a smaller and presumably lower mass galaxy than LGS 3, 
the non-detection of HI might be expected in this recently star forming dI/dSph.  
At a distance of 345 kpc from M81, d0926+70 is outside the second turnaround radius of the group
and thus likely has not yet experienced the effects associated with a group environment
such as ram pressure stripping, tidal stripping, or strangulation which would remove the remaining 
gas supply and
shut off star formation.  After an early burst of star formation, it seems likely that this galaxy
was able to retain enough remaining and recycled gas to carry on with a slow trickle of star formation
as the low gas density allowed.

\begin{figure}[t]
\begin{centering}
\includegraphics[angle=0, totalheight=4.2in]{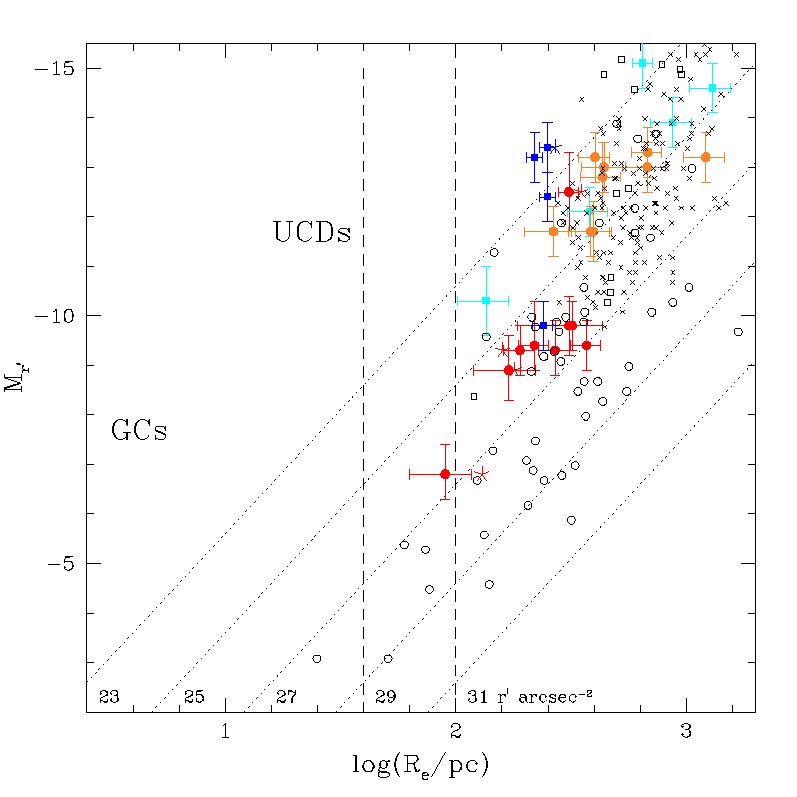}
\caption{Total  $r^{\prime}$ magnitude vs. size for the M81 dwarfs, color
coded by galaxy type.  Newly confirmed members from this work are shown as
blue squares (late type) and red circles (early type), previously known
M81 members are denoted by cyan squares and orange circles (late and early types,
respectively).  
Sizes and magnitudes are determined from our original CFHT/Megacam imaging.   
New size measurements from our ACS data are also included for 5 galaxies as colored asterisks.  
The tidal dwarf, d0959+68, is absent since the size measurement is highly 
uncertain.  
Also included are Local Group dwarf galaxies (open circles and squares
represent early and late types, respectively) and members of the Hydra and Centaurus
clusters (crosses).  See text. 
The rough locations of globular clusters and Fornax UCDs on this plane are
shown.  Few objects are known
which fall within the size gap region between globular clusters and galaxies represented 
here by the two dashed lines.  Lines of constant
effective surface brightness are also displayed.
\label{remr}}
\end{centering}
\end{figure}

\begin{figure}[t]
\begin{centering}
\includegraphics[angle=0, totalheight=4.0in]{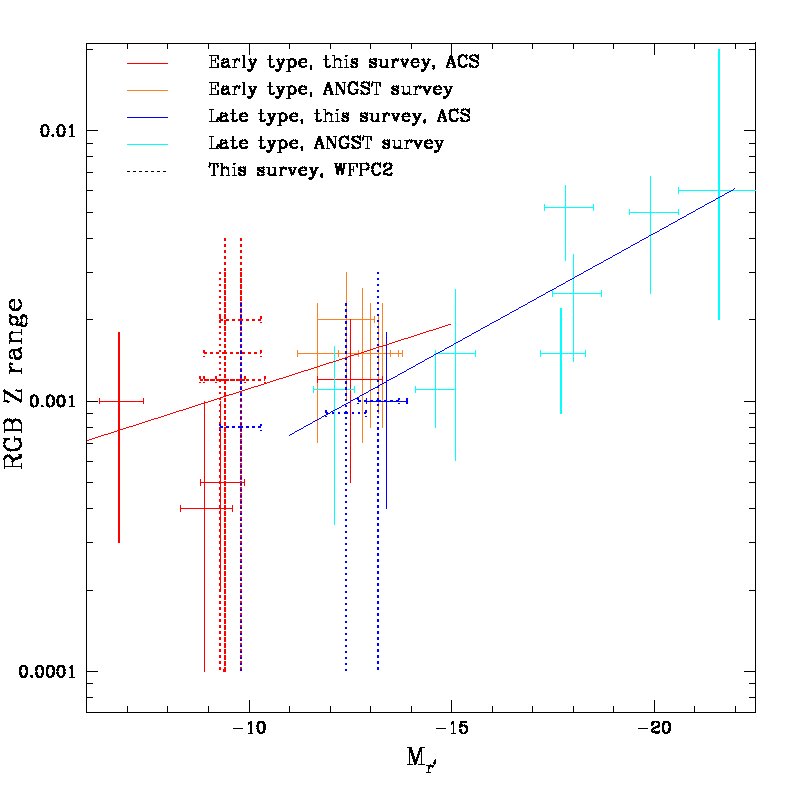}
\caption{Metallicity of the RGB as determined from isochrones (see text)
vs. total $r^{\prime}$ magnitude for M81 dwarfs.  Magnitudes were measured from our original
MegaCam survey data.  Error bars are supplied for magnitude measurement uncertainties.
Bars along the y-axis denote the range of metallicity spanned by $\pm 1\sigma$ of the photometric
uncertainties in color.
Much of the spread in the RGB is due to photometric uncertainties,  
although part of this range may be intrinsic and due to multiple stellar populations with a range of
ages and metallicities.  For M81 galaxies not part of our HST
follow-up, we use the data from the ANGST survey \citep{angst} to estimate metallicity.
Excluding measurements derived from WFPC2 data which have significantly larger photometric 
uncertanties, best fit relations are shown for the early and late type galaxies.
\label{zmr}}
\end{centering}
\end{figure}

\begin{figure}[t]
\begin{centering}
\includegraphics[angle=0, totalheight=4.0in]{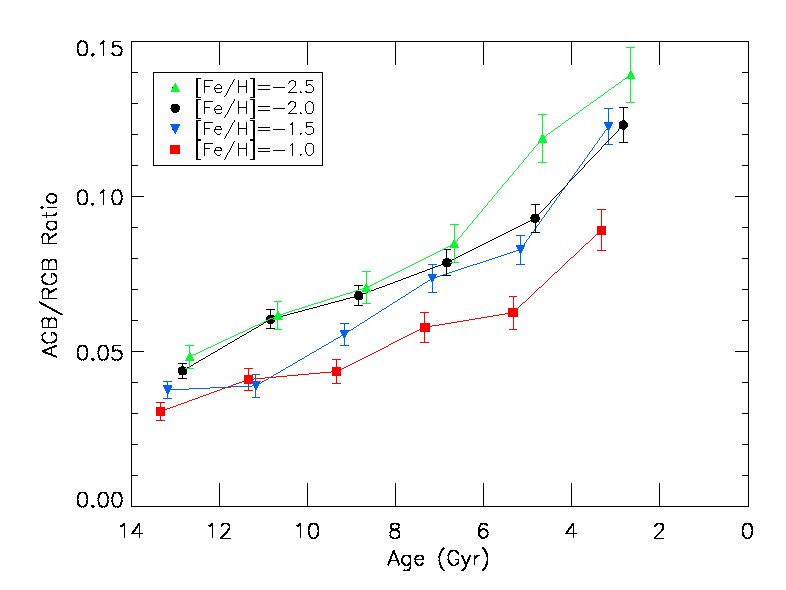}
\caption{Ratio of the number of AGB to RGB stars (within the upper 2 magnitudes of the
RGB) versus age in 2 Gyr bins (the horizontal offset of the symbols within bins is
for clarity only).  The ratio is omitted for the most recent age bins because other 
populations provide stonger constraints on the SFH as well as add complexity to the 
problem of defining an RGB or AGB star on the simulated CMDs.
\label{agbrgb}}
\end{centering}
\end{figure}

BCDs are generally compact, starbursting dwarf galaxies although their classification 
has assumed many definitions.  \citet{gil} have attempted to compile these 
into a single set of criteria. However, these depend on B,R, and K magnitudes.  Other
qualifications have required absolute blue magnitudes fainter than
M$_B > -18.15$ mag, diameters less
than 1 kpc, strong emission lines superposed on a blue
continuum \citep{tm81}, and spectra exhibiting high-excitation emission lines and low H$\alpha$
luminosities (L$_{H\alpha} < 10^{41}$ ergs s$^{-1}$ \citep{gal96}).  Our measurements satisfy many of
these criteria (see \citet{sambit} for L$_{H\alpha}$ measurements for these galaxies), although
in some respects resemble post-starburst galaxies displaying strong Balmer absorption lines in
their spectra.  The CMDs shown in this work provide strong evidence for
recent star formation with the presence of upper main sequence stars. In Figure \ref{remr},
the BCDs are found at the highest surface brightnesses for dwarfs.  They are also found 
just faintward of a boundary separating majority brighter late types from majority fainter early types.
\citet{sambit} find H$\alpha$ and FUV fluxes that indicate star formation within
the past $10^8$ years along with some very low level instantaneous star formation, in agreement
with the CMDs.  Intermediate aged AGB stars are also present. 

The question remains as to why these objects are undergoing a star
bursting phase.   The distances of d0958+66, d1028+70, and d1012+64 from M81 are 179, 228, and 337 kpc,
respectively.  For d1012+64 and d1028+70, this is close to or outside the nominal group second turnaround radius.  DDO 82,
another M81 group member with a BCD-like component, is also outside the second turnaround radius
at 329 kpc. A fifth object outside our survey region, VII Zw 403, is also considered an isolated BCD in this group.
Meanwhile d0958+66 is the one BCD within the 2nd turnaround radius and the one with
perhaps the least recent star formation event.  
As suggested in Paper 1, the excess of 4-5 BCDs within this group may
indicate that the termination of star formation is due to processes that are weak in this environment
such as ram pressure stripping, or, perhaps more likely, that these galaxies are on first infall orbits and processes
such as strangulation which play a role even in such poor groups have yet to act on these
galaxies or, in the case of d0958+66, may have recently started to act.  With a relative 
radial velocity with respect to M81 of $+90$ km s$^{-1}$ and a distance slightly larger than M81,
d0958+66 may have already passed near the center of the group and the interaction with the
group potential may have caused the recent cessation of star formation.  The next closest
BCD to M81, d1028+70, has a relative radial velocity with respect to M81 of $-69$ km s$^{-1}$
and a slightly larger distance, perhaps suggesting this object is on first approach. 

It may be possible that post-starbursting phase, these objects will fade enough to resemble 
the M81 dSphs at similar sizes.   Using our trick of estimating total magnitudes
based on model isochrones and scaling of stellar luminosity functions, and assuming something
about the ages and metallicities of various populations within the galaxies, we roughly
calculate how much the BCDs would fade if we simply removed all the young and intermediate
stars.  In the case of d1012+64 and d1028+70, we had only assumed a single old population
and single young population. Simply removing the young population would produce only 0.12 - 0.25 
mag of fading in $I$.  For d0958+66 we assumed an additional intermediate aged population.  
Removing these as well, we find $\sim 0.3$ mag fading.  Star formation in these galaxies
of course may have been continuous or have occurred in more than 2-3 bursts so these
may be underestimates.  Simply removing the contribution of all visible blue and AGB stars
from the visible component of each BCD finds 0.2, 0.5, and 0.6 mag dimming for d1012+64, d1028+70,
and d0958+66, respectively.  For d1028+70, this should be enough to push
this galaxy into the surface brightness / size / magnitude range of the brighter dSphs, within
the uncertainties (Figure \ref{remr}).  For the other two, at least 1-2 mag of fading would be necessary, although if we have 
underestimated the contribution of younger sources to the total light of the galaxy this may be 
possible.  A further issue is that $\sim2-3$ mag dimming would be required to put these objects on 
the same metallicity - brightness relation (Figure \ref{zmr}) as the dSphs, although given the
large scatter in these relations, such significant dimming might not be necessary.  

Our brightest new dSph, d0944+71, is quite similar to these BCDs in several respects. 
In the magnitude-size plane (Figure \ref{remr}), it is found at similar brightness and size, with a surface
brightness only slightly lower than that of the lowest surface brightness BCD, d1028+70. 
It has an RGB approximate metallicity of Z $= 0.0015$, similar to BCD d1012+64, and has a
similar size.  Although we can confidently say that this galaxy has a predominantly old
stellar population and has not recently formed stars, there are a handful of stellar 
detections blueward of the red sequence which could belong to a young $> 150$ Myr population
and which are found 
spatially coincident with the central region of the galaxy.  Assuming these are real,
this galaxy may still produce a frosting of star formation.  
In addition, this dSph lies at large distance from the center of the group, at a similar 
distance as the BCDs.    It may be that this object has already undergone a first pass
through the group and this is responsible for having halted most star formation.

The other dSphs are all much fainter, by several magnitudes, and have lower surface
brightness. These fall within an area of size and magnitude parameter space that 
previously had been unexplored in the M81 group due to the difficulty in detecting such
faint and small objects.   For the most part, these galaxies appear to have pure old
populations although a couple display evidence for intermediate aged AGB stars as well.  

The smallest object in our sample, dSph d0944+69, was originally measured in our MegaCam data
to have a size of 90 pc, which falls within the 40-100 pc size gap between most dwarf
galaxies and globular clusters \citep{gil07}. This can be seen in Figure \ref{remr} as the point
much smaller and fainter than any of the other new detections.  The size gap region is bounded
by two dashed vertical lines, and d0944+69 falls between them, albeit towards the galaxy scale
boundary with a $1 \sigma$ upper limit in size extending above
100 pc.  Current measurements using the ACS data and the bright RGB
population places the size at 130 pc.  This is a factor 1.4 times larger.
Using models and assuming a pure old stellar population, we estimate a total absolute $F814W$ magnitude
from the ACS imaging of  $-9.1$.  For such old, red objects, $V-I \sim 1.0$,
suggesting M$_V \sim -8.1$.  If our initial total $r^{\prime}$ magnitude measurement
was increased in the same proportion as the larger ACS size measurement, 
the expected brightness would increase
from $M_{r^{\prime}} = -6.8^{0.5}_{0.6}$ to $-7.4$.  Assuming a color $V-R > 0.3$
and taking the transformation $r^{\prime} = V - 0.84(V-R) + 0.13$ \citep{figdss96}, we
find $M_V > -7.2$.  This object is therefore similar in magnitude to the
brighter ultra-faint dwarfs recently discovered in the Local Group.

We cannot say too much about the
faintest and smallest dwarfs as the upper RGB is very poorly populated and no other stellar
populations appear within the depth of our observations.  There is a hint of elongation
visible in d1015+69 with ellipticity 0.18 while d0944+69 and d0939+71 both appear to be
rounder.  With a lack of any younger AGB or MS stars visible in the CMDs we find evidence
for pure old stellar populations.   The metallicity for d1015+69 and d0939+71
is constrained between Z $\sim0.0001$ to 0.001, while for d0944+69, it may be 
$\sim0.001$.  These are therefore metal poor objects, although not extreme cases.

Compared with the Local Group, we find a larger fraction of late types in M81, including BCDs
and tidal dwarfs, types unknown in the Local Group.
Around the Milky Way, only the Magellanic dwarfs are not considered dSph.  The population
around Andromeda is more varied, with a known population comprised of $\sim25$ dSphs, 1 transition
object, 1 dI, 3 dEs, 1 cE, and a low mass spiral \citep{mcc}.  The excess of the tidal dwarfs and
possibly BCDs can perhaps be explained by a dynamically younger and/or strongly interacting system in M81.  
Across both the M81 and Local groups, brighter dwarfs tend to be dIs while the majority 
fainter than M$_{r^{\prime}} \sim -14$ are dSph with only
a handful of star forming objects: the 3 M81 BCDs, several tidal dwarfs, and a few dIs and 
transition objects including d0926+70, Leo T, Leo P, Phoenix, LGS 3, and Antlia.  With the exception of the tidals, these all generally conform
to the morphology-density relation.  Our surface brightness limit is $\sim 2$ mag arcsec$^{-2}$ brighter 
than the limit for objects currently being discovered in the Local Group.  Based on the similaries between the 
groups, we can expect that a large number of faint dSph remain to be discovered around M81.

\begin{figure}[t]
\begin{centering}
\includegraphics[angle=270, totalheight=4.2in]{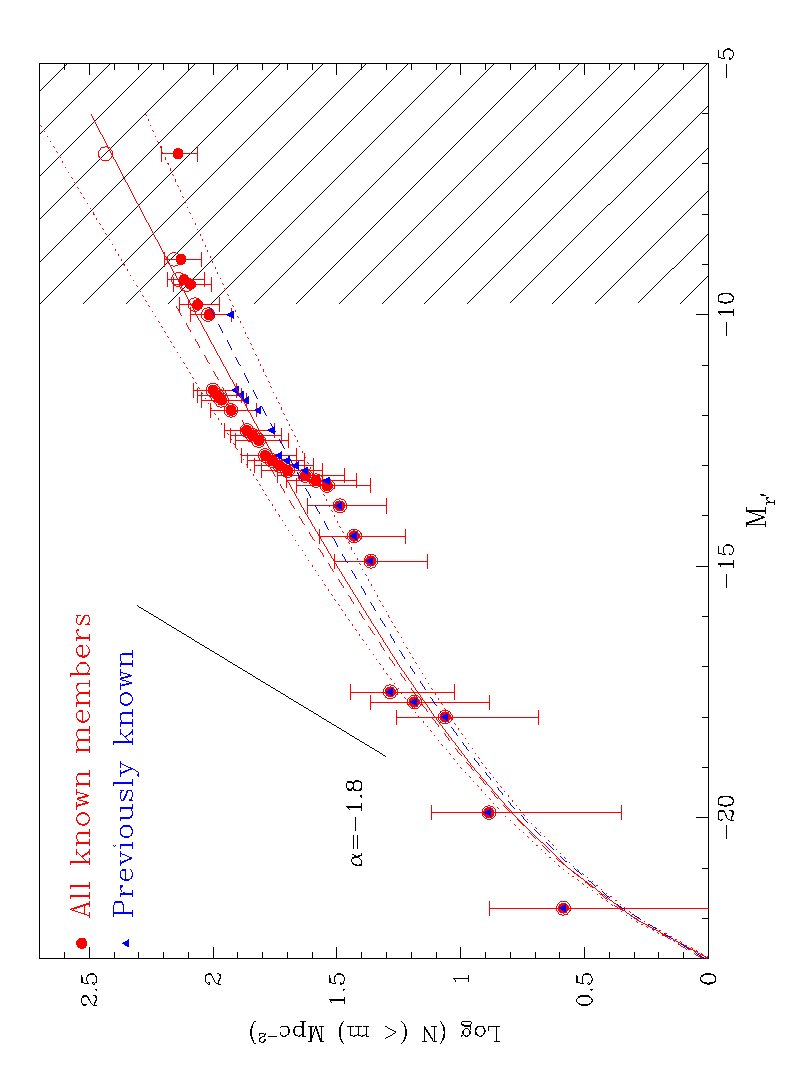}
\caption{The cumulative luminosity function for the M81 Group, using
corrected magnitudes assuming Sersic profiles.  Counts are normalized
to the area of the survey coverage. Large solid circles include all 
36 known members of the M81 Group within the surveyed region.  Open
circles are counts corrected for completeness.
Blue triangles are the previously known members.  The hatched region
denotes where the survey is incomplete. Schechter function fits to
previously known and all members, uncorrected for completeness, are
derived only to M$_{r} = -10$ (dashed blue and red lines, respectively).  
The solid line represents the best fit 
to completeness corrected counts with $\pm1 \sigma$ slopes shown
as dotted lines.  
\label{LFfin}}
\end{centering}
\end{figure}

\begin{figure}[t]
\begin{centering}
\plotone{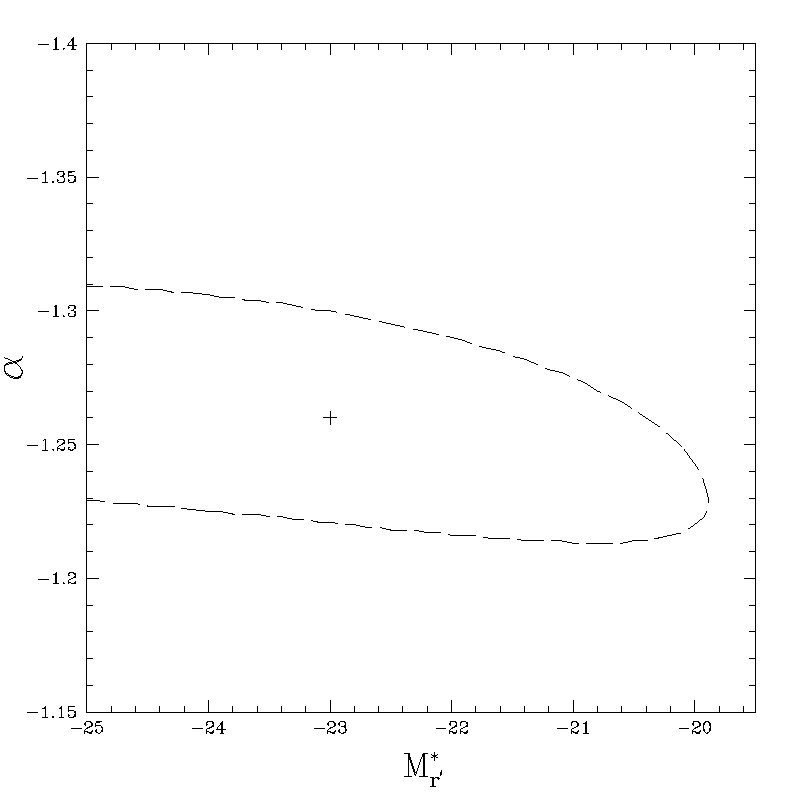}
\caption{The 1$\sigma$ error ellipses for the Schechter luminosity function parameters $\alpha$
and M$_{*}$ for the best fit to the final set of M81 members, with
correction for completeness.
\label{LFfinerr}}
\end{centering}
\end{figure}

We revisit the group luminosity function (LF) and faint end slope from our first paper.  In
Paper 1, we constructed both a differential and cumulative LF for known members, all candidates,
and most likely members.  We used total magnitudes corrected for light lost in the sky noise assuming
objects had surface brightness profiles that could be modeled with Sersic functions and
created artificial galaxies to test completeness limits.  Galaxy counts were normalized by the
areal coverage, which accounted for small gaps in the sky coverage, and were corrected 
for completeness at faint magnitudes.   Use of the cumulative LF was necessary due to the overall
small number counts of group members and hence large uncertainties within the magnitude bins.
A cumulative Schechter function was fit to the cumulative counts using maximum likelihood 
techniques with a Poisson estimator.   
We perform the same operations as in Paper 1 for the final set of 36 galaxies and display these
counts along with best cumulative Schechter function fit in 
Figure \ref{LFfin}.  The $1\sigma$ $\alpha-M_{*}$ error ellipse for the completeness corrected cumulative counts 
is shown in Figure \ref{LFfinerr}.   It can be seen that the exponential cutoff parameter $M_{*}$ is not at all 
constrained at the bright end, but the choice of $M_{*}$ will not greatly effect the value of the
logarithmic slope parameter $\alpha$.  Results for the faint-end slope are provided in Table \ref{LFtab}.

Since most of the galaxies we expected to be group members proved to be so,
results are not much different from the LF determined previously.  The hatched region
denotes where incompleteness sets in.  Blue triangles represent the original 22 known galaxies.
Including the new detections 
has not greatly changed the LF fit to our 90\% completeness limit M$_{r^{\prime}} = -10$.  
A fit to these counts to the limit M$_{r^{\prime}} < -10$ finds a faint-end slope of
$-1.26^{0.05}_{0.04}$ and $-1.28^{0.04}_{0.04}$ for power law and Schechter function fits, respectively,
only $-0.03$ steeper than found before our survey.  We find the same for fits to completeness corrected 
counts down to M$_{r^{\prime}} = -7$ with slopes of $-1.27^{0.04}_{0.04}$
and $-1.26^{0.04}_{0.04}$.  Thus, regardless of the numbers of ultra-faint dwarfs missed in our survey, to 
M$_{r^{\prime}} = -10$, we find a faint-end slope shallower than $-1.3$.
The steepest slope consistent with our data, shown by the $1\sigma$ fit 
(dotted line), is $-1.3$ whereas Cold Dark Matter theory predicts slopes for the mass function of $-1.8$ \citep{tt02}.  

Figure \ref{remr} displays M$_{r^{\prime}}$ vs log R$_e$ for the newly confirmed members with late
and early types distinguished by color, along with previously known M81 group members. Also plotted
are these quantities for the Local Group \citep{IH95, mm98, mci06, simon07,belo07,will1,bootes07,
zucker06,martin06,imi07,ifh08, mcc} along with Centaurus and Hydra cluster members \citep{mh11}.  
To transform $V$ magnitudes to $r^{\prime}$, we assume
$r^{\prime} = V - 0.84(V-R) + 0.13$ \citep{figdss96}.  Although
the M81 group points lie on top of those from other groups, they clearly fall along the upper envelope
of surface brightness for the full range of sizes currently observed for nearby dwarf galaxies.  From completeness testing in 
our original survey, we found that surface brightness affected detection
especially at fainter magnitudes. Of note, however, are a dearth
of any M81 galaxies in the region between $\mu = 25-27 r^{\prime}$ arcsec$^{-2}$ with M$_{r^{\prime}} < -10$
and log(R$_e$) $> 2.6$, and within a $\sim 1$ mag gap between the
majority of previously known objects and newly discovered objects.  
Detection limits for our original survey are discussed in Paper 1.  These were determined by simulating
galaxies as concentrations of resolved stars with a wide range of total magnitudes and stellar surface
densities.  It was determined that we were $\sim$90\% complete to M$_{r^{\prime}} = -11$ for 
log Re $< 3.2$, and to $-10$ for log Re $< 2.9$.  Thus, we do not expect to be missing significant
numbers of galaxies in this region. However, given the complete absence of objects detected in a parameter
space that one might expect to be populated suggests that some objects may have been missed.  Only a few
Local Group dwarfs are found in this size-magnitude region so it is conceivable that such galaxies are
present in the M81 group and were missed in our original survey because they lie in regions of 
higher extinction, reside on the far
side of M81 at greater distance, and/or fell directly within the large chip gap regions where effective
exposure times were half that of the majority of the survey.  At such faint magnitudes, it is also
possible that our magnitudes and sizes are underestimated.  Indeed, for some of our ACS observed
galaxies, including our smallest and faintest object, we do measure larger sizes in the new data.
Magnitudes are also expected to be brighter, but, for objects with $M_{r^{\prime}} \sim -9$, by not more than
several tenths of a magnitude based on the new $I-$band estimates.   Thus, although some of these
new detections may shift up and to the right roughly along lines of constant surface brightness, it is
unlikely that any of the known faint dwarfs could shift up into this currently empty 
region of magnitude-size space.
If an additional
$\sim6$ objects were included in the LF in the magnitude range $M_{r^{\prime}} = -10$ to $-12$ with 
$\mu = 25-27 r^{\prime}$ arcsec$^{-2}$, as one finds for the Local Group galaxies,
the resultant LF faint-end slope would increase insignificantly.

In Paper 1, we compared our faint-end slope measurements to that of the CenA group and Andromeda sub-group,
finding slopes for these of $-1.23^{0.04}_{0.1}$ and $-1.13^{0.06}_{0.06}$.  Since then, another $\sim12$
dwarfs have been discovered around Andromeda. 
Of these, only 3 have magnitudes as bright as M$_{V} < -10$, while most are in the realm of the ultra-faint
dwarfs with magnitudes fainter than M$_{V} \sim -8.0$.  One might expect a number of further discoveries 
out to greater distances from Andromeda in regions not yet surveyed \citep{mcc,martin13}, perhaps 
bringing the LF slope more in line with what we find for the M81 Group,
but given the higher completeness fraction at these brighter magnitudes, it
is doubtful that the faint-end slope for Andromeda will be revised to much steeper values.  

As we found in Paper 1, if the slope were $-1.8$ as predicted by cosmological structure formation models, 
between $-12 < M_{r^{\prime}} < -6$, we would be missing over 1700 galaxies
in our survey area. This is over an order of magnitude greater than we 
would expect based on our simulations.   The agreement in faint-end slope for these
nearby groups would therefore suggest that this discrepancy is real, necessitating the use of other 
explanations to account for the fewer observed dwarf galaxies at these faint magnitudes.
Incorporating the effects of feedback and star formation efficiency along with 
accounting for the suppression of gas infall into the low mass halos of the
forming galaxies by reionization in the early universe 
may help reconcile theory and observations.  

\section{Summary and Conclusions}\label{conc}

From TRGB distance measurements based on two color imaging with HST/ACS and WFPC2, we have 
confirmed 14 new members of the M81 Group,
out of 22 original candidates, within a 65 square degree region encompassing the
second turnaround radius of the group.  Of the 12 candidates originally expected to
be group members based on degree of resolution in CFHT/MegaCam imaging, 11 proved to be so, 
along with another 3 more uncertain cases.
The new members include 1 tidal dwarf with a young stellar population, 3 BCDs with ongoing
or recent star formation in addition to intermediate and old stellar populations, 1 dI/dSph transition
object with an intermediate aged population, and 9 dSph with predominantly or pure old stellar 
populations.  The absolute
$r^{\prime}$ magnitudes range from -13.3 to -6.8.  Half-light radii range from 370 to 
90 pc for galaxies well fit with Sersic profiles. These are typical sizes for
faint dwarf galaxies.   This brings the total membership of the M81 group within this region
to at least 36 galaxies, including several tidal dwarfs found in the HI streams between the interacting
giants M81, M82, and NGC 3077. Additional tidal objects and clumps of star formation have also 
been identified along the HI streams and within the M81 tidal debris field \citep{td08,mi10,kkkm}.
Overall, the M81 Group harbors a larger population of late types than the Local Group
and appears to be dynamically younger.
However, aside from the BCDs and tidal dwarf, the majority of the new detections
are found to be dSph with pure old stellar populations, 
more in line with the properties of Local Group faint dwarfs.

Using isochrones based on Padova models we have investigated the stellar populations
of each galaxy.  The RGB for the dSphs has a broad spread, but this is primarily due to photometric
errors at the faint magnitude limits of the observations. 
The $\pm 1\sigma$ allowed metallicity ranges for the RGBs is provided in Table \ref{Tstellpop}, but typically 
we find Z $\sim 0.001$ for these galaxies.  Only a
handful of these dSphs host some evidence for younger intermediate aged populations.  The brightest
of these, d0944+71 with M$_{r^{\prime}} = -12.4$, may have experienced a minor episode of star
formation as recently as 150 Myr ago, but otherwise exhibits a primarily old
stellar population.   Interestingly, it lies at large distance from M81, $\sim335$ kpc, outside the
second turnaround radius for the group, and at much larger distance than expected based on 
the morphology - density relation.  The faintest dSph with M$_{r^{\prime}} = -6.8$ displays evidence 
for a pure old stellar population.  This object has properties impinging on the
size and brightness range of the ultra-faint dwarf galaxies currently being
discovered in the Local Group.   The new ACS $F814W$ estimates find this
object to be slightly brighter and larger than measured previously in MegaCam imaging,
with M$_{F814W} = -9.1$ and size 130 pc, making this more of a borderline case between the 
ultra-faint and classical dwarfs.

One dwarf irregular, d0926+70, was identified, although this object may also be considered a
transition dwarf.  It exhibits only a small amount of recent star formation located near the
core of the galaxy.  It also appears to have an irregular elongated shape and is located near 
the periphery of the group.  Based on the CMD, there is some evidence for star formation as recently as $\sim 100$ Myr ago
along with evidence for an older intermediate aged population. The metallicity 
appears to be low for all populations within this galaxy, with no evidence for any significant metal enrichment.  
Although not
detected in this galaxy, HI may simply exist at levels below the detection limit at M$_{HI} < 3.1 \times 10^5$ M$_{\odot}$.  
With an effective radius
240 pc and M$_{r^{\prime}} = -9.8$, this is one of the smallest galaxies with recent star formation known.

The 3 BCDs are found to have ongoing or very recent star formation.  All recent
star formation is confined to the galaxy core regions.  The BCD with the least recent
star formation, d0958+66 with most star formation having occurred over 60 Myr ago, is the closest to the group core.  
A large fraction of the stellar population in this object appears to have ages within the range 80 Myr - 3 Gyr.  
Perhaps the initial approach and tidal influence
exerted by the group potential induced the recent burst of star formation in all 3 BCDs. 
Lying outside the second turnaround radius of the group, d1028+70 may have had a period
of star formation more recently than 60 Myr.  At a distance 0.1 Mpc greater than M81 and with a 
relative radial velocity of -69 km s$^{-1}$, this galaxy may be on first approach.
We investigate whether the BCDs could
fade enough to eventually resemble typical dSphs.  For d1012+64 and d0958+66 we find that 
these galaxies would need to experience $\sim2$ magnitudes of fading to reach typical surface brightnesses for
dSphs of the same size and have metallicities consistent with the metallicity - brightness
relation for dSph.  

One tidal dwarf was also identified, confirming a candidate previously detected by \citet{durrell} and 
adding to several previously known. \citet{mi10} have since identified another two tidal dwarf candidates.  
These three additions lie in projection on the
M81 - NGC 3077 stream and appear to consist entirely of young blue main sequence and supergiant stars.  The
RGB visible in the CMD for d0959+68 may actually be due to red giant stars found along the stream, as we see no evidence for
a concentration of these at the location of this object in the ACS imaging. Rather this object appears to have formed all its stars since
around the time of the tidal stream formation.   It is not known if there is enough mass for this object to 
remain bound.

With these new detections, we investigate the faint-end of the galaxy luminosity function of this group, 90\% complete
down to M$_{r^{\prime}} = -9.8$.  The inclusion of these 14 newly confirmed members increases
the faint-end slope only marginally from $\sim -1.24$ to $\sim -1.27$. This analysis applies a correction
for incompleness in our survey down to $-7$.  Although this slope is much shallower than cosmological predictions, it is
consistent with what is found in other nearby groups.

\acknowledgments
We thank the anonymous referee for a thorough reading our manuscript and for providing very 
helpful suggestions that have improved this paper.
Support for Hubble Space Telescope Program numbers 11126 and 11584 were provided by NASA through a grant from the 
Space Telescope Science Institute, which is operated by the Association of Universities for
Research in Astronomy, Incorporated, under NASA contract NAS5-26555.  IDK was partially supported
by RFBR-DFG grant 12-02-91338.  The authors would like to 
acknowledge Andy Dolphin, Luca Rizzi, and Dmitry Makarov for providing much of the
software used for the reduction and analysis of these data.

\bibliographystyle{apj}
\bibliography{m81cmd}

\begin{thebibliography}{82}
\expandafter\ifx\csname natexlab\endcsname\relax\def\natexlab#1{#1}\fi

\bibitem[{{Begum} {et~al.}(2008){Begum}, {Chengalur}, {Karachentsev},
  {Sharina}, \& {Kaisin}}]{begum}
{Begum}, A., {Chengalur}, J.~N., {Karachentsev}, I.~D., {Sharina}, M.~E., \&
  {Kaisin}, S.~S. 2008, \mnras, 386, 1667

\bibitem[{{Bellazzini} {et~al.}(2001){Bellazzini}, {Ferraro}, \&
  {Pancino}}]{bell01}
{Bellazzini}, M., {Ferraro}, F.~R., \& {Pancino}, E. 2001, \apj, 556, 635

\bibitem[{{Belokurov} {et~al.}(2007){Belokurov}, {Zucker}, {Evans}, {Kleyna},
  {Koposov}, {Hodgkin}, {Irwin}, {Gilmore}, {Wilkinson}, {Fellhauer},
  {Bramich}, {Hewett}, {Vidrih}, {De Jong}, {Smith}, {Rix}, {Bell}, {Wyse},
  {Newberg}, {Mayeur}, {Yanny}, {Rockosi}, {Gnedin}, {Schneider}, {Beers},
  {Barentine}, {Brewington}, {Brinkmann}, {Harvanek}, {Kleinman}, {Krzesinski},
  {Long}, {Nitta}, \& {Snedden}}]{belo07}
{Belokurov}, V., {Zucker}, D.~B., {Evans}, N.~W., {Kleyna}, J.~T., {Koposov},
  S., {Hodgkin}, S.~T., {Irwin}, M.~J., {Gilmore}, G., {Wilkinson}, M.~I.,
  {Fellhauer}, M., {Bramich}, D.~M., {Hewett}, P.~C., {Vidrih}, S., {De Jong},
  J.~T.~A., {Smith}, J.~A., {Rix}, H.-W., {Bell}, E.~F., {Wyse}, R.~F.~G.,
  {Newberg}, H.~J., {Mayeur}, P.~A., {Yanny}, B., {Rockosi}, C.~M., {Gnedin},
  O.~Y., {Schneider}, D.~P., {Beers}, T.~C., {Barentine}, J.~C., {Brewington},
  H., {Brinkmann}, J., {Harvanek}, M., {Kleinman}, S.~J., {Krzesinski}, J.,
  {Long}, D., {Nitta}, A., \& {Snedden}, S.~A. 2007, \apj, 654, 897

\bibitem[{{Bertelli} {et~al.}(1994){Bertelli}, {Bressan}, {Chiosi}, {Fagotto},
  \& {Nasi}}]{b94}
{Bertelli}, G., {Bressan}, A., {Chiosi}, C., {Fagotto}, F., \& {Nasi}, E. 1994,
  \aaps, 106, 275

\bibitem[{{Blanton} {et~al.}(2005){Blanton}, {Lupton}, {Schlegel}, {Strauss},
  {Brinkmann}, {Fukugita}, \& {Loveday}}]{bls05}
{Blanton}, M.~R., {Lupton}, R.~H., {Schlegel}, D.~J., {Strauss}, M.~A.,
  {Brinkmann}, J., {Fukugita}, M., \& {Loveday}, J. 2005, \apj, 631, 208

\bibitem[{{Boylan-Kolchin} {et~al.}(2012){Boylan-Kolchin}, {Bullock}, \&
  {Kaplinghat}}]{bk12}
{Boylan-Kolchin}, M., {Bullock}, J.~S., \& {Kaplinghat}, M. 2012, \mnras, 422,
  1203

\bibitem[{{Brown} {et~al.}(2012){Brown}, {Tumlinson}, {Geha}, {Kirby},
  {VandenBerg}, {Mu{\~n}oz}, {Kalirai}, {Simon}, {Avila}, {Guhathakurta},
  {Renzini}, \& {Ferguson}}]{brown12}
{Brown}, T.~M., {Tumlinson}, J., {Geha}, M., {Kirby}, E.~N., {VandenBerg},
  D.~A., {Mu{\~n}oz}, R.~R., {Kalirai}, J.~S., {Simon}, J.~D., {Avila}, R.~J.,
  {Guhathakurta}, P., {Renzini}, A., \& {Ferguson}, H.~C. 2012, \apjl, 753, L21

\bibitem[{{Caon} {et~al.}(1993){Caon}, {Capaccioli}, \& {D'Onofrio}}]{ccd93}
{Caon}, N., {Capaccioli}, M., \& {D'Onofrio}, M. 1993, \mnras, 265, 1013

\bibitem[{{Chabrier}(2001)}]{chab01}
{Chabrier}, G. 2001, \apj, 554, 1274

\bibitem[{{Chiaberge}(2012)}]{chia12}
{Chiaberge}, M. 2012, {A new accurate CTE photometric correction formula for
  ACS/WFC}, Tech. rep.

\bibitem[{{Chiboucas} {et~al.}(2009){Chiboucas}, {Karachentsev}, \&
  {Tully}}]{Ch1}
{Chiboucas}, K., {Karachentsev}, I.~D., \& {Tully}, R.~B. 2009, \aj, 137, 3009

\bibitem[{{Dalcanton} {et~al.}(2009){Dalcanton}, {Williams}, {Seth}, {Dolphin},
  {Holtzman}, {Rosema}, {Skillman}, {Cole}, {Girardi}, {Gogarten},
  {Karachentsev}, {Olsen}, {Weisz}, {Christensen}, {Freeman}, {Gilbert},
  {Gallart}, {Harris}, {Hodge}, {de Jong}, {Karachentseva}, {Mateo}, {Stetson},
  {Tavarez}, {Zaritsky}, {Governato}, \& {Quinn}}]{angst}
{Dalcanton}, J.~J., {Williams}, B.~F., {Seth}, A.~C., {Dolphin}, A.,
  {Holtzman}, J., {Rosema}, K., {Skillman}, E.~D., {Cole}, A., {Girardi}, L.,
  {Gogarten}, S.~M., {Karachentsev}, I.~D., {Olsen}, K., {Weisz}, D.,
  {Christensen}, C., {Freeman}, K., {Gilbert}, K., {Gallart}, C., {Harris}, J.,
  {Hodge}, P., {de Jong}, R.~S., {Karachentseva}, V., {Mateo}, M., {Stetson},
  P.~B., {Tavarez}, M., {Zaritsky}, D., {Governato}, F., \& {Quinn}, T. 2009,
  \apjs, 183, 67

\bibitem[{{Davidge}(2008)}]{td08}
{Davidge}, T.~J. 2008, \pasp, 120, 1145

\bibitem[{{de Jong} {et~al.}(2008){de Jong}, {Rix}, {Martin}, {Zucker},
  {Dolphin}, {Bell}, {Belokurov}, \& {Evans}}]{dJ08}
{de Jong}, J.~T.~A., {Rix}, H.-W., {Martin}, N.~F., {Zucker}, D.~B., {Dolphin},
  A.~E., {Bell}, E.~F., {Belokurov}, V., \& {Evans}, N.~W. 2008, \aj, 135, 1361

\bibitem[{{Dekel} \& {Silk}(1986)}]{ds86}
{Dekel}, A. \& {Silk}, J. 1986, \apj, 303, 39

\bibitem[{{Dolphin}(2000{\natexlab{a}})}]{dolphin00}
{Dolphin}, A.~E. 2000{\natexlab{a}}, \pasp, 112, 1397

\bibitem[{{Dolphin}(2000{\natexlab{b}})}]{hstphot00}
---. 2000{\natexlab{b}}, \pasp, 112, 1383

\bibitem[{{Dolphin} {et~al.}(2005){Dolphin}, {Weisz}, {Skillman}, \&
  {Holtzman}}]{dolphin05}
{Dolphin}, A.~E., {Weisz}, D.~R., {Skillman}, E.~D., \& {Holtzman}, J.~A. 2005,
  ArXiv Astrophysics e-prints

\bibitem[{{Dotter} {et~al.}(2008){Dotter}, {Chaboyer}, {Jevremovi{\'c}},
  {Kostov}, {Baron}, \& {Ferguson}}]{dotter08}
{Dotter}, A., {Chaboyer}, B., {Jevremovi{\'c}}, D., {Kostov}, V., {Baron}, E.,
  \& {Ferguson}, J.~W. 2008, \apjs, 178, 89

\bibitem[{{Durrell} {et~al.}(2004){Durrell}, {Decesar}, {Ciardullo},
  {Hurley-Keller}, \& {Feldmeier}}]{durrell}
{Durrell}, P.~R., {Decesar}, M.~E., {Ciardullo}, R., {Hurley-Keller}, D., \&
  {Feldmeier}, J.~J. 2004, in IAU Symposium, Vol. 217, Recycling Intergalactic
  and Interstellar Matter, ed. P.-A. {Duc}, J.~{Braine}, \& E.~{Brinks}, 90--+

\bibitem[{{Fitzpatrick}(1999)}]{fitz}
{Fitzpatrick}, E.~L. 1999, \pasp, 111, 63

\bibitem[{{Fukugita} {et~al.}(1996){Fukugita}, {Ichikawa}, {Gunn}, {Doi},
  {Shimasaku}, \& {Schneider}}]{figdss96}
{Fukugita}, M., {Ichikawa}, T., {Gunn}, J.~E., {Doi}, M., {Shimasaku}, K., \&
  {Schneider}, D.~P. 1996, \aj, 111, 1748

\bibitem[{{Gallego} {et~al.}(1996){Gallego}, {Zamorano}, {Rego}, {Alonso}, \&
  {Vitores}}]{gal96}
{Gallego}, J., {Zamorano}, J., {Rego}, M., {Alonso}, O., \& {Vitores}, A.~G.
  1996, \aaps, 120, 323

\bibitem[{{Gil de Paz} {et~al.}(2003){Gil de Paz}, {Madore}, \&
  {Pevunova}}]{gil}
{Gil de Paz}, A., {Madore}, B.~F., \& {Pevunova}, O. 2003, \apjs, 147, 29

\bibitem[{{Gilmore} {et~al.}(2007){Gilmore}, {Wilkinson}, {Wyse}, {Kleyna},
  {Koch}, {Evans}, \& {Grebel}}]{gil07}
{Gilmore}, G., {Wilkinson}, M.~I., {Wyse}, R.~F.~G., {Kleyna}, J.~T., {Koch},
  A., {Evans}, N.~W., \& {Grebel}, E.~K. 2007, \apj, 663, 948

\bibitem[{{Girardi} {et~al.}(2000){Girardi}, {Bressan}, {Bertelli}, \&
  {Chiosi}}]{Girardi00}
{Girardi}, L., {Bressan}, A., {Bertelli}, G., \& {Chiosi}, C. 2000, \aaps, 141,
  371

\bibitem[{{Girardi} {et~al.}(2008){Girardi}, {Dalcanton}, {Williams}, {de
  Jong}, {Gallart}, {Monelli}, {Groenewegen}, {Holtzman}, {Olsen}, {Seth},
  {Weisz}, \& {ANGST/ANGRRR Collaboration}}]{gir08}
{Girardi}, L., {Dalcanton}, J., {Williams}, B., {de Jong}, R., {Gallart}, C.,
  {Monelli}, M., {Groenewegen}, M.~A.~T., {Holtzman}, J.~A., {Olsen}, K.~A.~G.,
  {Seth}, A.~C., {Weisz}, D.~R., \& {ANGST/ANGRRR Collaboration}. 2008, \pasp,
  120, 583

\bibitem[{{Grebel}(1997)}]{grebel97}
{Grebel}, E.~K. 1997, in Reviews in Modern Astronomy, Vol.~10, Reviews in
  Modern Astronomy, ed. R.~E. {Schielicke}, 29--60

\bibitem[{{Groenewegen}(2006)}]{groenewegen}
{Groenewegen}, M.~A.~T. 2006, \aap, 448, 181

\bibitem[{{Hidalgo} {et~al.}(2011){Hidalgo}, {Aparicio}, {Skillman}, {Monelli},
  {Gallart}, {Cole}, {Dolphin}, {Weisz}, {Bernard}, {Cassisi}, {Mayer},
  {Stetson}, {Tolstoy}, \& {Ferguson}}]{hid11}
{Hidalgo}, S.~L., {Aparicio}, A., {Skillman}, E., {Monelli}, M., {Gallart}, C.,
  {Cole}, A., {Dolphin}, A., {Weisz}, D., {Bernard}, E.~J., {Cassisi}, S.,
  {Mayer}, L., {Stetson}, P., {Tolstoy}, E., \& {Ferguson}, H. 2011, \apj, 730,
  14

\bibitem[{{Ibata} {et~al.}(2007){Ibata}, {Martin}, {Irwin}, {Chapman},
  {Ferguson}, {Lewis}, \& {McConnachie}}]{imi07}
{Ibata}, R., {Martin}, N.~F., {Irwin}, M., {Chapman}, S., {Ferguson}, A.~M.~N.,
  {Lewis}, G.~F., \& {McConnachie}, A.~W. 2007, \apj, 671, 1591

\bibitem[{{Ibata} {et~al.}(2013){Ibata}, {Lewis}, {Conn}, {Irwin},
  {McConnachie}, {Chapman}, {Collins}, {Fardal}, {Ferguson}, {Ibata}, {Mackey},
  {Martin}, {Navarro}, {Rich}, {Valls-Gabaud}, \& {Widrow}}]{ibata13}
{Ibata}, R.~A., {Lewis}, G.~F., {Conn}, A.~R., {Irwin}, M.~J., {McConnachie},
  A.~W., {Chapman}, S.~C., {Collins}, M.~L., {Fardal}, M., {Ferguson},
  A.~M.~N., {Ibata}, N.~G., {Mackey}, A.~D., {Martin}, N.~F., {Navarro}, J.,
  {Rich}, R.~M., {Valls-Gabaud}, D., \& {Widrow}, L.~M. 2013, \nat, 493, 62

\bibitem[{{Irwin} \& {Hatzidimitriou}(1995)}]{IH95}
{Irwin}, M. \& {Hatzidimitriou}, D. 1995, \mnras, 277, 1354

\bibitem[{{Irwin} {et~al.}(2008){Irwin}, {Ferguson}, {Huxor}, {Tanvir},
  {Ibata}, \& {Lewis}}]{ifh08}
{Irwin}, M.~J., {Ferguson}, A.~M.~N., {Huxor}, A.~P., {Tanvir}, N.~R., {Ibata},
  R.~A., \& {Lewis}, G.~F. 2008, \apjl, 676, L17

\bibitem[{{Jacobs} {et~al.}(2012){Jacobs}, {Tully}, {Shaya}, \& {Rizzi}}]{bj12}
{Jacobs}, B., {Tully}, R.~B., {Shaya}, E.~J., \& {Rizzi}, L. 2012, in American
  Astronomical Society Meeting Abstracts, Vol. 219, American Astronomical
  Society Meeting Abstracts

\bibitem[{{Jacobs} {et~al.}(2009){Jacobs}, {Rizzi}, {Tully}, {Shaya},
  {Makarov}, \& {Makarova}}]{brad09}
{Jacobs}, B.~A., {Rizzi}, L., {Tully}, R.~B., {Shaya}, E.~J., {Makarov}, D.~I.,
  \& {Makarova}, L. 2009, \aj, 138, 332

\bibitem[{{Jacobs} {et~al.}(2011){Jacobs}, {Tully}, {Rizzi}, {Karachentsev},
  {Chiboucas}, \& {Held}}]{brad11}
{Jacobs}, B.~A., {Tully}, R.~B., {Rizzi}, L., {Karachentsev}, I.~D.,
  {Chiboucas}, K., \& {Held}, E.~V. 2011, \aj, 141, 106

\bibitem[{{Jerjen}(1995)}]{j95}
{Jerjen}, H. 1995, Ph.D.~Thesis

\bibitem[{{Karachentsev} {et~al.}(2011){Karachentsev}, {Kaisina}, {Kaisin}, \&
  {Makarova}}]{kkkm}
{Karachentsev}, I., {Kaisina}, E., {Kaisin}, S., \& {Makarova}, L. 2011,
  \mnras, 415, L31

\bibitem[{{Karachentsev} {et~al.}(2002){Karachentsev}, {Dolphin}, {Geisler},
  {Grebel}, {Guhathakurta}, {Hodge}, {Karachentseva}, {Sarajedini}, {Seitzer},
  \& {Sharina}}]{K2002}
{Karachentsev}, I.~D., {Dolphin}, A.~E., {Geisler}, D., {Grebel}, E.~K.,
  {Guhathakurta}, P., {Hodge}, P.~W., {Karachentseva}, V.~E., {Sarajedini}, A.,
  {Seitzer}, P., \& {Sharina}, M.~E. 2002, \aap, 383, 125

\bibitem[{{Klypin} {et~al.}(1999){Klypin}, {Kravtsov}, {Valenzuela}, \&
  {Prada}}]{klypin99}
{Klypin}, A., {Kravtsov}, A.~V., {Valenzuela}, O., \& {Prada}, F. 1999, \apj,
  522, 82

\bibitem[{{Larson}(1974)}]{larson74}
{Larson}, R.~B. 1974, \mnras, 169, 229

\bibitem[{{Loidl} {et~al.}(2001){Loidl}, {Lan{\c c}on}, \&
  {J{\o}rgensen}}]{loidl}
{Loidl}, R., {Lan{\c c}on}, A., \& {J{\o}rgensen}, U.~G. 2001, \aap, 371, 1065

\bibitem[{{Lynden-Bell}(1976)}]{lb76}
{Lynden-Bell}, D. 1976, \mnras, 174, 695

\bibitem[{{Makarov} {et~al.}(2006){Makarov}, {Makarova}, {Rizzi}, {Tully},
  {Dolphin}, {Sakai}, \& {Shaya}}]{mmrt06}
{Makarov}, D., {Makarova}, L., {Rizzi}, L., {Tully}, R.~B., {Dolphin}, A.~E.,
  {Sakai}, S., \& {Shaya}, E.~J. 2006, \aj, 132, 2729

\bibitem[{{Makarova} {et~al.}(2002){Makarova}, {Grebel}, {Karachentsev},
  {Dolphin}, {Karachentseva}, {Sharina}, {Geisler}, {Guhathakurta}, {Hodge},
  {Sarajedini}, \& {Seitzer}}]{mak02}
{Makarova}, L.~N., {Grebel}, E.~K., {Karachentsev}, I.~D., {Dolphin}, A.~E.,
  {Karachentseva}, V.~E., {Sharina}, M.~E., {Geisler}, D., {Guhathakurta}, P.,
  {Hodge}, P.~W., {Sarajedini}, A., \& {Seitzer}, P. 2002, \aap, 396, 473

\bibitem[{{Marigo} \& {Girardi}(2007)}]{mg07}
{Marigo}, P. \& {Girardi}, L. 2007, \aap, 469, 239

\bibitem[{{Marigo} {et~al.}(2008){Marigo}, {Girardi}, {Bressan}, {Groenewegen},
  {Silva}, \& {Granato}}]{marigo08}
{Marigo}, P., {Girardi}, L., {Bressan}, A., {Groenewegen}, M.~A.~T., {Silva},
  L., \& {Granato}, G.~L. 2008, \aap, 482, 883

\bibitem[{{Martin} {et~al.}(2006){Martin}, {Ibata}, {Irwin}, {Chapman},
  {Lewis}, {Ferguson}, {Tanvir}, \& {McConnachie}}]{martin06}
{Martin}, N.~F., {Ibata}, R.~A., {Irwin}, M.~J., {Chapman}, S., {Lewis}, G.~F.,
  {Ferguson}, A.~M.~N., {Tanvir}, N., \& {McConnachie}, A.~W. 2006, \mnras,
  371, 1983

\bibitem[{{Martin} {et~al.}(2013){Martin}, {Slater}, {Schlafly}, {Morganson},
  {Rix}, {Bell}, {Laevens}, {Bernard}, {Ferguson}, {Finkbeiner}, {Burgett},
  {Chambers}, {Hodapp}, {Kaiser}, {Kudritzki}, {Magnier}, {Morgan}, {Price},
  {Tonry}, \& {Wainscoat}}]{martin13}
{Martin}, N.~F., {Slater}, C.~T., {Schlafly}, E.~F., {Morganson}, E., {Rix},
  H.-W., {Bell}, E.~F., {Laevens}, B.~P.~M., {Bernard}, E.~J., {Ferguson},
  A.~M.~N., {Finkbeiner}, D.~P., {Burgett}, W.~S., {Chambers}, K.~C., {Hodapp},
  K.~W., {Kaiser}, N., {Kudritzki}, R.-P., {Magnier}, E.~A., {Morgan}, J.~S.,
  {Price}, P.~A., {Tonry}, J.~L., \& {Wainscoat}, R.~J. 2013, \apj, 772, 15

\bibitem[{{Mateo}(1998)}]{mm98}
{Mateo}, M.~L. 1998, \araa, 36, 435

\bibitem[{{McConnachie}(2012)}]{mcc}
{McConnachie}, A.~W. 2012, \aj, 144, 4

\bibitem[{{McConnachie} \& {Irwin}(2006)}]{mci06}
{McConnachie}, A.~W. \& {Irwin}, M.~J. 2006, \mnras, 365, 1263

\bibitem[{{Metz} {et~al.}(2008){Metz}, {Kroupa}, \& {Libeskind}}]{Metz08}
{Metz}, M., {Kroupa}, P., \& {Libeskind}, N.~I. 2008, \apj, 680, 287

\bibitem[{{Misgeld} \& {Hilker}(2011)}]{mh11}
{Misgeld}, I. \& {Hilker}, M. 2011, \mnras, 414, 3699

\bibitem[{{Moore} {et~al.}(1999){Moore}, {Ghigna}, {Governato}, {Lake},
  {Quinn}, {Stadel}, \& {Tozzi}}]{moore99}
{Moore}, B., {Ghigna}, S., {Governato}, F., {Lake}, G., {Quinn}, T., {Stadel},
  J., \& {Tozzi}, P. 1999, \apjl, 524, L19

\bibitem[{{Mouhcine} \& {Ibata}(2010)}]{mi10}
{Mouhcine}, M. \& {Ibata}, R. 2010, ArXiv e-prints

\bibitem[{{Pawlowski} {et~al.}(2012){Pawlowski}, {Pflamm-Altenburg}, \&
  {Kroupa}}]{Pawl12}
{Pawlowski}, M.~S., {Pflamm-Altenburg}, J., \& {Kroupa}, P. 2012, \mnras, 423,
  1109

\bibitem[{{Pietrinferni} {et~al.}(2004){Pietrinferni}, {Cassisi}, {Salaris}, \&
  {Castelli}}]{basti}
{Pietrinferni}, A., {Cassisi}, S., {Salaris}, M., \& {Castelli}, F. 2004, \apj,
  612, 168

\bibitem[{{Press} {et~al.}(1992){Press}, {Teukolsky}, {Vetterling}, \&
  {Flannery}}]{press92}
{Press}, W.~H., {Teukolsky}, S.~A., {Vetterling}, W.~T., \& {Flannery}, B.~P.
  1992, {Numerical recipes in FORTRAN. The art of scientific computing}
  (Cambridge: University Press, |c1992, 2nd ed.)

\bibitem[{{Ricotti}(2010)}]{ric09}
{Ricotti}, M. 2010, Advances in Astronomy, 2010

\bibitem[{{Rizzi} {et~al.}(2007){Rizzi}, {Tully}, {Makarov}, {Makarova},
  {Dolphin}, {Sakai}, \& {Shaya}}]{luca07}
{Rizzi}, L., {Tully}, R.~B., {Makarov}, D., {Makarova}, L., {Dolphin}, A.~E.,
  {Sakai}, S., \& {Shaya}, E.~J. 2007, \apj, 661, 815

\bibitem[{{Roychowdhury} {et~al.}(2012){Roychowdhury}, {Chengalur},
  {Chiboucas}, {Karachentsev}, {Tully}, \& {Kaisin}}]{sambit}
{Roychowdhury}, S., {Chengalur}, J.~N., {Chiboucas}, K., {Karachentsev}, I.~D.,
  {Tully}, R.~B., \& {Kaisin}, S.~S. 2012, \mnras, 426, 665

\bibitem[{{Schlafly} \& {Finkbeiner}(2011)}]{schlafly}
{Schlafly}, E.~F. \& {Finkbeiner}, D.~P. 2011, \apj, 737, 103

\bibitem[{{Schlegel} {et~al.}(1998){Schlegel}, {Finkbeiner}, \&
  {Davis}}]{sfd98}
{Schlegel}, D.~J., {Finkbeiner}, D.~P., \& {Davis}, M. 1998, \apj, 500, 525+

\bibitem[{{Secker} \& {Harris}(1997)}]{sh97}
{Secker}, J. \& {Harris}, W.~E. 1997, \pasp, 109, 1364

\bibitem[{{Shaya} \& {Tully}(2013)}]{st13}
{Shaya}, E.~J. \& {Tully}, R.~B. 2013, ArXiv e-prints

\bibitem[{{Simon} \& {Geha}(2007)}]{simon07}
{Simon}, J.~D. \& {Geha}, M. 2007, \apj, 670, 313

\bibitem[{{Thoul} \& {Weinberg}(1996)}]{tw96}
{Thoul}, A.~A. \& {Weinberg}, D.~H. 1996, \apj, 465, 608

\bibitem[{{Thuan} \& {Martin}(1981)}]{tm81}
{Thuan}, T.~X. \& {Martin}, G.~E. 1981, \apj, 247, 823

\bibitem[{{Trentham} {et~al.}(2005){Trentham}, {Sampson}, \& {Banerji}}]{tsb05}
{Trentham}, N., {Sampson}, L., \& {Banerji}, M. 2005, \mnras, 357, 783

\bibitem[{{Trentham} \& {Tully}(2002)}]{tt02}
{Trentham}, N. \& {Tully}, R.~B. 2002, \mnras, 335, 712

\bibitem[{{Tully} {et~al.}(2008){Tully}, {Shaya}, {Karachentsev}, {Courtois},
  {Kocevski}, {Rizzi}, \& {Peel}}]{tully08}
{Tully}, R.~B., {Shaya}, E.~J., {Karachentsev}, I.~D., {Courtois}, H.~M.,
  {Kocevski}, D.~D., {Rizzi}, L., \& {Peel}, A. 2008, \apj, 676, 184

\bibitem[{{Tully} \& {Trentham}(2008)}]{tt08}
{Tully}, R.~B. \& {Trentham}, N. 2008, \aj, 135, 1488

\bibitem[{{Tully}(2010)}]{tully10}
{Tully}, R.~B.~B. 2010, in Galaxies and their Masks, ed. D.~L. {Block}, K.~C.
  {Freeman}, \& I.~{Puerari}, 347

\bibitem[{{Walker}(2013)}]{mattw12}
{Walker}, M. 2013, {Dark Matter in the Galactic Dwarf Spheroidal Satellites},
  ed. T.~D. {Oswalt} \& G.~{Gilmore} (Dordrecht: Springer Science+Business
  Media, |c2013), 1039

\bibitem[{{Walsh} {et~al.}(2007){Walsh}, {Jerjen}, \& {Willman}}]{bootes07}
{Walsh}, S.~M., {Jerjen}, H., \& {Willman}, B. 2007, \apjl, 662, L83

\bibitem[{{Wang} {et~al.}(2012){Wang}, {Frenk}, {Navarro}, {Gao}, \&
  {Sawala}}]{wang12}
{Wang}, J., {Frenk}, C.~S., {Navarro}, J.~F., {Gao}, L., \& {Sawala}, T. 2012,
  \mnras, 424, 2715

\bibitem[{{Willman} {et~al.}(2005){Willman}, {Blanton}, {West}, {Dalcanton},
  {Hogg}, {Schneider}, {Wherry}, {Yanny}, \& {Brinkmann}}]{will1}
{Willman}, B., {Blanton}, M.~R., {West}, A.~A., {Dalcanton}, J.~J., {Hogg},
  D.~W., {Schneider}, D.~P., {Wherry}, N., {Yanny}, B., \& {Brinkmann}, J.
  2005, \aj, 129, 2692

\bibitem[{{Yun}(1999)}]{yun99}
{Yun}, M.~S. 1999, in IAU Symposium, Vol. 186, Galaxy Interactions at Low and
  High Redshift, ed. J.~E. {Barnes} \& D.~B. {Sanders}, 81

\bibitem[{{Yun} {et~al.}(1994){Yun}, {Ho}, \& {Lo}}]{yun94}
{Yun}, M.~S., {Ho}, P.~T.~P., \& {Lo}, K.~Y. 1994, \nat, 372, 530

\bibitem[{{Zucker} {et~al.}(2006){Zucker}, {Belokurov}, {Evans}, {Kleyna},
  {Irwin}, {Wilkinson}, {Fellhauer}, {Bramich}, {Gilmore}, {Newberg}, {Yanny},
  {Smith}, {Hewett}, {Bell}, {Rix}, {Gnedin}, {Vidrih}, {Wyse}, {Willman},
  {Grebel}, {Schneider}, {Beers}, {Kniazev}, {Barentine}, {Brewington},
  {Brinkmann}, {Harvanek}, {Kleinman}, {Krzesinski}, {Long}, {Nitta}, \&
  {Snedden}}]{zucker06}
{Zucker}, D.~B., {Belokurov}, V., {Evans}, N.~W., {Kleyna}, J.~T., {Irwin},
  M.~J., {Wilkinson}, M.~I., {Fellhauer}, M., {Bramich}, D.~M., {Gilmore}, G.,
  {Newberg}, H.~J., {Yanny}, B., {Smith}, J.~A., {Hewett}, P.~C., {Bell},
  E.~F., {Rix}, H.-W., {Gnedin}, O.~Y., {Vidrih}, S., {Wyse}, R.~F.~G.,
  {Willman}, B., {Grebel}, E.~K., {Schneider}, D.~P., {Beers}, T.~C.,
  {Kniazev}, A.~Y., {Barentine}, J.~C., {Brewington}, H., {Brinkmann}, J.,
  {Harvanek}, M., {Kleinman}, S.~J., {Krzesinski}, J., {Long}, D., {Nitta}, A.,
  \& {Snedden}, S.~A. 2006, \apjl, 650, L41

\end{thebibliography}

\clearpage


\begin{deluxetable}{rrrrrrrr}
\tabletypesize{\scriptsize}
\tablewidth{0pt}
\tablecaption{Observation Summary\label{tabobs}}
\tablehead{
\colhead{Galaxy} &
\colhead{Camera} &
\colhead{$F606_{texp}$ (s)\tablenotemark{a}} &
\colhead{$F814_{texp}$ (s)\tablenotemark{a}} &
\colhead{$\alpha$} &
\colhead{$\delta$ (J2000.0)} &
\colhead{PGC \#} &
\colhead{Other} \\
}

\startdata

d0926+70 & WFPC2 & 1000 & 1000 & 09 26 27.9 & +70 30 24 & 5056943 &  \\
d0934+70 & WFPC2 & 1000 & 1000 & 09 34 03.7 & +70 12 57 & 5056931 &  \\
d0946+68 & WFPC2 & 1000 & 900 & 09 46 13.0 & +68 42 55 & &   \\
d0955+70 & WFPC2 & 1000 & 1000 & 09 55 13.6 & +70 24 29 & 5056934 &  \\
d0957+70 & WFPC2 & 1000 & 1000 & 09 57 12.4 & +70 12 35 &  \\
d0958+66 & WFPC2 & 1000 & 900 & 09 58 48.5 & +66 50 59 & 28826 & KUG 0945+670 \\
d0959+68\tablenotemark{b} & WFPC2   & 1000 & 900 & 09 59 33.1 & +68 39 25 & 5056936 &  \\
d1006+67 & WFPC2 & 1000 & 900 & 10 06 46.2 & +67 12 04 & 5056937 &  \\
d1009+70 & WFPC2 & 1000 & 1000 & 10 09 34.9 & +70 32 55 & &   \\
d1014+68 & WFPC2 & 1000 & 900 & 10 14 55.8 & +68 45 27 & 5056938 &  \\
d1016+69 & WFPC2 & 1000 & 900 & 10 16 18.3 & +69 29 45 & &   \\
d1028+70 & WFPC2 & 1000 & 1000 & 10 28 39.7 & +70 14 01 & 5056941 &  \\
d1041+70 & WFPC2 & 1000 & 1000 & 10 41 16.8 & +70 09 03 & 5056942 &  \\
d1048+70 & WFPC2 & 1000 & 1000 & 10 48 57.0 & +70 25 38 & &   \\
d0939+71 & ACS   & 1300 & 1250 & 09 39 15.9 & +71 18 42 & 5056932 &  \\
d0944+69 & ACS   & 1280 & 1230 & 09 44 22.5 & +69 12 40 & 5056933 &  \\
d0944+71 & ACS   & 1300 & 1250 & 09 44 34.4 & +71 28 57 & 5056944 &  \\
d0959+68\tablenotemark{b} & ACS   & 1280 & 1230 & 09 59 33.1 & +68 39 25 & 5056936 &  \\
d1012+64 & ACS   & 1250 & 1160 & 10 12 48.4 & +64 06 27 & 29735 & UGC 5497 \\
d1013+68 & ACS   & 1280 & 1230 & 10 13 11.7 & +68 43 45 & &   \\
d1015+69 & ACS   & 1280 & 1230 & 10 15 06.9 & +69 02 15 & 5056947 &  \\
d1019+69 & ACS   & 1280 & 1230 & 10 19 52.9 & +69 11 19 & &   \\
d1020+69 & ACS   & 1280 & 1230 & 10 20 25.0 & +69 11 50 & &   \\

\enddata
                                                                                
\tablenotetext{a}{Total integration from two exposures}
\tablenotetext{b}{Observed with both WFPC2 and ACS}

\end{deluxetable}




\begin{deluxetable}{rrrrrrrrrrrrrrl}
\rotate {}
\tabletypesize{\scriptsize}
\tablewidth{0pt}
\tablecaption{Distances and Photometric Properties of M81 Group Members\label{tabX}}
\tablehead{
\colhead{Galaxy} &
\colhead{Cam} &
\colhead{$r^{\prime}_{cor_{s}}$\tablenotemark{a}} &
\colhead{$\mu_{\circ}$\tablenotemark{a}} &
\colhead{$\langle \mu_{e} \rangle$\tablenotemark{a}} &
\colhead{E$_{B-V}$\tablenotemark{b}} &
\colhead{TRGB$_{meas}$} &
\colhead{TRGB$_o$} &
\colhead{$(V-I)_{tip}$} &
\colhead{M$_{tip}$\tablenotemark{c}} &
\colhead{DM} &
\colhead{D} &
\colhead{M$_{r^{\prime}_{cor_{s}}}$\tablenotemark{d}} &
\colhead{M$_I$\tablenotemark{e}} &
\colhead{Type} \\
\colhead{} &
\colhead{} &
\colhead{} &
\colhead{$r^{\prime}$} &
\colhead{$r^{\prime}$} &
\colhead{} &
\colhead{$F814W$} &
\colhead{} &
\colhead{} &
\colhead{$F814W$} &
\colhead{} &
\colhead{(Mpc)} &
\colhead{} &
\colhead{} &
\colhead{}  \\
 }

\startdata

d0926+70 & W & 17.9 & 26.0 & 26.2 &  0.19 & $23.92^{0.12}_{0.15}$ & $23.63^{0.12}_{0.15}$ & $1.00\pm0.07$ &  $-4.03$ & $27.66^{0.12}_{0.15}$ & $3.4^{0.2}_{0.2}$ & $-9.8^{0.5}_{0.5}$ & $-11.4$ & dI/dSph \\
d0934+70 & W & 18.0 & 26.2 & 26.4 &  0.25 & $23.77^{0.70}_{0.08}$ & $23.38^{0.70}_{0.08}$ & $1.00\pm0.07$ &  $-4.02$ & $27.40^{0.70}_{0.08}$ & $3.0^{1.1}_{0.2}$ & $-9.4^{0.5}_{0.9}$ & $-10.5$ & dSph \\
d0939+71 & A & 18.5 & 25.5 & 26.1 &  0.04 & $23.83^{0.32}_{0.22}$ & $23.77^{0.32}_{0.22}$ & $1.02\pm0.03$ &  $-4.10$ & $27.87^{0.32}_{0.22}$ & $3.7^{0.5}_{0.4}$ & $-9.4^{0.6}_{0.4}$ & $-9.6$ & dSph \\
d0944+69 & A & 21.1 & 25.9 & 26.7 &  0.09 & $23.93^{0.37}_{0.19}$ & $23.79^{0.37}_{0.19}$ & $1.05\pm0.05$ &  $-4.10$ & $27.89^{0.37}_{0.19}$ & $3.8^{0.6}_{0.3}$ & $-6.8^{0.5}_{0.6}$ & $-9.1$ & dSph \\
d0944+71 & A & 15.2 & 23.4 & 23.9 &  0.03 & $23.60^{0.06}_{0.06}$ & $23.55^{0.06}_{0.06}$ & $1.13\pm0.07$ &  $-4.08$ & $27.63^{0.06}_{0.06}$ & $3.4^{0.1}_{0.1}$ & $-12.4^{0.8}_{0.8}$ & $-13.2$ & dSph \\
d0955+70 & W & 17.9 & 26.6 & 27.1 &  0.16 & $23.92^{0.35}_{0.31}$ & $23.67^{0.35}_{0.31}$ & $1.05\pm0.07$ & $-4.02$  & $27.69^{0.35}_{0.31}$ & $3.4^{0.6}_{0.4}$ & $-9.8^{0.6}_{0.6}$  & $-10.6$ & dSph \\
d0958+66 & W & 14.7 &  21.6 & 22.3 &  0.07 & $23.98^{0.05}_{0.05}$ & $23.88^{0.05}_{0.05}$ & $1.00\pm0.07$ &  $-4.03$ & $27.91^{0.06}_{0.06}$ & $3.8^{0.1}_{0.1}$ & $-13.2^{0.5}_{0.5}$ & $-13.7$ & BCD \\
d0959+68\tablenotemark{f} & A & 16.2 & 26.0 & 25.5 &  0.08 & $24.13^{0.15}_{0.15}$ & $24.01^{0.15}_{0.15}$ & $1.10\pm0.10$ &  $-4.09$ & $28.10^{0.15}_{0.15}$ & $4.2^{0.3}_{0.3}$ & $-11.9^{0.8}_{0.8}$ & $-11.8$ & dI/tdl \\
d1006+67 & W & 18.0 & 26.2 & 26.5 &  0.06 & $23.89^{0.14}_{0.12}$ & $23.79^{0.14}_{0.12}$ & $1.12\pm0.04$ & $-4.01$  & $27.80^{0.14}_{0.12}$ & $3.6^{0.2}_{0.2}$ & $-9.8^{0.5}_{0.5} $  & $-10.3$ & dSph \\
d1012+64 & A & 14.5 & 21.5 & 22.3 &  0.02 & $23.77^{0.05}_{0.05}$ & $23.74^{0.05}_{0.05}$ & $1.10\pm0.04$ & $-4.09$  & $27.83^{0.06}_{0.06}$ & $3.7^{0.1}_{0.1}$ & $-13.3^{0.5}_{0.5}$ & $-13.9$ &  BCD \\
d1014+68 & W & 18.5 &  27.2 & 27.3 &  0.05 & $23.97^{0.18}_{0.20}$ & $23.89^{0.18}_{0.20}$ & $1.00\pm0.07$ &  $-4.03$ & $27.92^{0.18}_{0.20}$ & $3.8^{0.3}_{0.3}$ & $-9.4^{0.5}_{0.5}$ & $-10.5$ & dSph \\
d1015+69 & A & 19.1 & 25.5 & 26.0 &  0.05 & $23.91^{0.14}_{0.12}$ & $23.83^{0.14}_{0.12}$ & $1.00\pm0.04$ &  $-4.11$ & $27.94^{0.14}_{0.12}$ & $3.9^{0.3}_{0.2}$ & $-8.8^{0.6}_{0.6}$ & $-9.5$ &  dSph \\
d1028+70 & W & 15.5 & 22.4 & 23.3 &  0.04 & $23.96^{0.05}_{0.05}$ & $23.90^{0.05}_{0.05}$ & $1.03\pm0.05$ &  $-4.02$ & $27.92^{0.06}_{0.06}$ & $3.8^{0.1}_{0.1}$ & $-12.4^{0.5}_{0.5}$ & $-13.1 $ & BCD \\
d1041+70 & W & 18.5 & 26.1 & 26.2 &  0.06 & $23.91^{0.14}_{0.16}$ & $23.82^{0.14}_{0.16}$ & $1.03\pm0.08$ &  $-4.02$ & $27.84^{0.14}_{0.16}$ & $3.7^{0.2}_{0.3}$ & $-9.3^{0.5}_{0.5}$ & $-10.7$ &  dSph \\

\enddata
                                                                                
\tablenotetext{a}{Magnitude and surface brightness measurements from CFHT/MegaCam r' imaging \citep{Ch1}}
\tablenotetext{b}{From \citet{sfd98} dust maps; this was converted to extinction in each bandpass using the \citet{schlafly} recalibration of the dust maps.}
\tablenotetext{c}{Absolute magnitude of the tip of the RGB using the calibration from Rizzi et al. (2007)}
\tablenotetext{d}{Integrated magnitude from curve of growth fitting with cumulative Sersic function based on MegaCam imaging, extinction corrected \citep{Ch1}}
\tablenotetext{e}{Estimated magnitudes assuming model stellar populations for each galaxy, see text. Uncertainties are typically 0.4 mag for ACS data and 0.6 mag for WFPC2.}
\tablenotetext{f}{TRGB measurements may apply only to tidal stream stars and not to the star forming 
concentration proposed to be a tidal dwarf.}

\end{deluxetable}


\begin{deluxetable}{lrrrrl}
\tabletypesize{\scriptsize}
\tablewidth{0pt}
\tablecaption{Projected and physical separation of group galaxies from M81\label{M81sep}}
\tablehead{
\colhead{Galaxy} &
\colhead{Projected} &
\colhead{Physical} &
\colhead{M$_{r^{\prime}}$} &
\colhead{R$_e$} &
\colhead{Type} \\
\colhead{} &
\colhead{deg} &
\colhead{Mpc} &
\colhead{} &
\colhead{kpc} &
\colhead{} \\
}

\startdata
M81        &   0.00 & 0.00  & -20.6 & 1.45 & Sb   \\
KDG61      &   0.49 & 0.043 & -13.3 & 0.68 & dSph   \\
BK5N       &   1.16 & 0.071 & -11.7 & 0.40 & dSph   \\
N2976      &   1.38 & 0.094 & -18.0 & 0.77 & Sc/pec   \\
IKN        &   1.32 & 0.095 &(-12.4)&      & dSph   \\
M82        &   0.62 & 0.099 & -19.9 & 1.05 & Irr   \\
FM1        &   0.98 & 0.101 & -11.7 & 0.39 & dSph   \\
KDG64      &   1.63 & 0.119 & -13.2 & 0.40 & dSph   \\
F8D1       &   1.87 & 0.127 & -13.2 & 1.22 & dSph   \\
d0944+69   &   1.01 & 0.127 & -6.8  & 0.09 & dSph   \\
Garlnd     &   0.82 &(0.135)&       &      & pec/tdl   \\
d1014+68   &   1.77 & 0.156 & -9.4  & 0.37 & dSph   \\
HolmIX     &   0.18 &(0.156)& -13.9 & 0.87 & dI/tdl   \\
KK77       &   1.64 & 0.158 & -13.0 & 0.68 & dSph   \\
d1006+67   &   2.14 & 0.163 & -9.8  & 0.32 & dSph   \\
d0939+71   &   2.63 & 0.173 & -9.4  & 0.27 & dSph   \\
KDG63      &   2.67 & 0.177 & -13.0 & 0.44 & dSph   \\
d0958+66   &   2.24 & 0.179 & -13.2 & 0.22 & BCD   \\
N3077      &   0.78 & 0.181 & -17.8 & 0.70 & Irr   \\
DDO78      &   3.18 & 0.227 &(-12.8)& 0.43 & dSph   \\
d1028+70   &   3.10 & 0.228 & -12.4 & 0.25 & BCD   \\
d1015+69   &   1.75 & 0.237 & -8.8  & 0.16 & dSph   \\
A0952      &   0.27 &(0.238)&       &      & pec/tdl   \\
d0955+70   &   1.34 & 0.260 & -9.8  & 0.31 & dSph   \\
d1041+70   &   4.12 & 0.265 & -9.3  & 0.19 & dSph   \\
HS117      &   3.01 & 0.308 & -12.1 & 0.38 & dI   \\
I2574      &   3.04 & 0.322 & -17.7 & 2.71 & SABm   \\
DDO82      &   3.39 & 0.329 & -15.1 & 0.65 & Im/BCD   \\
d0944+71   &   2.59 & 0.335 & -12.4 & 0.30 & dSph   \\
d1012+64   &   5.24 & 0.337 & -13.5 & 0.25 & BCD   \\
d0926+70   &   2.89 & 0.345 & -9.8  & 0.24 & dI/dSph   \\
HolmI      &   2.48 & 0.363 & -14.6 & 1.30 & dI   \\
BK3N       &   0.18 &(0.475)& -10.3 & 0.14 & dI/tdl   \\
BK6N       &   4.81 & 0.488 & -11.7 & 0.27 & dSph   \\
d0959+68   &   0.54 &(0.508)& -11.9 & 0.54 & dI/tdl   \\
d0934+70   &   2.19 &(0.705)& -9.4  & 0.22 & dSph   \\
                                                                                          
\enddata

\tablecomments{Values listed in parenthesis are highly uncertain, especially
for suspected tidal dwarfs where distances measured from RGB stars may apply to 
nearby galaxies/tidal streams rather than the tidal dwarf itself.} 
                                                                                          
\end{deluxetable}



\begin{deluxetable}{rlrrrrrrrrrrrr}
\rotate {}
\tabletypesize{\scriptsize}
\tablewidth{0pt}
\tablecaption{Structural Properties of M81 Group Members\label{tabS}}
\tablehead{
\colhead{Galaxy} &
\colhead{Cam} &
\colhead{{$\epsilon_{CFHT}$}\tablenotemark{a}} &
\colhead{$\epsilon_{HST}$\tablenotemark{b}} &
\colhead{{PA$_{CFHT}$}\tablenotemark{a}} &
\colhead{PA$_{HST}$\tablenotemark{b}} &
\colhead{$\alpha$} &
\colhead{$\delta$\tablenotemark{c} } &
\colhead{R$_{e, CFHT}$\tablenotemark{d}} &
\colhead{R$_{e, CFHT}$} &
\colhead{n$_{CFHT}$} &
\colhead{R$_{e, HST}$\tablenotemark{e}} &
\colhead{R$_{e, HST}$} &
\colhead{n$_{HST}$} \\
\colhead{} &
\colhead{} &
\colhead{} &
\colhead{} &
\colhead{} &
\colhead{} &
\colhead{(J2000.0)} &
\colhead{(J2000.0)} &
\colhead{arcsec} &
\colhead{kpc} &
\colhead{} &
\colhead{arcsec} &
\colhead{kpc} &
\colhead{} \\
 }

\startdata

d0939+71 & A &  0.05 &  0.15 &  -80.9 &    3.1  &     9:39:16.01 &  71:18:41.00  &    15.3 &  0.27 & 0.66 &   $8.9\pm0.9$ & $0.16\pm0.02$ & 0.75 \\
d0944+69 & A &  0.09 &  0.04 &   -9.7 &   34.3  &     9:44:22.49 &  69:12:37.94  &    4.9  &  0.09 & 0.78 &  $7.2\pm1.3$ & $0.13\pm0.05$ & 1.28 \\
d0944+71 & A &  0.20 &  0.11 &  -27.4 &   3.6  &     9:44:34.37 &  71:28:55.60  &    18.3 &  0.30 & 0.61 &  $21.4\pm0.4$ & $0.35\pm0.02$ & 0.67 \\
d0959+68 & A &  0.10 &  0.22 &   35.1 &   73.5  &     9:59:34.90 &  68:39:25.78  &    26.6 &  0.48\tablenotemark{f} & 0.21  & $20.8\pm2.8$ & $0.37\pm0.08$\tablenotemark{f} & 0.34 \\
d1012+64 & A &  0.15 &  0.14 &   17.0 &    4.4  &    10:12:48.41 &  64:06:26.21  &    13.9 &  0.25 & 0.88 &  $14.9\pm0.8$ & $0.31\pm0.02$ & 0.67 \\
d1015+69 & A &  0.07 &  0.17 &  -49.7 &  -41.0  &    10:15:06.89 &  69:02:13.81  &     8.5 &  0.16 & 0.68 &  $9.6\pm1.6$ & $0.18\pm0.05$ & 0.99 \\
d0926+70 & W & 0.18  &  0.26 &  -23.9 &    0.3  &     9:26:27.94 &  70:30:18.79  &    14.3 &  0.24 & 0.42 &       &      & \\
d0934+70 & W & 0.07  &  0.15 &   63.5 &  -82.9  &     9:34:03.22 &  70:12:58.32  &    15.3 &  0.22 & 0.42  &       &      & \\
d0955+70 & W & 0.04  &  0.08 &   86.8 &   84.8  &     9:55:14.14 &  70:24:25.96  &    18.6 &  0.31 & 0.61  &       &      & \\
d0958+66 & W & 0.36  &  0.23 &   -6.9 &   -7.5  &     9:58:48.74 &  66:50:57.34  &    11.6 &  0.22 & 0.71 &       &      & \\
d1006+67 & W & 0.05  &  0.17 &   86.1 &   50.2  &    10:06:46.80 &  67:11:59.78  &    18.1 &  0.32 & 0.44 &       &      & \\
d1014+69 & W & 0.01  &  0.08 &   -8.9 &  -34.6  &    10:14:55.80 &  68:45:29.71  &    19.7 &  0.37 & 0.29  &       &      & \\
d1028+70 & W & 0.20  &  0.20 &  -54.8 &  -67.0  &    10:28:39.98 &  70:13:59.84  &    13.4 &  0.25 & 0.88  &       &      & \\
d1041+70 & W & 0.18  &  0.28 &   26.0 &   26.0  &    10:41:18.14 &  70:09:13.75  &    10.9 &  0.19 & 0.30  &       &      & \\

\enddata
                                                                                

\tablenotetext{a}{Measurements based on the second moments of the light distribution in MegaCam $r^\prime$ imaging \citep{Ch1}, including both resolved and unresolved components.}
\tablenotetext{b}{Measurements based on the second moments of the distribution of resolved RGB stars only.}
\tablenotetext{c}{Centroid based on the resolved RGB population in HST imaging.}
\tablenotetext{d}{From integrated Sersic profile fits to curves of growth in MegaCam $r^\prime$ imaging \citep{Ch1}. Values listed are the geometric mean of the semi-major and -minor axes.}
\tablenotetext{e}{Based on Sersic profile fits to number counts in HST $F814W$-band imaging of resolved stars likely to be galaxy members. This assumes the mean brightness of stars is independent of radius, as expected for RGB stars.  Measurements are for data taken with ACS only. Radii listed are the geometric mean of the major and minor
axes.}
\tablenotetext{f}{For D = 3.69 Mpc. Actual TRGB measured distance is 4.2 Mpc.}

\end{deluxetable}


\begin{deluxetable}{rrrrrrrrrr}
\rotate {}
\tabletypesize{\scriptsize}
\tablewidth{0pt}
\tablecaption{Summary of Stellar Populations\label{Tstellpop}}
\tablehead{
\colhead{Name} &
\colhead{Instr.} &
\colhead{M$_{r^{\prime}}$} &
\colhead{R$_e$} &
\colhead{Type} &
\colhead{Age Young} &
\colhead{Age Int.} &
\colhead{AGB/RGB\tablenotemark{a}} &
\colhead{Z$_{RGB}$\tablenotemark{b}} &
\colhead{Notes\tablenotemark{c}} \\
\colhead{} &
\colhead{} &
\colhead{} &
\colhead{(kpc)} &
\colhead{} &
\colhead{(Myr)} &
\colhead{AGB (Gyr)} &
\colhead{(Gyr)} &
\colhead{} &
\colhead{} \\
}
\startdata
d1012+64 & A & -13.3 & 0.25 & BCD     & $\bf{< 60}$ &  \bf{0.5 - 4}  &  5-8 &  0.0004 - 0.0018  & M$_{HI} = 1.04\times10^6$ M$_\odot$, v$_r = 190$ km s$^{-1}$  \\
d0958+66 & W & -13.2 & 0.22 & BCD     & $\bf{\geq 60}$ & \bf{0.5 - 4} & $\sim3$  &  0.0001 - 0.003 & M$_{HI} = 0.73\times10^6$ M$_\odot$, v$_r = 90$ km s$^{-1}$ \\
d1028+70 & W & -12.4 & 0.25 & BCD     & $\bf{< 60}$ & \bf{1-4} & 5-9   &  0.0001 - 0.0023 & M$_{HI} = 1.3\times10^6$ M$_\odot$, v$_r = -69$ km s$^{-1}$ \\
d0944+71 & A & -12.4 & 0.30 & dSph    & ($\geq 150$) &  ($\sim2$) &  old  & 0.0005 - 0.002 & M$_{HI} < 0.31\times10^6$ M$_\odot$ \\
d0959+68 & A & -11.9 & 0.48 & dI/tdl  & $\bf{< 80}$ &       &     &  0.0001 - 0.0035 &  \\
d0926+70 & W & -9.8  & 0.24 & dI/dSph & ($\geq 100$) & $\bf{\geq 0.5}$ & 4-7  &  0.0001 - 0.0023 & M$_{HI} < 0.31\times10^6$ M$_\odot$ \\
d0955+70 & W & -9.8  & 0.31 & dSph    &        &      &        &   0.0001 - 0.003 & \\
d1006+67 & W & -9.8  & 0.32 & dSph    &        &      &      &  0.0004 - 0.004 &  \\
d0934+70 & W & -9.4  & 0.22 & dSph    &        & ($\sim1$)&  & 0.0001 - 0.004 &  \\
d0939+71 & A & -9.4  & 0.27 & dSph    &        &      &       & 0.0002 - 0.001 &  \\
d1014+68 & W & -9.4  & 0.37 & dSph    &        &   ($\sim1$) &     & 0.0001 - 0.003 &  \\
d1041+70 & W & -9.3  & 0.19 & dSph    &        &      &     & 0.0001 - 0.003 &  \\            
d1015+69 & A & -8.8  & 0.16 & dSph    &        &      &       &  0.0001 - 0.001 &  \\
d0944+69 & A & -6.8  & 0.09 & dSph    &        &      &       & 0.0003 - 0.0018 & \\

\enddata

\tablecomments{Values listed for ages and metallicities are approximate ranges determined from Padova model isochrones, see text.  
Prominent young or intermediate populations are denoted in boldface, while more tentative identifications are listed in
parentheses.}
\tablenotetext{a}{Composite age estimate for the older populations, see text. These estimates are best used as relative age indicators.}
\tablenotetext{b}{The range of metallicities which span the $\pm 1\sigma$ width of the RGB, assuming a 12.5 Gyr age for the old stellar population.}
\tablenotetext{c}{HI measurements taken from \citet{sambit}}
                                                                                          
\end{deluxetable}

\begin{deluxetable}{lrr}
\tabletypesize{\scriptsize}
\tablewidth{0pt}
\tablecaption{Faint-end slope measurements\label{LFtab}}
\tablehead{
\colhead{Sample} &
\colhead{$\alpha$ ($r^{\prime}_{diff}$)} &
\colhead{$\alpha$ ($r^{\prime}_{cum}$)} \\
}
\startdata
Previously known & $-1.23^{+0.08}_{-0.05}$  & $-1.25^{+0.06}_{-0.06}$ \\
All ($M_{r^{\prime}} < -10$) & $-1.26^{+0.05}_{-0.04}$ &  $-1.28^{+0.04}_{-0.05}$ \\
All (completeness corr.) & $-1.27^{+0.04}_{-0.04}$ & $-1.26^{+0.04}_{-0.04}$ \\
                                                                                          
\enddata
                                                                                          
\end{deluxetable}

\clearpage


\clearpage

\clearpage

\clearpage

\clearpage

\clearpage 

\clearpage

\end{document}